\documentclass[aps,showpacs,nofootinbib,preprintnumbers,amsmath,amssymb]{revtex4}

\usepackage{graphicx}
\usepackage{dcolumn}
\usepackage{bm}
\usepackage{epsfig}
\usepackage[usenames]{color}
\usepackage{hyperref} 

\def\eq#1{{Eq.~(\ref{#1})}}
\def\fig#1{{Fig.~\ref{#1}}}
\newcommand{\ben}{\begin{eqnarray*}}
\newcommand{\een}{\end{eqnarray*}}

\newcommand{\amu}{\alpha_\mu}

\newcommand{\stackeven}[2]{{{}_{\displaystyle{#1}}\atop\displaystyle{#2}}}

\newcommand{\gsim}{\stackeven{>}{\sim}}
\newcommand{\as}{\alpha_s}

\newcommand{\bas}{{\bar\alpha}_s}
\newcommand{\bam}{{\bar\alpha}_\mu}

\newcommand{\dhd}{{\textstyle d}
\lower.03ex\hbox{\kern-0.38em$^{\scriptstyle-}$}\kern-0.05em{}}
\newcommand{\dbar}{{\textstyle \delta}
\lower.03ex\hbox{\kern-0.38em$^{\scriptstyle-}$}\kern-0.05em{}}
\newcommand{\half}{{1\over 2}}

\newcommand{\call}{{\cal L}}

\newcommand{\calu}{{\cal U}}

\begin{document}

\title{$\gamma^* \gamma^*$ Cross Section at NLO and Properties of the
  BFKL Evolution at Higher Orders}

\author{Giovanni~A.~Chirilli,\footnote{chirilli.1@asc.ohio-state.edu}
  Yuri~V.~Kovchegov\footnote{kovchegov.1@asc.ohio-state.edu}}

\affiliation{Department of Physics, The Ohio State University,
  Columbus, OH 43210, USA}

\begin{abstract}
  We obtain a simple analytic expression for the high energy $\gamma^*
  \gamma^*$ scattering cross section at the next-to-leading order in
  the logarithms-of-energy power counting. To this end we employ the
  eigenfunctions of the NLO BFKL equation constructed in our previous
  paper. We also construct the eigenfunctions of the NNLO BFKL kernel
  and obtain a general form of the solution for the NNLO BFKL
  equation, which confirms the ansatz proposed in our previous paper.
\end{abstract}


\maketitle



\section{Introduction}

Understanding the high energy dynamics of strong interactions is
important both for the description of the strong interaction data
reported by the current and future accelerators worldwide, and to
improve our theoretical understanding of quantum chromodynamics
(QCD). Significant progress towards achieving these goals has been
accomplished in the recent decades by the physics of saturation and Color
Glass Condensate (CGC) (see 
\cite{Balitsky:2001gj, Jalilian-Marian:2005jf,Weigert:2005us,Iancu:2003xm,Gelis:2010nm,KovchegovLevin}
for reviews). This approach is based on the existence of an intrinsic
hard momentum scale which characterizes hadronic and nuclear wave
functions at high energy --- the saturation scale $Q_s$
\cite{Gribov:1984tu}. At small values of Bjorken $x$ and/or for large
nuclei the saturation scale $Q_s$ is much larger than the QCD
confinement scale, justifying the use of the small-coupling expansion
since $\as (Q_s^2) \ll 1$, and allowing for first-principles
calculations of many high energy scattering cross sections and of
other related observables.

Calculations of observables in the saturation/CGC framework consist of
two steps: (i) first one has to calculate the observables in the
quasi-classical Glauber--Mueller (GM) \cite{Mueller:1989st} /
McLerran--Venugopalan (MV) approximation
\cite{McLerran:1993ni,McLerran:1993ka,McLerran:1994vd} and then (ii)
evolve the result with energy $s$ using the non-linear
Balitsky--Kovchegov (BK)
\cite{Balitsky:1996ub,Balitsky:1998ya,Kovchegov:1999yj,Kovchegov:1999ua}
or Jalilian-Marian--Iancu--McLerran--Weigert--Leonidov--Kovner
(JIMWLK)
\cite{Jalilian-Marian:1997dw,Jalilian-Marian:1997gr,Iancu:2001ad,Iancu:2000hn}
evolution equations. It appears that to successfully describe and,
sometimes, predict a host of deep inelastic scattering (DIS), hadronic
and nuclear collisions data (see
e.g. \cite{Albacete:2010sy,ALbacete:2010ad,Lappi:2013zma}) one needs
to include the running-coupling corrections
\cite{Balitsky:2006wa,Gardi:2006rp,Kovchegov:2006vj,Kovchegov:2006wf,Albacete:2007yr}
into the leading-$\ln s$ BK and JIMWLK evolution
equations. (Henceforth we will refer to the leading-$\ln s$ BK and
JIMWLK equations either as leading-order (LO) BK and JIMWLK
equations.)

The next-to-leading order (NLO) corrections to both the BK
\cite{Balitsky:2008zz} and JIMWLK
\cite{Balitsky:2013fea,Kovner:2013ona} evolution equations are known,
but have not yet been implemented in phenomenological applications. At
the same time keeping the higher-order corrections under theoretical
control is essential for improving the precision of the saturation/CGC
physics predictions. Perhaps more importantly, an agreement of any
theory with the data is established only once the uncertainty due to
higher-order corrections is understood and quantified, demonstrating
explicitly that higher-order corrections do not significantly affect
the existing agreement of the theory with the data. It is, therefore,
very important to understand both the structure and magnitude of the
higher-order corrections to the small-$x$ evolution equations.

Since no exact analytic solutions for the nonlinear LO BK and JIMWLK
equations are known, it is easier to start organizing the higher-order
corrections for the linear Balitsky--Fadin--Kuraev--Lipatov (BFKL)
\cite{Kuraev:1977fs,Kuraev:1976ge,Balitsky:1978ic} evolution equation,
which predated both the BK and JIMWLK equations by over two decades:
in the linear (low parton density) regime both the BK and JIMWLK
equations reduce to BFKL. The LO BFKL equation's solution is
well-known and was constructed almost simultaneously with the equation
itself \cite{Balitsky:1978ic}. The kernel of the BFKL equation is
known up to the next-to-leading order
\cite{Fadin:1998py,Ciafaloni:1998gs}. In the fifteen years since the
construction of the NLO BFKL equation numerous efforts have been made
to understand the features of its solution: non-Regge behavior was
found in the amplitude resulting from the NLO BFKL equation
\cite{Kovchegov:1998ae,Levin:1998pka,Armesto:1998gt,Ciafaloni:2001db};
it was also shown that the NLO correction to the BFKL intercept is
large and negative
\cite{Fadin:1998py,Ross:1998xw,Andersen:2003an,Andersen:2003wy},
though higher-order collinear divergences are likely to reduce the
size of this correction
\cite{Salam:1998tj,Ciafaloni:1999yw,Ciafaloni:2003rd}.

More recently the eigenfunctions of the NLO BFKL kernel were
constructed in our previous paper \cite{Chirilli:2013kca} by using a
perturbative expansion around the eigenfunctions of the LO BFKL
kernel. This method yields a systematic way of constructing any-order
BFKL kernel eigenfunctions, therefore allowing one to find the
solution of any-order BFKL equation. (See \cite{Grabovsky:2013gta} for
an alternative, though perhaps related way of solving the NLO BFKL
equation.) In the general form of the BFKL solution found in
\cite{Chirilli:2013kca} the higher-order corrections enter through the
perturbative expansions in the intercept {\sl and} in the
eigenfunctions. This is an essential difference between the BFKL
solution in QCD and in a conformal field theory, such as the ${\cal N}
=4$ super-Yang-Mills (SYM) theory: in the latter, conformal symmetry
uniquely fixes the eigenfunctions of the BFKL kernel to be the
eigenfunctions of the Casimir operators of the M\"{o}bius group,
$E^{n, \nu}$ \cite{Lipatov:1985uk}, such that the higher-order
corrections enter the SYM BFKL solution only through perturbative
expansion in the intercept
\cite{Kotikov:2000pm,Balitsky:2008rc,Balitsky:2009xg,Balitsky:2009yp}. Note
also a similar difference from the solution of the
Dokshitzer--Gribov--Lipatov--Altarelli--Parisi (DGLAP) evolution
equation \cite{Dokshitzer:1977sg,Gribov:1972ri,Altarelli:1977zs},
where the higher-order corrections in the strong coupling $\as$ only
enter in the anomalous dimensions of the quark and gluon distribution
operators
\cite{Gross:1973id,Georgi:1951sr,Christ:1972ms,Altarelli:1977zs,Floratos:1977au,Vogt:2004mw}.
The eigenfunctions found in \cite{Chirilli:2013kca} allowed us to
construct a simple form of the NLO BFKL Green function and to
conjecture an ansatz for the next-to-next-to-leading order (NNLO) BFKL
Green function in the same reference.

The aim of the present paper is twofold. One goal is to construct the
NNLO BFKL kernel eigenfunctions in QCD by finding perturbative
corrections to the NLO BFKL kernel eigenfunctions of
\cite{Chirilli:2013kca}. This would allow us to verify the NNLO BFKL
Green function ansatz proposed in \cite{Chirilli:2013kca}. The second
goal is to build on the existing work
\cite{Balitsky:2012bs,Babansky:2002my,Balitsky:2009yp} along with, of
course, \cite{Chirilli:2013kca}, to construct the cross section for
high-energy $\gamma^* \gamma^*$ scattering at NLO. We use the standard
BFKL power counting in which
\begin{align}
  \label{counting}
  \as \, Y \sim {\cal O} (1),
\end{align}
where $Y \sim \ln s$ is the rapidity variable and $\as$ is the strong
coupling constant. In this power counting the LO $\gamma^* \gamma^*$
scattering mediated by the LO BFKL exchange is ${\cal O} (\as^2)$,
while the NLO correction we construct below is ${\cal O} (\as^3)$. The
resulting NLO $\gamma^* \gamma^*$ scattering cross section would not
only allow for a comparison of the prediction with the existing and
future experimental data from linear colliders, but would also allow
one to see the structure of higher-order corrections to the $\gamma^*
\gamma^*$ scattering by obtaining a general form of this cross section
(in the linear regime). 

The outline of the paper is as follows: we begin in
Sec.~\ref{sec:azim} by generalizing the NLO BFKL kernel Green
functions to the azimuthal-angle dependent case, not considered
originally in \cite{Chirilli:2013kca}. This generalization is rather
straightforward for a reader familiar with the technique of
\cite{Chirilli:2013kca}, and results in the eigenfunctions $H_{n,\nu}
(k)$ given in \eq{eigenf_n_nu}. We continue in Sec.~\ref{sec:gamgam}
where we construct the NLO $\gamma^* \gamma^*$ scattering cross
section. The answer is particularly simple and is given in
Eqs.~\eqref{ggNLOcrossec} and \eqref{ggNLOcrossecLL} as a convolution
of the LO+NLO evolution with the LO+NLO impact factors and LO+NLO
daughter dipole--dipole scattering. It appears that higher-order
corrections, in the linear regime considered, only enter the
expression for the $\gamma^* \gamma^*$ cross section either through
the intercept of the evolution, the impact factor or the daughter
dipole--dipole scattering amplitude. The $\gamma^* \gamma^*$ cross
section is plotted in Figs.~\ref{NLOsigma1LT} and
\ref{NLOsigma5LT}. Finally, the NNLO BFKL kernel eigenfunctions are
constructed via a direct calculation in Sec.~\ref{sec:nnlo} and are
presented in \eq{NNLOeigef} below (along with Eqs.~(\ref{eq:f1}) and
(\ref{f2fin})). The corresponding eigenvalue is given by
\eq{eigNNLO}. Using the eigenfunctions to construct the NNLO BFKL
Green function (\ref{NNLOsolu2}) we reproduce (and, therefore,
confirm) the ansatz proposed in \cite{Chirilli:2013kca}. (Indeed our
NNLO BFKL Green function \eqref{NNLOsolu2} depends on the NNLO BFKL
intercept $\chi_2 (\nu)$, which is not known at present.) We present
conclusions and outlook in Sec.~\ref{sec:concl}.


\section{Generalizing NLO BFKL Eigenfunctions to the
  Azimuthally-Dependent Case}
\label{sec:azim}

The NLO BFKL eigenfunction were constructed in \cite{Chirilli:2013kca}
in the azimuthal-angle independent approximation. However, those
results are easy to generalize to the case with non-trivial azimuthal
angular dependence. 

Consider an arbitrary-order BFKL equation
\begin{align}
  \label{eq:BFKL}
  \partial_Y G \! \left( {\vec k}_\perp, {\vec k'}_\perp, Y \right) =
  \int d^2 q \, K \! \left({\vec k}_\perp, {\vec q}_\perp \right) \, G
  \! \left( {\vec q}_\perp, {\vec k'}_\perp, Y\right)
\end{align}
for the Green function $G \! \left({\vec q}_\perp, {\vec k'}_\perp, Y
\right)$. The initial condition is
\begin{align}
  \label{eq:init}
  G \! \left( {\vec k}_\perp , {\vec k'}_\perp , Y=0 \right) =
  \delta^2 \! \left({\vec k}_\perp - {\vec k'}_\perp \right).
\end{align}
Here $Y$ is the rapidity variable and ${\vec k}_\perp , {\vec
  k'}_\perp$ and ${\vec q}_\perp$ denote transverse momenta.

The BFKL kernel can be expanded in the powers of the strong coupling
$\amu$, such that
\begin{align}
  \label{eq:kernel}
  K \!\left( {\vec k}_\perp , {\vec q}_\perp \right) =
  \bar{\alpha}_\mu \, K^{\rm LO} \!\left( {\vec k}_\perp, {\vec
      q}_\perp \right) + \bar{\alpha}_\mu^2 \, K^{\rm NLO} \!\left(
    {\vec k}_\perp, {\vec q}_\perp \right) + {\cal O} (\bam^3)
\end{align}
with
\begin{align}
  \label{eq:bam}
  \bam \equiv {\alpha_\mu \, N_c \over \pi},
\end{align}
where $\mu$ is an arbitrary renormalization scale and $N_c$ is the
number of colors.

To find the NLO BFKL eigenfunctions we need to know the action of the
LO+NLO kernel defined by
\begin{align}
  \label{eq:LONLO}
  K^{{\rm LO}+ {\rm NLO}}\!\left({\vec k}_\perp, {\vec q}_\perp
  \right) \equiv \bar{\alpha}_\mu \, K^{\rm LO}\! \left({\vec
      k}_\perp, {\vec q}_\perp \right) + \bar{\alpha}_\mu^2 \, K^{\rm
    NLO} \!  \left( {\vec k}_\perp, {\vec q}_\perp \right)
\end{align}
on the LO BFKL eigenfunctions $q^{-1 + 2 \, i \, \nu} \, e^{i \, n \,
  \phi_q}$, where $q = |{\vec q}_\perp|$, $\phi_q$ is the azimuthal
angle of $ {\vec q}_\perp$ with respect to some chosen direction in
the transverse plane, $\nu$ is a real parameter and $n$ is an
integer. This projection was calculated in
\cite{Kotikov:2000pm,Balitsky:2012bs} yielding
\begin{align}
  \label{conf-proj}
  \int d^2 q \, & K^{{\rm LO}+ {\rm NLO}}\!\left({\vec k}_\perp, {\vec
      q}_\perp \right) \ q^{-1 + 2 \, i \, \nu} \, e^{i \, n \,
    \phi_q} \notag \\ & = \left[ \bar{\alpha}_\mu \, \chi_0 (n, \nu) -
    \bar{\alpha}^2_\mu \, \beta_2 \, \chi_0(n, \nu) \, \ln{k^2\over
      \mu^2} + \frac{i}{2} \, \bam^2 \, \beta_2 \, \chi'_0 (n, \nu) +
    \bar{\alpha}_\mu^2 \, \chi_1 (n, \nu) \right] \, k^{-1 + 2 \, i \,
    \nu} \, e^{i \, n \, \phi_k},
\end{align}
where $k = |{\vec k}_\perp|$ and $\phi_k$ is the azimuthal angle of
vector ${\vec k}_\perp$ with respect to the same chosen direction in
the transverse plane. The LO BFKL kernel eigenvalue is
\begin{align}
  \label{eq:LOeig}
  \chi_0(n, \nu)= 2 \, \psi(1) - \psi \left( \frac{1 + |n|}{2} + i \,
    \nu \right) - \psi \left( \frac{1 + |n|}{2} - i \, \nu \right)
\end{align}
with integer $n$ and real $\nu$, and where $\psi (z) = d \ln \Gamma
(z)/dz$ as usual. The prime denotes derivatives with respect to $\nu$,
such that $\chi'_0 (n, \nu) = \partial \chi_0(n, \nu)/\partial \nu$.

The one-loop running of the strong coupling is defined as
\begin{align}
  \label{eq:beta2}
  \bas (Q^2) = \frac{\bam}{1 + \bam \, \beta_2 \, \ln
    \frac{Q^2}{\mu^2}}, \ \ \ \beta_2 = {11 \, N_c - 2 \, N_f\over 12
    \, N_c},
\end{align}
where $N_f$ is the number of quark flavors.

The real part of the projection of the NLO BFKL kernel on the LO BFKL
eigenfunctions is given by \cite{Kotikov:2000pm,Balitsky:2012bs}
\begin{align}
  \label{chi_1nu}
  & \chi_1(n, \nu) = - \beta_2 \, \frac{\chi_0^2 (n, \nu)}{2} +
  \frac{5}{3} \, \beta_2 \, \chi_0 (n, \nu) + \frac{1}{3} \, \left( 1
    - \frac{\pi^2}{4} \right) \, \chi_0 (n, \nu) + \frac{3}{2} \,
  \zeta (3) + \frac{1}{4} \, \chi''_0 (n, \nu) - \frac{1}{2} \, \Phi
  (n, \nu) - \frac{1}{2} \, \Phi (n, - \nu) \notag \\ & + \frac{\pi^2
    \, \sinh (\pi \, \nu)}{8 \, \nu \, \cosh^2 (\pi \, \nu) } \,
  \left\{ - \delta_{n0} \, \left[ 3 + \left( 1 + \frac{N_f}{N_c^3}
      \right) \frac{11 + 12 \, \nu^2}{16 \, (1+\nu^2)} \right] +
    \delta_{n2} \, \left( 1 + \frac{N_f}{N_c^3} \right) \, \frac{1 + 4
      \, \nu^2}{32 \, (1+\nu^2)} \right\} ,
\end{align}
where $\chi''_0 (n, \nu) = \partial^2 \chi_0(n, \nu)/\partial \nu^2$
and
\begin{align}
  \label{eq:Phi_def2}
   \Phi(n,\nu)~=~&\int_0^1\!{dt\over 1+t}~t^{{n-1\over 2}+ i \, \nu}
  \Bigg\{{\pi^2\over 12}-{1\over 2}\psi'\Big({n+1\over 2}\Big)
  -{\rm Li}_2(t)-{\rm Li}_2(-t)&
  \nonumber\\
    &
  -~\Big[\psi(n+1)-\psi(1)+\ln(1+t)
  \hspace{-1mm}+\sum_{k=1}^\infty{(-t)^k\over k+n}\Big]\ln t
  -\sum_{k=1}^\infty{t^k\over (k+n)^2}[1-(-1)^k]\Bigg\}.& 
\end{align}

To construct the eigenfunctions of $K^{{\rm LO}+ {\rm NLO}}$ one could
follow the steps employed in \cite{Chirilli:2013kca} for the
azimuthally-symmetric case. One may notice, however, that for $n \neq
0$ the construction of eigenfunctions is identical to the $n=0$ case
of \cite{Chirilli:2013kca} with the simple replacements
\begin{align}
  \label{eq:repl}
  k^{-1 + 2 \, i \, \nu} \to k^{-1 + 2 \, i \, \nu} \, e^{i \, n \,
    \phi_k}, \ \ \ \chi_0 (\nu) \equiv \chi_0 (0, \nu) \to \chi_0 (n,
  \nu), \ \ \ \chi_1 (\nu) \equiv \chi_1 (0, \nu) \to \chi_1 (n, \nu).
\end{align}
Performing the substitutions \eqref{eq:repl} in Eq.~(47) of Ref.
\cite{Chirilli:2013kca} we obtain the eigenfunctions of the LO+NLO
BFKL kernel in the azimuthal angle-dependent case
\begin{align} 
  H_{n, \nu} ({\vec k}_\perp) = k^{-1 + 2\, i \, \nu} \, e^{i \, n \, \phi_k}
  \, \left[ 1 + \bar{\alpha}_\mu \, \beta_2 \left( i \, {\chi_0 (n,
        \nu) \over 2 \, \chi'_0 (n, \nu)} \, \ln^2{k^2\over \mu^2} +
      \frac{1}{2} \left( 1 - \frac{\chi_0 (n, \nu) \, \chi_0'' (n,
          \nu)}{\chi_0' (n, \nu)^2} \right) \ln{k^2\over \mu^2}
    \right) \right].
          \label{eigenf_n_nu}
\end{align}

Similar to \cite{Chirilli:2013kca} one can show that the functions
$H_{n,\nu} ({\vec k}_\perp)$ satisfy the completeness
\begin{align}
  \label{eq:compl}
  \sum_{n = - \infty}^\infty \ \int\limits_{-\infty}^\infty \frac{d
    \nu}{2 \, \pi^2} \, H_{n, \nu} ({\vec k}_\perp) \, H^*_{n, \nu}
  ({\vec k'}_\perp) = \delta^2 \! \left( {\vec k}_\perp - {\vec
      k'}_\perp \right)
\end{align}
and orthogonality 
\begin{align}
  \label{eq:ortho}
  \int d^2 k \, H_{n, \nu} ({\vec k}_\perp) \, H^*_{n', \nu'} ({\vec
    k}_\perp) = 2 \, \pi^2 \, \delta_{n n'} \, \delta (\nu - \nu')
\end{align}
relations up to (and including) ${\cal O} (\amu)$, which is the
accuracy of the NLO expansion for the eigenfunctions. (The asterisk
denotes complex conjugation.) The completeness relation
\eqref{eq:compl} is already satisfied by the LO eigenfunctions and is
not affected by the perturbative corrections used to construct the NLO
BFKL eigenfunctions \eqref{eigenf_n_nu}. Therefore, the space of
functions upon which the eigenfunctions $H_{n, \nu} ({\vec k}_\perp)$
form a complete basis is the same as the one of the power-like (LO
BFKL) eigenfunctions.

By analogy with \cite{Chirilli:2013kca} the eigenvalues of $K^{{\rm
    LO}+ {\rm NLO}}$ corresponding to the eigenfunctions $H_{n,\nu}
({\vec k}_\perp)$ are
\begin{align}
  \label{eq:eig}
  \Delta (n, \nu) = \bar{\alpha}_\mu \, \chi_0(n, \nu) +
  \bar{\alpha}_\mu^2 \, \chi_1 (n, \nu),
\end{align}
such that the Green function in \eq{eq:BFKL} is
\begin{align}
  \label{eq:Green}
  G \left({\vec k}_\perp, {\vec k'}_\perp, Y \right) = \sum_{n = -
    \infty}^\infty \ \int\limits_{-\infty}^{\infty} {d\nu \over
    2\pi^2} \, e^{\left[\bar{\alpha}_\mu \, \chi_0(n, \nu) +
      \bar{\alpha}_\mu^2 \, \chi_1 (n, \nu) \right] \, Y} \, H_{n,
    \nu} ({\vec k}_\perp) \, H^*_{n, \nu} ({\vec k'}_\perp).
\end{align}

This accomplishes the generalization of the NLO BFKL eigenfunctions to
the case of non-trivial azimuthal dependence. Below, when discussing
the azimuthally-symmetric $n=0$ case we will use the simplified
notation
\begin{align}
  \label{eq:simpleH}
  H_\nu (k) = H_{\frac{1}{2} + i \, \nu} (k) \equiv H_{0 , \nu} ({\vec
    k}_\perp),
\end{align}
which also connects to the notation used in \cite{Chirilli:2013kca}.


\section{$\gamma^* \gamma^*$ Scattering Cross Section at NLO}
\label{sec:gamgam}

\begin{figure}
  \includegraphics[width=2.0in]{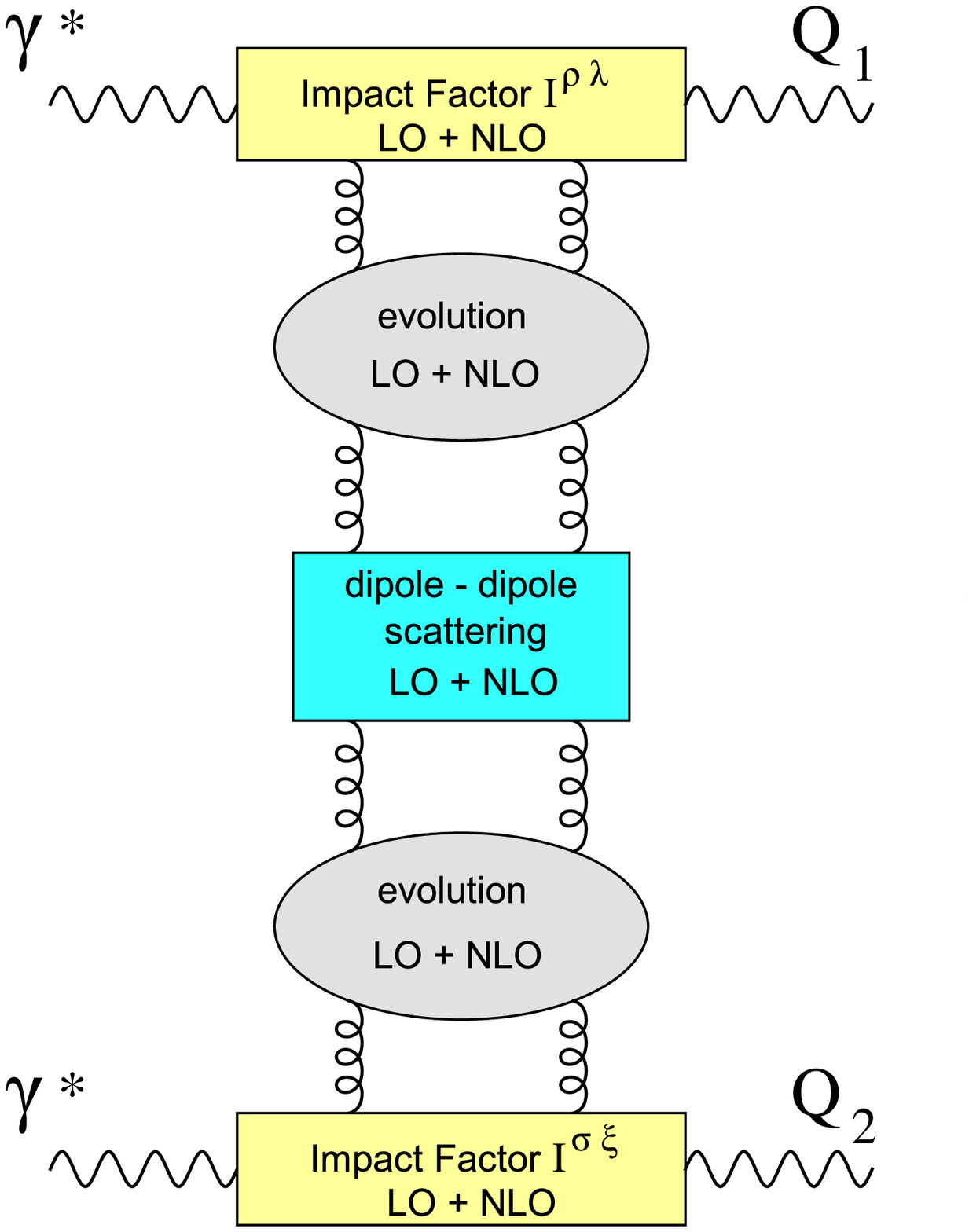}
  \includegraphics[width=4.0in]{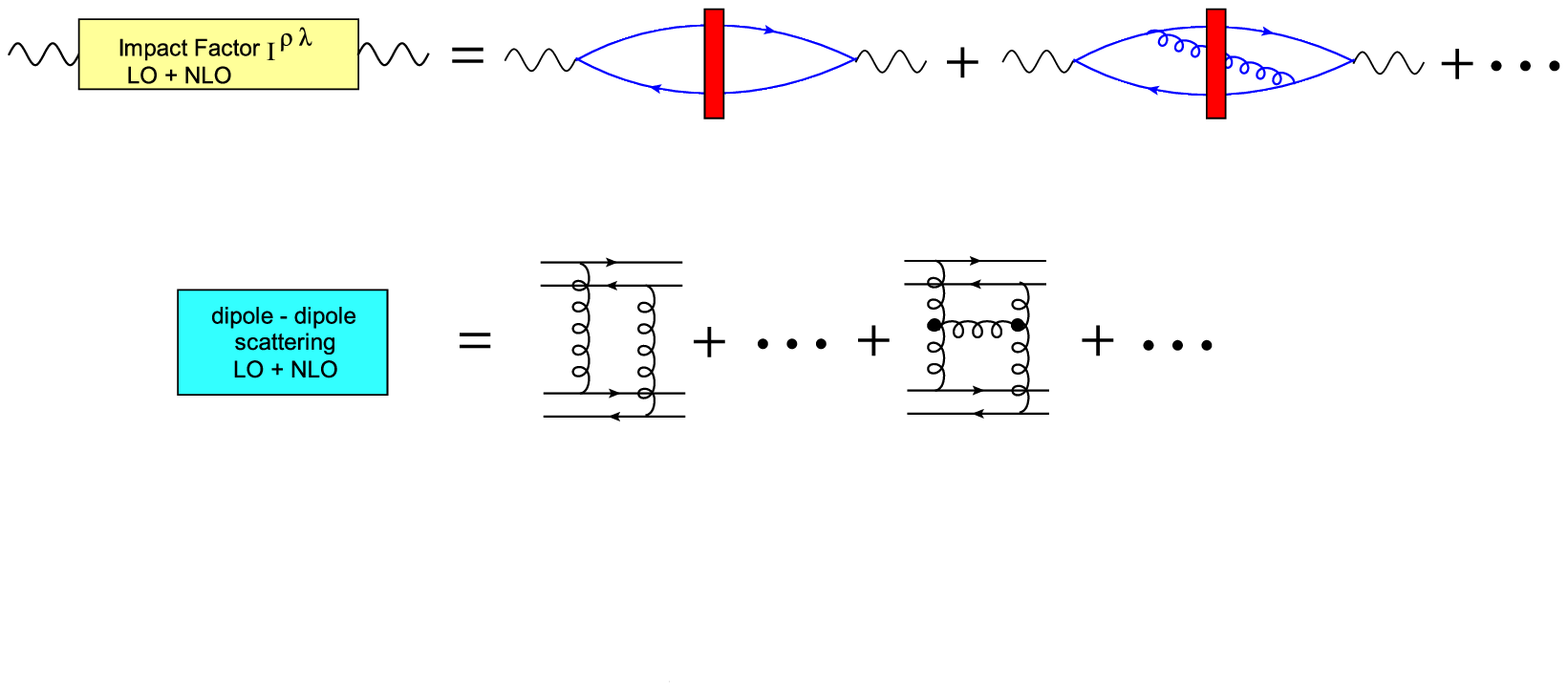}
  \caption{The left panel contains a diagrammatic representation of
    the $\gamma^* \gamma^*$ scattering amplitude factorized into the
    impact factors, small-$x$ evolution, and the (energy-independent)
    daughter dipole--dipole scattering. This structure of the
    factorization is suggested by the high-energy OPE applied to the
    electromagnetic currents. In the right panel, the structure of the
    LO+NLO impact factor is illustrated by some of the diagrams in the
    shock-wave formalism originally calculated at NLO in
    \cite{Balitsky:2010ze, Balitsky:2012bs}. Examples of diagrams
    contributing to the daughter dipole--dipole scattering at LO and
    NLO are also shown in the right panel.}
\label{gg_graph} 
\end{figure}

The three main ingredients for the determination of the NLO cross
section for $\gamma^* \gamma^*$ scattering are the impact factor
(light-cone wave function squared), calculated up to NLO
\cite{Balitsky:2012bs}, the solution of the NLO BFKL evolution
equation \cite{Chirilli:2013kca}, and the energy-independent forward
scattering amplitude at NLO of the two fundamental ``daughter''
color-dipoles (quark--antiquark pairs) produced by the two virtual
photons through the evolution before the interaction
\cite{Babansky:2002my,Balitsky:2009yp}. The convolution of these three
contributions yields the total inclusive $\gamma^* \gamma^*$
scattering cross section. \fig{gg_graph} provides a diagrammatic
representation of the factorization of the $\gamma^*\gamma^*$
scattering cross section.

The Operator Product Expansion (OPE) at high-energy
\cite{Balitsky:1996ub,Balitsky:2009xg} provides a natural framework
were the scattering amplitude factorizes in these three main
contributions. In the next subsection we review the logic of the OPE
at high-energy before performing the actual calculation of the LO+NLO
$\gamma^* \gamma^*$ scattering cross section in the two subsequent
subsections.


\subsection{OPE at high-energy}
In this section we review the high-energy OPE \cite{Balitsky:1996ub}
which we will employ to calculate the NLO $\gamma^* \gamma^*$ cross
section below.

Let us consider the forward scattering amplitude of two virtual
photons with four-momenta $q_1$ and $q_2$ and polarizations
$\lambda_1$ and $\lambda_2$ represented by four electromagnetic
currents
\begin{align} 
  \label{4j}
  {\cal A}_{\lambda_1 \lambda_2} (q_1, q_2) = -i \,
  \varepsilon^{\lambda_1 \, *}_{\rho_1} (q_1) \,
  \varepsilon^{\lambda_1}_{\sigma_1} (q_1) \, \varepsilon^{\lambda_2
    \, *}_{\rho_2} (q_2) \, \varepsilon^{\lambda_2}_{\sigma_2} (q_2)
  \,
  \int d^2z_\perp dz^- dz^+ \int d^4x \int d^4 y \, e^{iq_1\cdot x+iq_2\cdot y} \nonumber \\
  \langle 0| {\rm T} \, j^{\rho_1} (x^++z^+, x^-, {\vec x}_\perp +
  {\vec z}_\perp) \, j^{\sigma_1} (z^+,0, {\vec z}_\perp) \,
  j^{\rho_2} (y^+,y^-+z^-, {\vec y}_\perp) \, j^{\sigma_2} (0,z^-,
  0_\perp) |0\rangle.
\end{align}
Here $\varepsilon^\lambda_\rho (q)$ are the gluon polarization
vectors, $x^\pm = (x^0 + x^3)/\sqrt{2}$, and $j^\sigma$ are the
electromagnetic currents.

Since we are interested in the $\gamma^* \gamma^*$ cross section at
high energy (in the Regge limit), we factorize (\ref{4j}) in the
following way
\begin{align}
  {\cal A}_{\lambda_1 \lambda_2} (q_1, q_2) & = -i \,
  \varepsilon^{\lambda_1 \, *}_{\rho_1} (q_1) \,
  \varepsilon^{\lambda_1}_{\sigma_1} (q_1) \, \varepsilon^{\lambda_2
    \, *}_{\rho_2} (q_2) \, \varepsilon^{\lambda_2}_{\sigma_2} (q_2)
  \, \int d^2 z_\perp \, {\cal N}^{-1} \int {\cal D}A \, e^{i \, S(A)}
  \, {\rm det}(i\nabla)
  \nonumber\\
  & \times\int dz^+ \int d^4 x \, e^{iq_1\cdot x} \, \langle {\rm T}
  \, j^{\rho_1} (x^++z^+, x^-, {\vec x}_\perp + {\vec z}_\perp) \,
  j^{\sigma_1} (z^+,0, {\vec z}_\perp)\rangle_A
  \nonumber\\
  & \times\int dz^- \int d^4 y \, e^{iq_2\cdot y} \, \langle {\rm T}
  \, j^{\rho_2} (y^+, y^-+z^-, {\vec y}_\perp) \, j^{\sigma_2} (0,z^-,
  0_\perp)\rangle_A \ ,
\label{fact4j}
\end{align}
where $S(A)$ is the part of the QCD action which depends only on the
gluon field $A_\mu$, ${\rm det}(i\nabla)$ is the fermionic determinant
with $\nabla$ the covariant derivative, angle brackets $\langle \ldots
\rangle_A$ denote the expectation value in the background gluon field
$A_\mu$, and the normalization factor $\cal N$ is
\begin{align}
  \label{eq:Norm}
  {\cal N} = \int {\cal D}A \, e^{i \, S(A)} \, {\rm det}(i\nabla).
\end{align}
In the following calculation we omit the factor $\left( \sum_f e_f^2
\right)^2$ (where $e_f$ labels the electromagnetic charge of the quark
with flavor $f$), which we will reinstate in the expressions for the
cross sections.

\begin{figure}
  \includegraphics[width=4.0in]{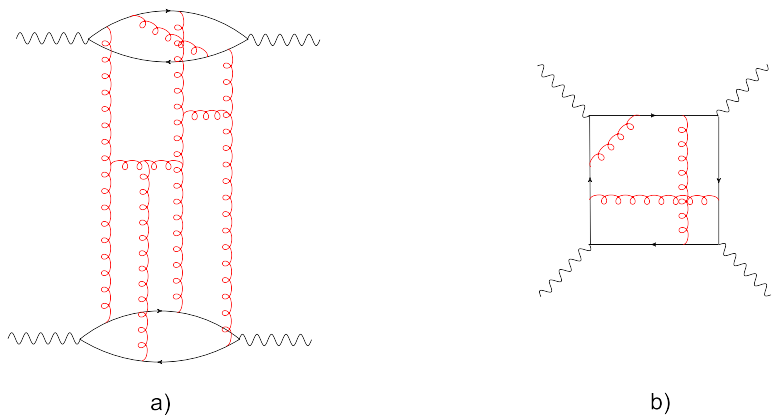}
  \caption{Left panel depicts a Feynman diagram that can be
    obtained from \eq{fact4j} which contributes to the high-energy
    asymptotics of $\gamma^* \gamma^*$ scattering. The right panel
    shows an example of the "box" diagram that can not be generated
    from \eq{fact4j}: box diagrams are suppressed at high energies by
    a power of energy and, therefore, can be neglected. However, 
    as we will see in the numerical results section, such types of 
    diagrams become relevant, and therefore not anymore negligible 
    at low rapidity where energy is not high enough to suppress 
    such contributions.}
\label{graph-fact4j} 
\end{figure}

The factorization of \eq{fact4j} is simply due to the fact that in the
high-energy asymptotics the dominant contribution to the $\gamma^*
\gamma^*$ forward scattering amplitude should contain at least a
single two-gluon exchange in the $t$-channel. A typical dominant
high-energy diagram that can be obtained from \eq{fact4j} is shown in
\fig{graph-fact4j}a. An energy-suppressed diagram of the type we
neglected in writing \eq{fact4j} is shown in \fig{graph-fact4j}b, and
belongs to the class of the so-called ``box'' diagrams, where there is
a quark $t$-channel exchange spanning the whole rapidity interval of
the scattering process. Such diagrams are suppressed by a power of
center-of-mass energy squared $s$ compared to the leading graphs in
\fig{graph-fact4j}a, and can be neglected in the high-energy
asymptotics.

Each term in the angle brackets $\langle \ldots \rangle_A$ in
Eq. (\ref{fact4j}) is evaluated using the OPE at high energy
\cite{Balitsky:1996ub} and the background field technique.  The
coefficient functions of the OPE of two electromagnetic currents at
high energy is the photon impact factor (PIF) as originally defined in
\cite{Balitsky:1978ic}. The PIF is convoluted with a matrix element of
an operator made out of Wilson lines along the light cone.
 
Since we are interested in the NLO $\gamma^* \gamma^*$ cross section,
we need the LO and NLO PIF. The calculation of NLO PIF has been
performed in \cite{Balitsky:2010ze,Balitsky:2012bs}. We will only
employ the results of that work in our calculation and refer the
reader to the original papers \cite{Balitsky:2010ze,Balitsky:2012bs}
for the details of the PIF calculation. We stress, however, that at
NLO the PIF can be contaminated by the rapidity-dependent (or
energy-dependent) terms. As explained in \cite{Balitsky:2009xg}, when
the high-energy OPE is performed in terms of composite Wilson-line
operators, by redefining the operator at hand all the energy dependent
terms can be \textit{shifted} into the matrix elements of Wilson
lines, thus leaving the NLO impact factor energy-scale invariant. This
is what has been done in the PIF calculation of
\cite{Balitsky:2010ze,Balitsky:2012bs}. In addition, the composite
Wilson lines operator obtained by the shifting procedure of
\cite{Balitsky:2010ze,Balitsky:2012bs} renders the impact factor
conformally invariant also at NLO: the energy-dependent terms clearly
would have broken conformal invariance. The high-energy OPE in terms
of composite Wilson line operator provides an operatorial and
systematic procedure to factorize scattering amplitudes into
coefficient functions which are energy-scale invariant and matrix
elements of composite Wilson line operators that encode the energy
dependence of the amplitude.

In $\gamma^* \gamma^*$ scattering at high energy, each of the two
virtual photons can be treated at the same time as the projectile and
as the target due to the symmetry of the process. We will work in a
frame in which the photons collide head-on, such that their
four-momenta are
\begin{align}
  \label{eq:q12}
  q_1^\mu = \left( q_1^+, - \frac{Q_1^2}{2 \, q_1^+}, {0}_\perp
  \right), \ \ \ q_2^\mu = \left( - \frac{Q_2^2}{2 \, q_2^-}, q_2^-,
    {0}_\perp \right)
\end{align}
with $q_1^+$ and $q_2^-$ very large. Further, we will work in a gauge
which reduces to the $A^+ =0$ light-cone gauge near the $x^- =0$ light
cone, and to the $A^- =0$ light-cone gauge near the $x^+ =0$ light
cone, that is, it reduces to the light-cone gauges for each of the
photons: examples of such gauges include the Coulomb ${\vec \nabla}
\cdot {\vec A} =0$ gauge and the $A^0=0$ temporal gauge. In such
gauge, each virtual photon, long before the interaction, splits into a
quark--antiquark pair which then traverses the gluonic field produced
by the target (the other $q\bar q$ pair): this interaction is
described by a color dipole made of two fundamental Wilson lines
scattering on another fundamental dipole. In DIS the small-$x$
evolution of Wilson lines is described by the Balitsky-JIMWLK
non-linear equation
\cite{Balitsky:1996ub,Jalilian-Marian:1997dw,Jalilian-Marian:1997gr,Iancu:2001ad,Iancu:2000hn};
in the case of a fundamental dipole in the large-$N_c$ limit the
evolution equation corresponds to the Balitsky-Kovchegov equation
\cite{Balitsky:1996ub,Balitsky:1998ya,Kovchegov:1999yj,Kovchegov:1999ua}
and its linearization corresponds to the BFKL equation
\cite{Kuraev:1977fs,Kuraev:1976ge,Balitsky:1978ic}. While it is not
presently clear which evolution equation describes Wilson-line
correlators in $\gamma^* \gamma^*$ scattering at high energy, it is
clear that in the linear regime (see Sec.~\ref{sec:nnlo} for the
definition of such regime) the evolution of a dipole operator is given
by the BFKL equation.

We are interested in the $\gamma^* \gamma^*$ scattering cross section
in the linear case where the resummation in the leading logarithmic
approximation in the Regge limit is achieved by the linearization of
the evolution equation for the color Wilson line operator defined as
\begin{align} 
  \label{col-dipo} {\cal U}^\eta ({\vec x}_\perp, {\vec y}_\perp) = 1
  - {1\over N_c}{\rm tr}\{U^\eta({\vec x}_\perp) \,
  U^{\eta\dagger}({\vec y}_\perp)\}
\end{align}
where $U^\eta({\vec x}_\perp)$ is the Wilson line operator in the
fundamental representation, which, for a projectile moving along the
$x^+$ light cone is defined by
\begin{align}
  \label{Wline_def}
  U^\eta({\vec x}_\perp) = {\rm P} \exp \left\{ i \, g
    \int\limits_{-\infty}^\infty dx^+ A^-(x^+, x^-=0, {\vec x}_\perp)
  \right\}
\end{align}
and $x$ and $y$ are the points of interaction of the quark and
anti-quark pair with the gluonic external field of the target; in the
center-of-mass frame the gluonic external field of the target reduces
to a shock wave due to Lorentz contraction and time dilation. The
Wilson line operator depends on the rapidity $\eta$, which may be
included in \eq{Wline_def} either by changing the slope of the Wilson
line to also include the $x^-$ direction or by a rigid cut-off imposed
on the longitudinal momenta of the gluons when calculating matrix
elements of the Wilson line operators.  In the DIS case the NLO
evolution of the color dipole \eqref{col-dipo} with respect to
rapidity $\eta$ has been calculated in \cite{Balitsky:2008zza} where
its linearization was proven to coincide with the known result for the
NLO BFKL equation \cite{Fadin:1998py,Ciafaloni:1998gs}.

At NLO the high-energy OPE is performed in terms of the composite
operator $[{\cal U}^a({\vec x}_\perp, {\vec y}_\perp)]^{\rm comp}$
which differs from the one in Eq.~(\ref{col-dipo}) by the addition of
a counter term that restores conformal symmetry present in the leading
order impact factor
\cite{Balitsky:2009xg,Balitsky:2010ze,Balitsky:2012bs}. Note also,
that the evolution parameter for the composite operator is not the
rapidity parameter $\eta$ from Eq.~(\ref{col-dipo}), but is the new
parameter $a$ which ensures that the evolution equation of the
composite operator $[{\cal U}^a({\vec x}_\perp, {\vec y}_\perp)]^{\rm
  comp}$ with respect to $a$ is the same as the one for ${\cal U}^\eta
({\vec x}_\perp, {\vec y}_\perp)$ with respect to $\eta$ at LO. At NLO
the evolution equation for $[{\cal U}^a({\vec x}_\perp, {\vec
  y}_\perp)]^{\rm comp}$ with respect to $a$ has a conformal kernel in
${\cal N}=4$ SYM; in QCD the NLO evolution kernel for $[{\cal
  U}^a({\vec x}_\perp, {\vec y}_\perp)]^{\rm comp}$ consists of a
conformally-invariant piece and a running-coupling term.

For DIS the momentum-space high-energy OPE of two electromagnetic
currents (in the linearized approximation, that is, in absence of
nonlinear saturation effects) is
\cite{Balitsky:2010ze,Balitsky:2012bs}
\begin{align}
  \label{eq:DIS}
  \int d z^+ \, d^2 z \, \int\! d^4x \, e^{i \, q \cdot x } \, &
  \langle {\rm T} \, j^\rho(x^++z^+, x^-, {\vec x}_\perp + {\vec
    z}_\perp) \, j^\sigma (z^+,0, {\vec z}_\perp )\rangle_A =
  \sqrt{s\over 2} \!\int\!{d^2 k \over (2\pi)^2 \,k^2} \, I^{\rho\sigma} (q, k)
  \int d^2 z \, k^2 \, \langle \calu^\eta ({\vec k}_\perp, {\vec
    z}_\perp) \rangle_A
\end{align}
where $I^{\rho\sigma} (q, k)$ is the photon impact factor and the
Wilson-line operator in the matrix element is (see Appendix~A for
details)
\begin{align}\label{Utr}
  \calu^\eta ({\vec k}_\perp, {\vec z}_\perp) =\int\! d^2 x \, e^{-i
    \, {\vec k}_\perp \cdot {\vec x}_\perp} \, {\cal U}^\eta ( {\vec
    x}_\perp + {\vec z}_\perp , {\vec z}_\perp).
\end{align} 
Note that in \eq{eq:DIS} the composite operator
$[\calu^a(\vec{k}_\perp,\vec{z}_\perp)]^{\rm comp}$ has to be used
instead of $\calu^\eta(\vec{k}_\perp,\vec{z}_\perp)$ at NLO and
beyond.  The background field in the averaging $\langle \ldots
\rangle_A$ is due to the target, which could be a proton or a nucleus.

The impact factor $I^{\rho\sigma} (q, k)$ is known up to NLO
\cite{Balitsky:2010ze,Balitsky:2012bs}. At the LO and NLO the impact
factor can be written as a Mellin transform
\begin{align}
  \label{IF_Mellin}
  I^{\rho\sigma}_{{\rm LO}+{\rm NLO}} (q, k) =
  \int\limits_{-\infty}^\infty d \nu \, \left({k^2\over
      Q^2}\right)^{\half-i \, \nu} \, {\tilde I}^{\rho\sigma}_{{\rm
      LO}+{\rm NLO}} (q, \nu)
\end{align}
with \cite{Balitsky:2010ze,Balitsky:2012bs}
\begin{align}
  {\tilde I}^{\rho\sigma}_{{\rm LO}+{\rm NLO}}(q, \nu) = & {N_c\over
    64} {(\pi\nu)^{-1}\sinh\pi\nu\over (1+\nu^2)\cosh^2\pi\nu} \Big\{
  \Big({9\over 4}+\nu^2\Big)\Big[1+{\amu \over\pi} + {\bam \over
    2}{\cal F}_1(\nu)\Big]P_1^{\rho\sigma} \notag \\ & + \Big({11\over
    4}+3\, \nu^2\Big)\Big[1+{\amu \over\pi} + {\bam \over 2}{\cal
    F}_2(\nu)\Big]P_2^{\rho\sigma} + \Big({1\over 8}+{\nu^2\over
    2}\Big) \Big[1+{\amu \over\pi} + {\bam \over 2}{\cal
    F}_3(\nu)\Big]P_3^{\rho\sigma}\Big\},
\label{NLOIFnu1}
\end{align}
where the functions ${\cal F}_i(\nu)$ with $i=1,2,3$ are defined in
\cite{Balitsky:2012bs}, $k = |{\vec k}_\perp|$, and\footnote{Note that
  $P_3^{\rho\sigma}$ depends only on the direction of ${\vec
    k}_\perp$, and not on its magnitude.}
\begin{subequations}\label{tensors}
\begin{align}
  &  P_1^{\rho\sigma} = g^{\rho\sigma} - {q^\rho \, q^\sigma \over q^2}
  \label{Pes} \\
  & P_2^{\rho\sigma} = {1\over q^2}\Big(q^\rho -{p_2^\rho \, q^2\over
    q \cdot
    p_2}\Big)\Big(q^\sigma-{p_2^\sigma \, q^2\over q\cdot p_2}\Big) \label{P2} \\
  & P_3^{\rho\sigma} 
  = 2\, {k_\perp^\rho k_\perp^\sigma \over k_\perp^2} +
  g_\perp^{\rho\sigma} .
\end{align}
\end{subequations}
with $p_2^\mu = (0, p_2^-, 0_\perp)$. Note that the tensor structure
${\cal P}_3^{\rho\sigma}$ in (\ref{NLOIFnu1}) will not contribute in
the case of total cross section of $\gamma^* \gamma^*$ scattering we
are about to investigate. This is due to the fact that scattering of
two unpolarized photons in the kinematics of \eqref{eq:q12} has no
preferred transverse direction: averaging ${\cal P}_3^{\rho\sigma}$
over the angles of $k_\perp^\rho$ gives zero. This conclusion also
holds for the scattering of longitudinally polarized photons and for
the scattering of transversely polarized photons summed over
transverse polarizations.

We now turn our attention back to $\gamma^* \gamma^*$ scattering.
Using Eqs.~(\ref{eq:DIS}) and (\ref{IF_Mellin}) in the scattering
amplitude (\ref{fact4j}) yields
\begin{align}
  {\cal A}_{\lambda_1 \lambda_2} (q_1, q_2) = & -i \,
  \varepsilon^{\lambda_1 \, *}_{\rho_1} (q_1) \,
  \varepsilon^{\lambda_1}_{\sigma_1} (q_1) \, \varepsilon^{\lambda_2
    \, *}_{\rho_2} (q_2) \, \varepsilon^{\lambda_2}_{\sigma_2}
  (q_2) \label{fact4j-1} \int d^2 z \\ & \times \left\langle \left[
      \sqrt{{s\over 2}}\int d\nu_1 {\tilde I}^{\rho_1 \sigma_1}_{{\rm
          LO}+{\rm NLO}}(q_1, \nu_1) \int {d^2 k_1\over (2\pi)^2\, k^2_1}
      \left({k^2_1\over Q^2_1}\right)^{\half-i\nu_1} k_1^2 \, 
      [{\cal U}^{a_1}
      ({\vec k}_{1 \perp} , {\vec z}_\perp)]^{\rm comp} \right] \right. \notag \\
  & \times \left. \left[ \sqrt{{s\over 2}}\int d\nu_2 {\tilde
        I}^{\rho_2 \sigma_2}_{\rm LO + NLO}(q_2, \nu_2) \int {d^2
        k_2\over (2\pi)^2\,k^2_2} \left({k_2^2\over Q_2^2}\right)^{\half-i\nu_2}
      k_2^2 \,[ {\cal U}^{a_2} ({\vec k}_{2 \perp} , 0_\perp )]^{\rm comp}
    \right] \right\rangle, \notag
\end{align}
where
\begin{align}
  \langle \ldots \rangle = {\cal N}^{-1} \, \int {\cal D}A \, e^{i \,
    S(A)}\, {\rm det}(i\nabla) \, \ldots \ 
\label{V-scatdipo}
\end{align}
and $a_1, \, a_2$ are fixed so that the NLO impact factors in
coordinate space are conformally invariant \cite{Balitsky:2012bs}; in
momentum space $a_1, \, a_2$ are related to the rapidities of the two
colliding virtual photons, $a_1 = s/Q_1^2$ and $a_2 = s/Q_2^2$ with
$s= (q_1+q_2)^2$ the center-of-mass energy squared. Note that
(cf. \eq{Utr} along with \eq{fact4j})
\begin{align}
  \label{eq:dd_corr}
  \langle [{\cal U}^{a_1} ({\vec k}_{1 \perp} , {\vec z}_\perp)]^{\rm
    comp} \, [{\cal U}^{a_2} ({\vec k}_{2 \perp} , 0_\perp)]^{\rm
    comp} \rangle = \int\!  d^2 x \, d^2 y \, e^{-i \, {\vec k}_{1
      \perp} \cdot {\vec x}_\perp -i \, {\vec k}_{2 \perp} \cdot {\vec
      y}_\perp } \, \langle [{\cal U}^{a_1} ( {\vec x}_\perp + {\vec
    z}_\perp , {\vec z}_\perp)]^{\rm comp} \ [{\cal U}^{a_2} ( {\vec
    y}_\perp , {0}_\perp)]^{\rm comp} \rangle.
\end{align}
Opening the square brackets in \eq{fact4j-1} we can write the
$\gamma^* \gamma^*$ forward scattering amplitude as
\begin{align}
  {\cal A}_{\lambda_1 \lambda_2} & (q_1, q_2) = - i \, {s\over 2} \,
  \varepsilon^{\lambda_1 \, *}_{\rho_1} (q_1) \,
  \varepsilon^{\lambda_1}_{\sigma_1} (q_1) \, \varepsilon^{\lambda_2
    \, *}_{\rho_2} (q_2) \, \varepsilon^{\lambda_2}_{\sigma_2} (q_2)
  \int {d^2 k_{1}\over (2\pi)^2 \,k^2_{1}} \int d\nu_1 \, {\tilde I}^{\rho_1
    \sigma_1}_{\rm LO+NLO}(q_1, \nu_1) \, \left({k_1^2\over
      Q_1^2}\right)^{\half-i \, \nu_1}
  \nonumber\\
  & \times \int {d^2 k_2\over (2\pi)^2\,k^2_2} \int d\nu_2 \, {\tilde I}^{\rho_2
    \sigma_2}_{\rm LO+ NLO}(q_2, \nu_2) \, \left({k_2^2\over
      Q_2^2}\right)^{\half-i \, \nu_2} \, \int d^2 z \, \langle k_1^2
  \, [{\cal U}^{a_1} ({\vec k}_{1 \perp} , {\vec z}_\perp)]^{\rm comp} \, k_2^2
  \, [{\cal U}^{a_2} ({\vec k}_{2 \perp} , 0_\perp)]^{\rm comp} \rangle .
\label{fact4j-2}
\end{align}

Using the OPE in terms of composite Wilson line operators, the forward
scattering amplitude (\ref{fact4j-2}) is now factorized into the
energy-scale invariant impact factors ${\tilde I}^{\rho \sigma}_{\rm
  LO+NLO}$ for each of the virtual photons, the evolution of the
composite Wilson line operators $[{\cal U}^a]^{\rm comp}$ convoluted
with each impact factor and the scattering amplitude of the composite
Wilson line operators $\langle [{\cal U}^{a_1} ({\vec k}_{1\perp} ,
{\vec z}_\perp)]^{\rm comp} \, [{\cal U}^{a_2} ({\vec k}_{2 \perp} ,
0_\perp)]^{\rm comp} \rangle$ defined in Eqs.~\eqref{V-scatdipo} and
\eqref{eq:dd_corr} which has to be calculated at NLO (without
evolution).  Notice that, the NLO impact factor calculated in
\cite{Bartels:2004bi,Bartels:2002uz,Bartels:2001mv}, which is provided
as a combination of numerical and analytic results, would differ from
the one calculated in \cite{Balitsky:2010ze, Balitsky:2012bs}. This is
because the NLO dipole-dipole scattering, which, in the OPE language,
is a universal factor independent of the process under consideration,
has to be calculated only once, while in
\cite{Bartels:2004bi,Bartels:2002uz,Bartels:2001mv} it is included in
the definition of the impact factor. Thus, in
\cite{Bartels:2004bi,Bartels:2002uz,Bartels:2001mv} the scattering
cross-sections are factorized in a different way and the contribution
of the dipole-dipole scattering has to be re-calculated as part of the
impact factor for every different process considered.


\subsection{LO $\gamma^*$$\gamma^*$ cross section}

Before proceeding with the NLO calculation, it is instructive to
rederive the LO $\gamma^* \gamma^*$ cross section. Such calculation
would also allow one to better understand the issues that have
prevented the calculation of $\gamma^* \gamma^*$ cross section at NLO
until now. The LO $\gamma^* \gamma^*$ cross section is defined as the
${\cal O} (\as^2)$ contribution in the power counting of
\eq{counting}.

Since at LO we do not need to use the composite operator \eq{fact4j-2} reduces to 
\begin{align}
  {\cal A}_{\lambda_1 \lambda_2} (q_1, q_2) = & - i \, {s\over 2} \,
  \varepsilon^{\lambda_1 \, *}_{\rho_1} (q_1) \,
  \varepsilon^{\lambda_1}_{\sigma_1} (q_1) \, \varepsilon^{\lambda_2
    \, *}_{\rho_2} (q_2) \, \varepsilon^{\lambda_2}_{\sigma_2} (q_2)
  \int {d^2 k_{1}\over (2\pi)^2 \, k^2_{1}} \int d\nu_1 \, {\tilde I}^{\rho_1
    \sigma_1}_{\rm LO}(q_1, \nu_1) \, \left({k_1^2\over
      Q_1^2}\right)^{\half-i \, \nu_1}
  \nonumber\\
  & \times \int {d^2 k_2\over (2\pi)^2\,k^2_2} \int d\nu_2 \, {\tilde I}^{\rho_2
    \sigma_2}_{\rm LO}(q_2, \nu_2) \, \left({k_2^2\over
      Q_2^2}\right)^{\half-i \, \nu_2} \, \int d^2 z \, \langle k_1^2
  \, {\cal U}^{\eta_1} ({\vec k}_{1 \perp} , {\vec z}_\perp) \, k_2^2
  \, {\cal U}^{\eta_2} ({\vec k}_{2 \perp} , 0_\perp) \rangle
\label{fact4jap}
\end{align}
with the LO impact factor \cite{Balitsky:2012bs}
\begin{align}
  {\tilde I}^{\rho\sigma}_{{\rm LO}}(q, \nu) = {N_c\over 64}
  {(\pi\nu)^{-1}\sinh\pi\nu\over (1+\nu^2)\cosh^2\pi\nu} \Big\{
  \Big({9\over 4}+\nu^2\Big)\Big[1+{\amu
    \over\pi}\Big]P_1^{\rho\sigma} + \Big({11\over 4}+3\,
  \nu^2\Big)\Big[1+{\amu \over\pi}\Big]P_2^{\rho\sigma} \notag \\ +
  \Big({1\over 8}+{\nu^2\over 2}\Big) \Big[1+{\amu
    \over\pi}\Big]P_3^{\rho\sigma}\Big\}.
\label{NLOIFnu}
\end{align}

We now need to calculate the dipole-dipole scattering amplitude
$\langle {\cal U}^{\eta_1} ({\vec k}_{1 \perp} , {\vec z}_\perp) \,
{\cal U}^{\eta_2} ({\vec k}_{2 \perp} , 0_\perp) \rangle$ at LO
including the high-energy leading logarithmic (LO BFKL)
resummation. Each operator $k_i^2 \, {\cal U}^{\eta_i}$ needs to be
evolved using the LO BFKL evolution. To do so we employ the fact that
transverse momentum powers are eigenfunctions of the LO BFKL. Using
Mellin representation (see Eqs.~\eqref{fnu} and \eqref{Ukz2} in
Appendix~A)
\begin{align}
  \label{eq:UMell} 
  \int d^2 z_\perp \, k^2 \, \calu^\eta ({k}, {\vec z}_\perp) =
  \int\limits_{-\infty}^\infty d\nu \, k^{-1 + 2 \, i \, \nu} \,
  f(\nu) \ \calu^\eta (\nu)
\end{align}
with
\begin{align}
  \label{eq:UUU}
  \calu^\eta (\nu) = \int d^2 z_\perp \, \calu^\eta (\nu, {\vec
    z}_\perp)
\end{align}
we write the LO BFKL evolution for the (Mellin moment of the) dipole
amplitude as
\begin{align}
  \label{eq:BFKL_U}
  {\cal U}^{\eta} (\nu) = e^{\bam \, \chi_0 (\nu) \, (\eta - \eta_0)}
  \, {\cal U}^{\eta_0} (\nu),
\end{align}
where $\eta_0$ is some initial rapidity value and $\chi_0 (\nu)$ is
defined in Eqs.~\eqref{eq:repl} and \eqref{eq:LOeig}.

Using Eqs.~\eqref{eq:UMell} and \eqref{eq:BFKL_U} in (\ref{fact4jap})
we have
\begin{align}
  {\cal A}_{\lambda_1 \lambda_2} (q_1, q_2) = & - {i\over 8} \, s \, 
  \, \varepsilon^{\lambda_1 \, *}_{\rho_1} (q_1) \,
  \varepsilon^{\lambda_1}_{\sigma_1} (q_1) \, \varepsilon^{\lambda_2
    \, *}_{\rho_2} (q_2) \, \varepsilon^{\lambda_2}_{\sigma_2} (q_2)
  \int d\nu_1 \, f (\nu_1) \, {\tilde I}^{\rho_1 \sigma_1}_{\rm
    LO}(q_1, \nu_1) \, \left(Q_1^2\right)^{-\half+i \, \nu_1}
  \nonumber\\
  & \times \int d\nu_2 \, f(\nu_2) \, {\tilde I}^{\rho_2
    \sigma_2}_{\rm LO}(q_2, \nu_2) \, \left(Q_2^2\right)^{-\half+i \,
    \nu_2} \, e^{\bam \, \chi_0 (\nu_1) \, (\eta_1 - \eta_0) + \bam \,
    \chi_0 (\nu_2) \, (\eta_0 - \eta_2)} \, \langle \, {\cal
    U}^{\eta_0} ({\nu}_{1}) \, {\cal U}^{\eta_0} (\nu_2) \rangle
  /S_\perp
\label{fact4jap-1}
\end{align}
where $\eta_{1} - \eta_{2} = \ln {s\over Q_1Q_2}$ with $s=
(q_1+q_2)^2$ the center-of-mass energy squared. Since each ${\cal
  U}^{\eta} (\nu)$ in \eq{fact4jap-1} contains an integral over impact
parameters (see \eqref{eq:UUU}), while in the $\gamma^* \gamma^*$
scattering at hand there is only one integration over the impact
parameter (see \eq{fact4jap}), we have to divide out the infinite
factor of $S_\perp = \int d^2 z$ from the correlator in
\eq{fact4jap-1}.

We now need the scattering of two color-dipoles at LO in Mellin space.
This calculation is provided in \cite{Balitsky:2009yp} but in ${\cal
  N}=4$ SYM theory. To obtain the QCD expression we have to modify
only the color factors getting
\begin{align}
  \langle \calu^{\eta_{0}}(\nu_1) \,
  \calu^{\eta_{0}}(\nu_2)\rangle/S_\perp =\left[-{4 \, \pi^2(N^2_c -
      1)\over N^2_c}\right]{\amu^2 \over \nu_1^2 \,
    (1+4\nu_1^2)^2}\delta(\nu_1+\nu_2).
\label{scadiponu}
\end{align}
Using (\ref{scadiponu}) in (\ref{fact4jap-1}) and making use of
(\ref{fnu}) and (\ref{ffnu}) yields
\begin{align} {\cal A}_{\lambda_1 \lambda_2} (q_1, q_2) = & i \, 2 \,
  s \, \pi^2 \varepsilon^{\lambda_1 \, *}_{\rho_1} (q_1) \,
  \varepsilon^{\lambda_1}_{\sigma_1} (q_1) \, \varepsilon^{\lambda_2
    \, *}_{\rho_2} (q_2) \, \varepsilon^{\lambda_2}_{\sigma_2} (q_2)
  \,{\amu^2\over Q_1Q_2} {N^2_c - 1 \over N^2_c} \notag \\
  \times & \, \int\limits_{-\infty}^\infty \!\!  d\nu \, {\tilde
    I}^{\rho_1 \sigma_1}_{\rm LO}(q_1, \nu) \, {\tilde I}^{\rho_2
    \sigma_2}_{\rm LO}(q_2, \nu) \,\left({Q_1^2\over
      Q_2^2}\right)^{i\nu} \, e^{\bam \, \chi_0 (\nu) \, (\eta_1 -
    \eta_2)},
\label{fact4jap-3}
\end{align}
where we have used the fact that ${\tilde I}^{\rho \sigma}_{\rm LO}(q,
-\nu) = {\tilde I}^{\rho \sigma}_{\rm LO}(q, \nu)$ and $\chi_0 (\nu) =
\chi_0 (-\nu)$.

To obtain the total LO $\gamma^* \gamma^*$ cross section we use the
optical theorem 
\begin{align}
  \sigma_{\rm tot}^{\gamma^*\gamma^*} (\lambda_1 , \lambda_2) = 2 \,
  \mbox{Im} \left[ \frac{{\cal A}_{\lambda_1 \lambda_2} (q_1, q_2)}{2
      \, s} \right]
\label{gg-xsect}
\end{align}
to get
\begin{align} \label{ggLOcrossec} \sigma^{\gamma^*\gamma^*}_{\rm LO}
  (\lambda_1 , \lambda_2) = & 2 \, \pi^2 \varepsilon^{\lambda_1 \,
    *}_{\rho_1} (q_1) \, \varepsilon^{\lambda_1}_{\sigma_1} (q_1) \,
  \varepsilon^{\lambda_2 \, *}_{\rho_2} (q_2) \,
  \varepsilon^{\lambda_2}_{\sigma_2} (q_2) \left( \sum_f e_f^2
  \right)^2 {\amu^2\over Q_1Q_2} {N^2_c - 1 \over N^2_c} \notag \\
  & \times \int\limits_{-\infty}^\infty d\nu \, {\tilde I}^{\rho_1
    \sigma_1}_{\rm LO}(q_1, \nu) \, {\tilde I}^{\rho_2 \sigma_2}_{\rm
    LO}(q_2, \nu) \,\left({Q_1^2 \over Q_2^2 }\right)^{i\nu} \,
  \left(\frac{s}{Q_1 \, Q_2} \right)^{\bam \chi_0(\nu)}
\end{align}
where we have replaced $\eta_1 - \eta_2$ by $\ln {s\over Q_1Q_2}$ and
reinstated the quark electromagnetic charges.

To calculate the NLO $\gamma^* \gamma^*$ cross section one could
repeat at the next order in $\bam$ the steps we have just performed to
get the LO cross section.  At NLO, however, it is well known that the
impact factor may get contributions of energy-dependent terms; hence,
if we want to preserve the factorization structure of the LO
scattering amplitude, we need to define the impact factor in such a
way that the NLO impact factor is not dependent on the energy
scale. This result is provided by the high-energy OPE in terms of the
composite Wilson line operator. The composite Wilson line operator is
the dipole operator defined in \eqref{col-dipo} with counterterms that
restore the conformal invariance violated by the energy-dependent
terms present at NLO \cite{Balitsky:2012bs}. The NLO impact factor is
defined as the coefficient function in front of the composite Wilson
line operator. At the moment, the high-energy OPE in terms of
composite Wilson line operator \cite{Balitsky:2009xg} represents the
only known way to systematically define and calculate the impact
factor at any order as the coefficient function in the OPE
language. The energy-independent NLO impact factor was found in
\cite{Balitsky:2012bs} allowing us to proceed with the calculation of
the NLO $\gamma^* \gamma^*$ cross section.

There is another issue which had to be resolved before the calculation
of the NLO $\gamma^* \gamma^*$ cross section is possible. One had to
solve the NLO BFKL equation in order to also include NLO evolution
into the cross section.  The NLO BFKL equation has been solved
recently in \cite{Chirilli:2013kca} by constructing the NLO BFKL
eigenfunctions and the corresponding eigenvalues.

We now have at our hand all the necessary ingredients to proceed to
the calculation of the NLO $\gamma^* \gamma^*$ cross section.


\subsection{Assembling the NLO $\gamma^* \gamma^*$ cross section}

We start with \eq{fact4j-2}, which, with the help of \eq{IF_Mellin} we
rewrite as
\begin{align}
  {\cal A}_{\lambda_1, \lambda_2} (q_1, q_2) = - i \, {s\over 2} \,
  \varepsilon^{\lambda_1 \, *}_{\rho_1} (q_1) \,
  \varepsilon^{\lambda_1}_{\sigma_1} (q_1) \, \varepsilon^{\lambda_2
    \, *}_{\rho_2} (q_2) \, \varepsilon^{\lambda_2}_{\sigma_2} (q_2) &
  \int {d^2 k_1\over (2\pi)^2 \,k_1^2} I^{\rho_1\sigma_1} (q_1,k_1) \int {d^2
    k_2\over (2\pi)^2\, k_2^2} I^{\rho_2\sigma_2} (q_2,k_2) \notag \\ & \times \,
  \langle k_1^2 \, [{\cal U}^{a_1} ({\vec k}_{1 \perp})]^{\rm comp} \, k_2^2 \,
 [{\cal U}^{a_2} ({\vec k}_{2 \perp} )]^{\rm comp} \rangle /S_\perp,
 \label{fact4j-2ap}
\end{align}
where 
\begin{align}
  \label{eq:Ukdef}
  [{\cal U}^a ({\vec k}_{\perp})]^{\rm comp} = \int d^2 z \, [{\cal U}^a
  ({\vec k}_{\perp} , {\vec z}_\perp)]^{\rm comp}.
\end{align}

As shown in \cite{Balitsky:2012bs}, linearization of the NLO evolution
equation for the operator
\begin{align} 
  \label{L-dipole} {\cal L}^a (k) \equiv \frac{k^2 \, [{\cal
      U}^a ({\vec k}_{\perp})]^{\rm comp}}{\as (k^2)}
\end{align}
coincides with the NLO BFKL equation obtained in
\cite{Fadin:1998hr}. Therefore, the eigenfunctions \eqref{eigenf_n_nu}
are the eigenfunctions of the kernel of the linearized equation for
${\cal L}^a (k)$.

Considering only the $n=0$
contribution which dominates in high energy scattering, we use the
completeness relation \eqref{eq:compl} for $n=0$,
\begin{align}
  \label{eq:compl0}
  \int\limits_{-\infty}^\infty \frac{d \nu}{2 \, \pi} \, H_\nu (k) \,
  H_{-\nu} (k') = \delta (k^2 - k'^2)
\end{align}
to rewrite \eq{fact4j-2ap} as
\begin{align}
  {\cal A}_{\lambda_1, \lambda_2} (q_1, q_2) = - i \, {s\over 2} \,
  \varepsilon^{\lambda_1 \, *}_{\rho_1} (q_1) \,
  \varepsilon^{\lambda_1}_{\sigma_1} (q_1) \, \varepsilon^{\lambda_2
    \, *}_{\rho_2} (q_2) \, \varepsilon^{\lambda_2}_{\sigma_2} (q_2)
  \, \int \frac{d \nu_1}{2 \pi} \, \frac{d \nu_2}{2 \pi} \,
  I^{\rho_1\sigma_1} (q_1,\nu_1) \, I^{\rho_2\sigma_2} (q_2,\nu_2)
  \notag \\ \times \left\langle \int\limits_0^\infty d k'^2_1
    H_{-\nu_1} (k'_1) \, {\cal L}^{a_1} (k'_1) \, \int\limits_0^\infty
    d k'^2_2 H_{-\nu_2} (k'_2) \, {\cal L}^{a_2} (k'_2) \right\rangle
  /S_\perp
\label{NLO_A1}
\end{align}
with
\begin{align}
  \label{eq:Iqnu_def}
  I^{\rho \sigma} (q,\nu) = \int \frac{d^2 k}{(2\pi)^2\,k^2} \, I^{\rho \sigma}
  (q,k) \, \as (k^2) \, H_\nu (k)
\end{align}
a projection of the impact factor on the LO+NLO BFKL eigenfunctions.

Since, as we have just mentioned, the functions $H_\nu (k)$ are the
eigenfunctions of the evolution equation for ${\cal L}^a (k)$
\cite{Balitsky:2012bs}, the linearized evolution of ${\cal L}^a (k)$
for $n=0$ can be written as
\begin{align} 
  \label{Lsol} 2 \, a \, {d\over d a}{{\cal L}^a (\nu)} = \Delta(\nu)
  \, {\cal L}^a (\nu)
\end{align}
where
\begin{align}
  \label{eq:int0}
  \Delta(\nu) = \bam \, \chi_0(0,\nu) + \bam^2 \, \chi_1(0,\nu)
\end{align}
and 
\begin{align} 
  \label{Leige} {\cal L}^a (\nu) \equiv \int d k^2 \, H_{-\nu}(k)
  \, {\cal L}^a (k).
\end{align}
We thus rewrite \eq{NLO_A1} as (cf. \eq{fact4jap-1})
\begin{align}
  {\cal A}_{\lambda_1, \lambda_2} (q_1, q_2) = - i \, {s\over 2} \,
  \varepsilon^{\lambda_1 \, *}_{\rho_1} (q_1) \,
  \varepsilon^{\lambda_1}_{\sigma_1} (q_1) \, \varepsilon^{\lambda_2
    \, *}_{\rho_2} (q_2) \, \varepsilon^{\lambda_2}_{\sigma_2} (q_2)
  \, \int \frac{d \nu_1}{2 \pi} \, \frac{d \nu_2}{2 \pi} \,
  I^{\rho_1\sigma_1} (q_1,\nu_1) \, I^{\rho_2\sigma_2} (q_2,\nu_2)
  \notag \\ \times e^{\frac{1}{2} \, \Delta(\nu_1) \, \ln{a_1\over
      a_0} + \frac{1}{2} \, \Delta(\nu_2) \ln ({a_0 \, a_2})}
  \left\langle {\cal L}^{a_0} (\nu_1) \, {\cal L}^{a_0} (\nu_2)
  \right\rangle /S_\perp.
\label{NLO_A2}
\end{align}
Here $(1/2) \ln a_0$ is an arbitrary rapidity for the LO+NLO
dipole--dipole scattering while $(1/2) \ln (a_1 \, a_2) = \ln {s\over
  Q_1Q_2} = \eta_1 - \eta_2$. Note that the expression \eqref{NLO_A2}
for the forward amplitude is quite general, and is likely to be valid
at any order in $\as$ within the linear evolution approximation.

We see that the eigenfunctions of LO+NLO BFKL allowed us to write down
the evolution effects explicitly in \eq{NLO_A2}. To complete the
calculation, we now need to find the projections $I^{\rho\sigma}
(q,\nu)$ of the impact factors onto the eigenfunctions $H_\nu (k)$
(see \eq{eq:Iqnu_def}) and to construct the correlator $\left\langle
  {\cal L}^{\eta_0} (\nu_1) \, {\cal L}^{\eta_0} (\nu_2)
\right\rangle$ at the LO+NLO level. These quantities are calculated in
the Appendix~B. The result for the dipole--dipole forward scattering
amplitude is
\begin{align}\label{L0}
  \left\langle {\cal L}^{a_0} (\nu_1) \, {\cal L}^{a_0} (\nu_2)
  \right\rangle /S_\perp = - 4 \, (2 \, \pi)^4 \, \frac{N_c^2
    -1}{N_c^2} \, \left[ 1 + \bam \, F (\nu_1) \right] \, \delta
  (\nu_1 + \nu_2)
\end{align}
where the NLO correction to the eikonal dipole-dipole scattering was
calculated in \cite{Balitsky:2009yp} for ${\cal N} =4$ SYM, which we
modified to work for QCD obtaining\footnote{The expression for
  $F(\nu)$ found in \cite{Balitsky:2009yp} for ${\cal N} =4$ SYM has
  recently been challenged in \cite{Costa:2013zra} due to it
  apparently violating the principle of maximum
  transcendentality. Note that nothing in our discussion depends on a
  specific form of $F(\nu)$, and the final results in
  Eqs.~\eqref{gg-analytic}, \eqref{ggNLOcrossec} and
  \eqref{ggNLOcrossecLL} would still be valid for a different
  $F(\nu)$.}
\begin{align}\label{Fnu}
  {\rm Re} \left[ F(\nu) \right] = \chi_0(\nu)\left(2\gamma_E -
    {4\over 4\, \nu^2 +1} \right) + \frac{67}{18} - {\pi^2\over 6} -
  \frac{5 \, N_f}{9 \, N_c}.
\end{align}
Note that $\frac{67}{18} - {\pi^2\over 6}$ on the right-hand side of
\eq {Fnu} are the same terms as in the cusp anomalous dimension
\cite{Korchemsky:1987wg,Korchemskaya:1994qp}. The last term in
\eq{Fnu} is not present in
\cite{Korchemsky:1987wg,Korchemskaya:1994qp} since quark loops were
not considered in those works. Let us stress here, that it is
important to include the NLO correction to the ``daughter''
dipole-dipole scattering to obtain a complete expression for the
$\gamma^* \gamma^*$ cross section. Note the difference between the NLO
dipole-dipole scattering and the NLO corrections to the impact factor:
the latter include the ${\cal O} (\as)$ correction to the $\gamma^*
\to q {\bar q}$ light-cone wave function squared due to {\sl
  non-eikonal} $q \to qG$ splittings and mergers; the former includes
${\cal O} (\as^3)$ corrections to the leading-order (${\cal O}
(\as^2)$) scattering of two dipoles made out of pairs of {\sl eikonal}
Wilson lines. (Note that, while we put rapidities of the two operators
in \eq{L0} to be equal, this is done in order to exclude small-$x$
evolution corrections; the ``daughter'' dipole-dipole scattering is
still assumed to be high-energy.)

The calculation in Appendix~B for the LO+NLO impact factor projection
yields
\begin{align}\label{IprojH}
  I^{\rho\sigma} (q,\nu) =  {\amu\over 2} \, \left\{ 1 + \bam \,
    \beta_2 \, \left[ i \, \frac{\chi_0 (\nu)}{2 \, \chi'_0 (\nu)} \,
      (i \, \partial_{\nu} + \ln \mu^2)^2 - \left( \partial_\nu
        \frac{\chi_0 (\nu)}{2 \, \chi'_0 (\nu)} \right) \, (i
      \, \partial_{\nu} + \ln \mu^2) + i \, \partial_{\nu} + \ln \mu^2
    \right] \right\} \notag \\ \times \, Q^{-1 + 2 \, i \, \nu} \,
  {\tilde I}^{\rho\sigma}_{{\rm LO}+{\rm NLO}} (q, \nu)
\end{align}
where ${\tilde I}^{\rho\sigma}_{{\rm LO}+{\rm NLO}} (q, \nu)$ is the
Mellin transform of the LO+NLO impact factor defined in \eq{IF_Mellin}
and $\partial_{\nu} \equiv \partial/\partial \nu$.

Substituting Eqs.~\eqref{L0} and \eqref{IprojH} into \eq{NLO_A2} we
obtain (see Appendix~B for details)
\begin{align}
  & {\cal A}_{\lambda_1 \lambda_2} (q_1, q_2) = i \, 2 \, \pi^2 \, s
  \, \varepsilon^{\lambda_1 \, *}_{\rho_1} (q_1) \,
  \varepsilon^{\lambda_1}_{\sigma_1} (q_1) \, \varepsilon^{\lambda_2
    \, *}_{\rho_2} (q_2) \, \varepsilon^{\lambda_2}_{\sigma_2} (q_2)
  \, {N^2_c-1\over N^2_c} \, \frac{\as (Q_1^2) \, \as (Q_2^2)}{Q_1 \,
    Q_2} \, \int\limits_{-\infty}^\infty d\nu \, \left({Q_1^2\over
      Q_2^2}\right)^{i\nu} \label{gg-analytic} \\ & \times \,
  e^{[\bas(Q_1Q_2) \, \chi_0(\nu) + \bas^2(Q_1Q_2) \, \chi_1(\nu)] \,
    \frac{1}{2} \, \ln({a_1\, a_2})} \left\{ {\tilde I}_{\rm
      LO+NLO}^{\rho_1 \sigma_1}(q_1, \nu) \, {\tilde I}_{\rm
      LO+NLO}^{\rho_2 \sigma_2}(q_2, -\nu) \, \Bigg[ 1 + \bas(Q_1Q_2)
    F(\nu) \Bigg] \right. \notag \\ & \left. - \bas(Q_1Q_2) \, \beta_2
    \, \Bigg[ {\tilde I}_{\rm LO}^{\rho_1 \sigma_1}(q_1, \nu)
    \Big(i\partial_\nu {\tilde I}_{\rm LO}^{\rho_2 \sigma_2}(q_2, \nu)
    \Big) - {\tilde I}_{\rm LO}^{\rho_2 \sigma_2}(q_2,
    \nu)\Big(i\partial_\nu {\tilde I}_{\rm LO}^{\rho_1 \sigma_1}(q_1,
    \nu)\Big) \Bigg]\Bigg(1 + \bas(Q_1Q_2) \, {\chi_0(\nu)\over 4} \,
    \ln({a_1\, a_2}) \Bigg) \right\}. \notag
\end{align}
where $\chi_0 (\nu) = \chi_0 (0, \nu)$ and $\chi_1 (\nu) = \chi_1 (0,
\nu)$. We have also made use of the one-loop running coupling $\as
(Q^2) \approx \amu \, \left[ 1 - \bam \, \beta_2 \, \ln
  \frac{Q^2}{\mu^2} \right]$ with the NLO precision.  Note that the
term in the last line of \eq{gg-analytic} does not contribute if the
polarizations of the two virtual photons are either identical or
summed over.

The prefactor of \eq{gg-analytic} contains $\as (Q_1^2) \, \as
(Q_2^2)$: with the NLO precision this term is indistinguishable from
$\as^2 (Q_1 \, Q_2)$. However, we made this choice because it is known
that in the $Q_1 \gg Q_2$ regime only one of the couplings should
become very small while the other one would encode the
non-perturbative features of the ``target'' like in DIS. An explicit
higher-order calculation (see Appendix~A of \cite{Kovchegov:2007vf})
employing the Brodsky--Lepage--Mackenzie (BLM) prescription
\cite{Brodsky:1983gc} confirmed this result.

The $\gamma^*\gamma^*$ cross section is obtained from \eq{gg-analytic}
using the optical theorem. For transversely polarized photons the
cross section is\footnote{Note that $P_2^{\rho_2 \sigma_2}$ in
  ${\tilde I}_{\rm LO+NLO}^{\rho_2 \sigma_2}$ is given by \eq{P2} with
  $p_2$ replaced by $p_1^\mu = (p_1^+, 0, 0_\perp)$.}
\begin{align}
  \label{ggNLOcrossec}
  \sigma^{\gamma^*\gamma^*}_{\rm LO+NLO} & \, (TT) = \frac{1}{4}
  \sum_{\lambda_1, \, \lambda_2 = \pm 1} \,2 \, \pi^2 \,
  \varepsilon^{\lambda_1 \, *}_{\rho_1} (q_1) \,
  \varepsilon^{\lambda_1}_{\sigma_1} (q_1) \, \varepsilon^{\lambda_2
    \, *}_{\rho_2} (q_2) \, \varepsilon^{\lambda_2}_{\sigma_2} (q_2)
  \, {N^2_c-1\over N^2_c} \left( \sum_f e_f^2
  \right)^2 \frac{\as (Q_1^2) \, \as (Q_2^2)}{Q_1 \,
    Q_2} \, \int\limits_{-\infty}^\infty d\nu \, \left({Q_1^2\over
      Q_2^2}\right)^{i\nu} \notag \\ \times \, & \left(\frac{s}{Q_1 \,
      Q_2} \right)^{\bas(Q_1Q_2) \, \chi_0(\nu) + \bas^2(Q_1Q_2) \,
    \chi_1(\nu)} \, {\tilde I}_{\rm LO+NLO}^{\rho_1 \sigma_1}(q_1,
  \nu) \, {\tilde I}_{\rm LO+NLO}^{\rho_2 \sigma_2}(q_2, -\nu) \,
  \Bigg[ 1 + \bas(Q_1Q_2) {\rm Re}[F(\nu)] \Bigg],
\end{align}
while for the longitudinally polarized gluons we get
 \begin{align}
  \label{ggNLOcrossecLL}
  \sigma^{\gamma^*\gamma^*}_{\rm LO+NLO} & \, (LL) = 2 \, \pi^2 \,
  \varepsilon^{L \, *}_{\rho_1} (q_1) \, \varepsilon^{L}_{\sigma_1}
  (q_1) \, \varepsilon^{L \, *}_{\rho_2} (q_2) \,
  \varepsilon^{L}_{\sigma_2} (q_2) \, {N^2_c-1\over N^2_c} \left(
    \sum_f e_f^2 \right)^2 \frac{\as (Q_1^2) \, \as (Q_2^2)}{Q_1 \,
    Q_2} \, \int\limits_{-\infty}^\infty d\nu \, \left({Q_1^2\over
      Q_2^2}\right)^{i\nu} \notag \\ \times \, & \left(\frac{s}{Q_1 \,
      Q_2} \right)^{\bas(Q_1Q_2) \, \chi_0(\nu) + \bas^2(Q_1Q_2) \,
    \chi_1(\nu)} \, {\tilde I}_{\rm LO+NLO}^{\rho_1 \sigma_1}(q_1,
  \nu) \, {\tilde I}_{\rm LO+NLO}^{\rho_2 \sigma_2}(q_2, -\nu) \,
  \Bigg[ 1 + \bas(Q_1Q_2) {\rm Re}[F(\nu)] \Bigg],
\end{align}
where we have also used $\frac{1}{2} \ln (a_1\, a_2) = \ln
\frac{s}{Q_1 \, Q_2}$ along with the fact that $F (\nu) = F
(-\nu)$. The latter condition, while satisfied by $F (\nu)$ in
\eq{Fnu}, is also true in general, since it follows from the symmetry
of the dipole-dipole scattering under the interchange of the dipoles.
In Eqs.~\eqref{ggNLOcrossec} and \eqref{ggNLOcrossecLL}, ${\tilde
  I}_{\rm LO+NLO}^{\rho\sigma}(q, \nu)$ are given by (\ref{NLOIFnu1})
with $\bam\to \bas(Q_1Q_2)$ and ${\rm Re}[F(\nu)]$ is given by
\eq{Fnu} \cite{Balitsky:2009yp}. Note that ${\tilde I}_{\rm
  NLO}^{\rho\sigma}(q, \nu)\neq {\tilde I}_{\rm NLO}^{\rho\sigma}(q,
-\nu)$. In \eq{ggNLOcrossec} the transverse polarizations are
$\varepsilon^{\lambda, \, \mu} = (0,0, {\vec
  \varepsilon}_\perp^\lambda)$ with ${\vec \varepsilon}_\perp^\lambda
= (-1/\sqrt{2}) (\lambda,i)$. The longitudinal polarizations in
\eq{ggNLOcrossecLL} are $\varepsilon^{L, \, \mu} (q) = (q^+/Q, Q/2
q^+, 0_\perp)$.

Eqs.~\eqref{gg-analytic}, \eqref{ggNLOcrossec} and
\eqref{ggNLOcrossecLL} are the main results of this Section, giving us
the NLO forward scattering amplitude and the transverse (TT) and
longitudinal (LL) cross sections for the $\gamma^* \gamma^*$
scattering. Comparing Eqs.~\eqref{ggNLOcrossec},
\eqref{ggNLOcrossecLL} with \eqref{ggLOcrossec}, we make an important
observation that the structure of the NLO $\gamma^* \gamma^*$ cross
section is the same as that at the LO with the addition of the running
of the coupling and the corrections to the impact factors, intercept,
and the ``daughter'' dipole-dipole scattering. The $\gamma^* \gamma^*$
cross sections for different polarizations (e.g. TL or LT) can be
constructed using \eq{gg-analytic} and the optical theorem.

An equation similar to \eqref{ggNLOcrossec} and
\eqref{ggNLOcrossecLL}, but for the $\gamma^* \gamma^* \to \rho\rho$
process has been obtained in
\cite{Caporale:2008is,Ivanov:2006gt,Ivanov:2005gn} (see Eq.~(43) in
\cite{Ivanov:2005gn}). The approach of
\cite{Caporale:2008is,Ivanov:2006gt,Ivanov:2005gn} does not involve
constructing the NLO BFKL eigenfunctions: rather NLO corrections are
included into the $\gamma^* \gamma^* \to \rho\rho$ cross section, with
Eq.~(43) in \cite{Ivanov:2005gn} containing a power-of-energy form of
the NLO evolution correction as a conjecture for how higher-order
iterations of the NLO evolution corrections would come in. The power
of energy in Eq.~(43) of \cite{Ivanov:2005gn} agrees with the power of
$s/(Q_1\, Q_2)$ in our \eq{ggNLOcrossec}, with the latter being an
exact NLO BFKL eigenvalue resulting from construction of the NLO BFKL
eigenfunctions in \cite{Chirilli:2013kca}. The $\gamma^* \gamma^* \to
\rho\rho$ process considered in \cite{Ivanov:2005gn} is different (in
the impact factors) from the $\gamma^* \gamma^* \to \gamma^* \gamma^*$
process that Eqs.~\eqref{ggNLOcrossec} and \eqref{ggNLOcrossecLL} were
derived for. Therefore, a complete side-by-side comparison of the two
results is not possible at this point. Further work is needed to
relate our approach in this paper and in \cite{Chirilli:2013kca} to
that of \cite{Caporale:2008is,Ivanov:2006gt,Ivanov:2005gn}.


\subsection{Numerical result of the $\gamma^* \gamma^*$ cross-section}

The NLO $\gamma^* \gamma^*$ cross section we obtained in
Eqs.~\eqref{ggNLOcrossec} and \eqref{ggNLOcrossecLL} is an analytic
result: it is a convolution of the two photon impact factors at NLO
and the energy-dependent factor given by the exponentiation of the
eigenvalues of the NLO BFKL equation.  In this section we plot the
result of the cross section as a function of rapidity $Y = \ln (s/Q_1
Q_2)$ and for several values of the virtualities of the two scattering
photons.

\begin{figure}[ht]
\begin{center}
\includegraphics[width=0.45 \textwidth]{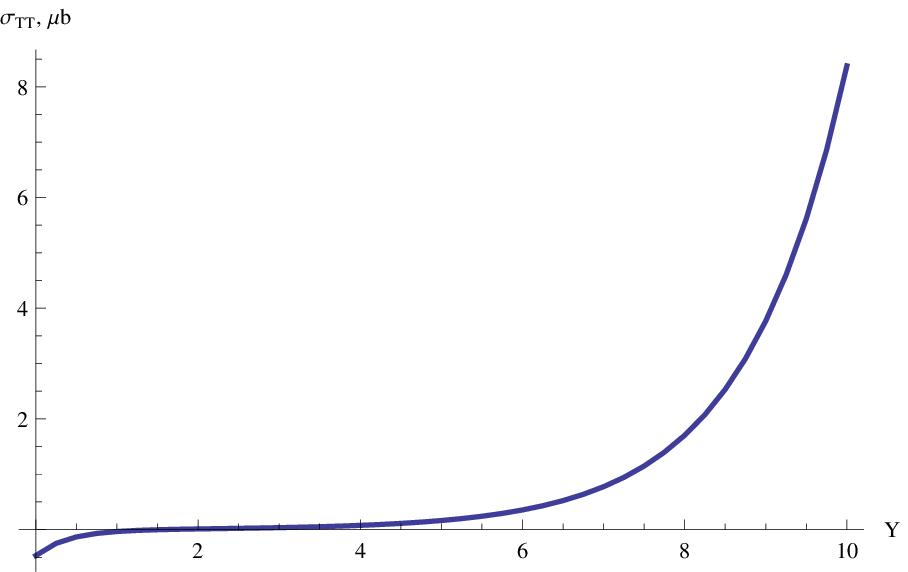} \ \ \
\includegraphics[width=0.45 \textwidth]{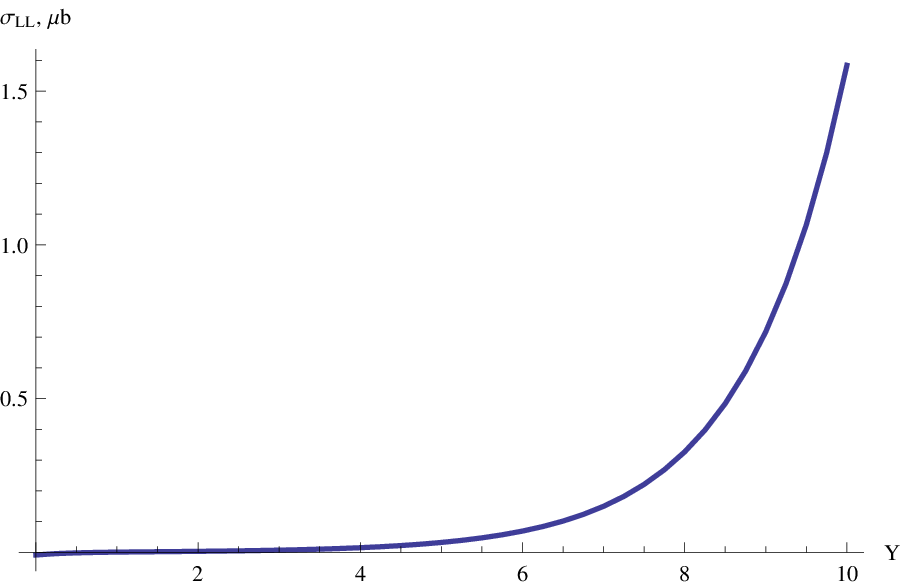}
\caption{NLO $\gamma^* \gamma^*$ cross section with $Q_1=Q_2=1$~GeV
  plotted as a function of rapidity $Y$. Virtual photons have
  transverse polarizations in the left panel and longitudinal
  polarizations in the right panel.}
\label{NLOsigma1LT}
\end{center}
\end{figure}

We plot the $\gamma^* \gamma^*$ cross section as a function of
rapidity for virtual photons with momenta $Q_1=Q_2 = 1$~GeV in
\fig{NLOsigma1LT}. The left panel of \fig{NLOsigma1LT} depicts the TT
cross section from \eq{ggNLOcrossec}, while the right panel contains
the LL cross section from \eq{ggNLOcrossecLL}. The one-loop coupling
constant in our results \eqref{ggNLOcrossec} and
\eqref{ggNLOcrossecLL} runs with the momenta of the virtual
photons. In \fig{NLOsigma1LT} we use \eq{eq:beta2} with $N_f=3$ and
$\Lambda_{\rm QCD}=250$~MeV; hence $\alpha_s (1 {\rm GeV}^2) \approx
0.50$.  Fitting the curves in \fig{NLOsigma1LT} with $e^{\Delta \, Y}$
we make an estimate of the effective intercept; we get $\Delta \approx
0.78$ for the TT cross section and $\Delta \approx 0.75$ for the LL
cross section. Compared to the LO BFKL intercept $\Delta_{LO} \approx
\frac{4 \as \, N_c}{\pi} \, \ln 2 \approx 1.33$ for $\as \approx 0.50$
we conclude that the NLO BFKL evolution indeed tends to significanlty
reduce the intercept compared to the LO result. However, as is clear
from \fig{NLOsigma1LT}, the LO+NLO cross section keeps growing with
rapidity even for rather large values of the coupling $\as \approx
0.50$. This appears to be in qualitative agreement with the numerical
solution of the NLO BFKL equation in \cite{Andersen:2003wy}. Note also
that the shape of our cross sections in \fig{NLOsigma1LT} is similar
to that for the $\gamma^* \gamma^* \to \rho\rho$ cross section from
\cite{Ivanov:2005gn}, though a detailed numerical comparison between
the two can not be carried out due to different scattering processes
considered.

\begin{figure}[ht]
\begin{center}
\includegraphics[width=0.45 \textwidth]{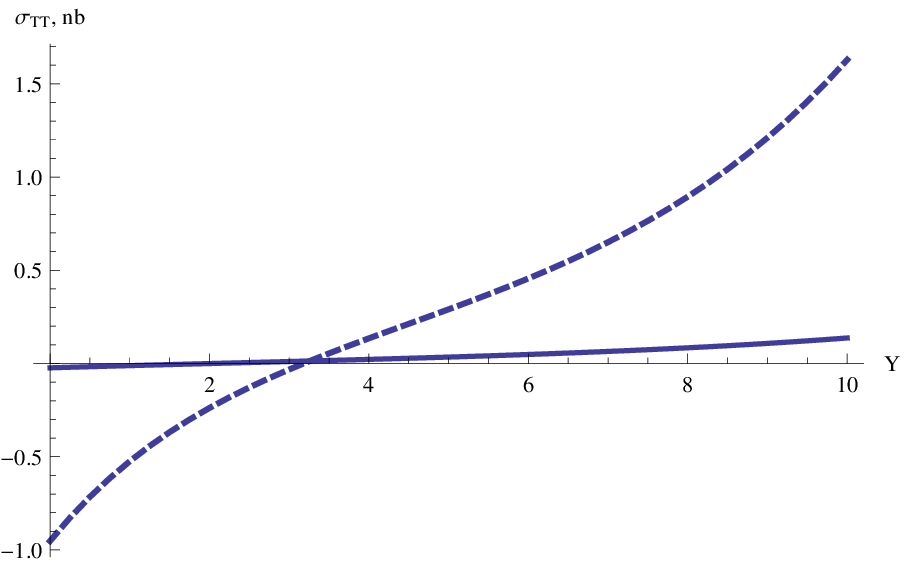} \ \ \
\includegraphics[width=0.45 \textwidth]{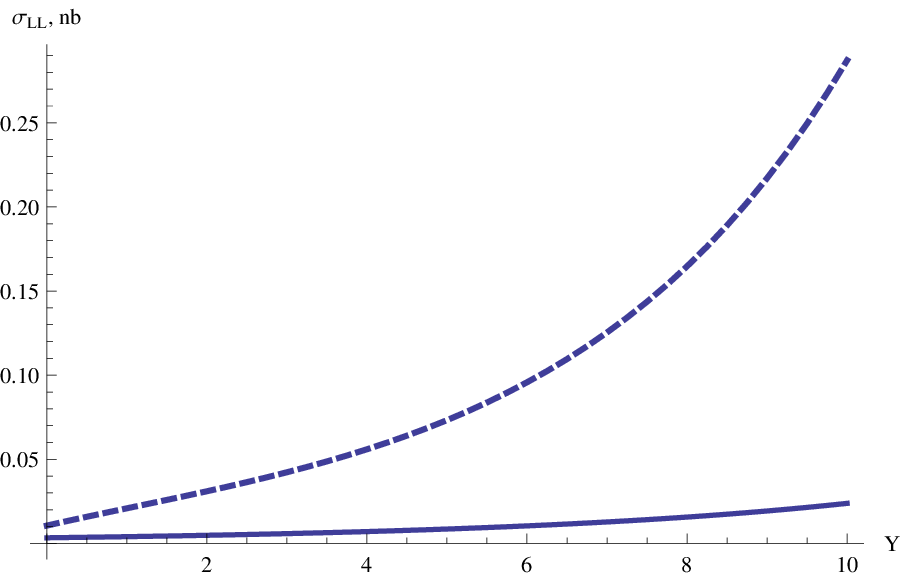}
\caption{NLO $\gamma^* \gamma^*$ cross section with $Q_1=Q_2=5$~GeV
  (dashed line) and $Q_1=Q_2=10$~GeV (solid line) plotted as functions
  of rapidity $Y$. Again the virtual photons have transverse
  polarizations in the left panel and longitudinal polarizations in
  the right panel.}
\label{NLOsigma5LT}
\end{center}
\end{figure}

In \fig{NLOsigma5LT} we plot the TT (left panel) and LL (right panel)
cross sections for $Q_1=Q_2=5$~GeV (dashed line) and $Q_1=Q_2= 10$~GeV
(solid line). In \fig{NLOsigma5LT} we use $N_f = 5$ such that
$\alpha_s ((5 {\rm GeV})^2) = 0.27$ and $\alpha_s ((10 {\rm GeV})^2)=
0.22$ respectively. The estimates of the intercepts are as follows:
$\Delta \approx 0.34$ for the TT cross section and $\Delta \approx
0.32$ for the LL cross section for $Q_1=Q_2=5$~GeV and $\Delta \approx
0.27$ for both the TT and LL cross sections at $Q_1=Q_2=
10$~GeV. Comparing this to $\Delta_{LO} \approx 0.73$ for
$Q_1=Q_2=5$~GeV and $\Delta_{LO} \approx 0.59$ for $Q_1=Q_2= 10$~GeV
we again see a reduction of the intercept at NLO, though again it
remains positive corresponding to a cross section growing with $Y$.

We notice that the all the plotted cross sections are negative for
small values of rapidity. Similar behavior has been observed in
\cite{Ivanov:2005gn}. In fact, negativity takes place without BFKL
evolution at $Y=0$: we see that the NLO correction to the daughter
dipole--dipole cross section (corresponding to the forward amplitude
in \eq{L0}) is numerically large and negative. Comparing the curves in
Figs.~\ref{NLOsigma1LT} and \ref{NLOsigma5LT} we see that negativity
tends to diminish with inclreasing $Q$/decreasing $\as$, which
supports this interpretation. We have verified that at very large
$Q_1, \, Q_2$ the cross sections become positive-definite for all
values of $Y$.

We would like to point out that at small rapidities, $Y \approx 0$,
there are corrections to the dipole--dipole cross section which are
inversely proportional to $s$ and are given by the diagrams like that
in \fig{graph-fact4j}b along with the subleading contributions to the
diagram in \fig{graph-fact4j}a. Such corrections were neglected in
deriving \eq{L0} \cite{Balitsky:2009yp}. However, such corrections are
likely to become important near $Y \approx 0$, modifying the
dipole--dipole cross section and possibly making it positive again
even for the range of $Q_1, Q_2$ considered here. Hence the negativity
of the cross sections in Figs.~\ref{NLOsigma1LT} and \ref{NLOsigma5LT}
at small $Y$ is not necessarily a sign of the failure of the
formalism, but rather an indication that energy-suppressed corrections
need to be added in that region. Since the NLO cross sections in
Eqs.~\eqref{ggNLOcrossec} and \eqref{ggNLOcrossecLL} are likely to
dominate over the energy-suppressed corrections at rapidities of the
order $Y \approx 2 \div 3$, we conclude that our NLO result gives a
reliable prediction for the $\gamma^* \gamma^*$ cross section only for
sufficiently large values of $Y$, e.g. for $Y \gsim 2$ in the
kinematics of the plots in Figs.~\ref{NLOsigma1LT} and
\ref{NLOsigma5LT}.

At very high rapidity ($Y=Y_U$) the unitarity corrections due to
nonlinear evolution will become important
\cite{Mueller:1994jq,Braun:1995hh,Mueller:1995gb,Salam:1995uy,Mueller:1996te};
hence the validity region of Eqs.~\eqref{ggNLOcrossec} and
\eqref{ggNLOcrossecLL} in rapidity is also limited from above, $Y <
Y_U$. At aymptotically small values of $\as$ a simple parametric
estimate (see e.g. \cite{Kovchegov:1999ua}) shows that unitarity
corrections become important at $Y \approx Y_U \sim (1/\as) \ln
(1/\as)$, which is lower than the rapidity at which NLO corrections
become large, $Y \sim 1/\as^2$ (obtained by requiring that $\as^2 \, Y
\sim 1$). However, for realistic values of $\as$ the situation is not
so clear-cut, and it is possible that the NLO corrections discussed
here would delay the onset of saturation effects.

Note also that if the cross sections in Figs.~\ref{NLOsigma1LT} and
\ref{NLOsigma5LT} are replotted using Re~$[F(\nu)]$ from \eq{Fnu}
without the second term in parenthesis on its right-hand side ($\sim
4/(4 \nu^2 +1)$), the existence of which was challenged in
\cite{Costa:2013zra} as violating the principle of maximum
transcedentality, they become positive-definite for all values of $Y$
since in this case the NLO correction to the dipole--dipole
cross section is also positive-definite. However, such cross sections
are not monotonic functions of $Y$ and tend to decrease at small $Y$
before the growth picks up at larger $Y$.


\section{Solving the NNLO BFKL Equation}
\label{sec:nnlo}


\subsection{Defining the BFKL Equation Beyond NLO}
\label{sec:bey}

This Section is dedicated to constructing a solution to the NNLO BFKL
equation in QCD. However, before we begin constructing the solution,
it appears necessary to clarify what we mean by the BFKL equation
beyond NLO.\footnote{We thank Jochen Bartels and Misha Braun for
  pointing out to us the need to better define BFKL evolution beyond
  next-to-leading order.}

Consider first a scattering of two quarkonia, say, resulting from
$\gamma^* \gamma^*$ decays as considered in the previous Section,
mediated by the LO BFKL evolution. In terms of parameters $\as$ and
$Y$ the corresponding cross section is proportional to
\begin{align}\label{LO_param}
  \sigma^{\gamma^* \gamma^*}_{LO} \sim \as^2 \, e^{\text{const} \, \as
    \, Y} \sim \as^2,
\end{align}
with the last transition done in the power counting of \eq{counting}
in which $\as \, Y \sim 1$. The NLO cross section of
Sec.~\ref{sec:gamgam}, with the NLO corrections included in the impact
factors, evolution, and daughter dipole-dipole scattering is
parametrically
\begin{align}\label{NLO_param}
  \sigma^{\gamma^* \gamma^*}_{LO+NLO} \sim (\as^2 + \as^3) \,
  e^{\text{const} \, (\as + \as^2) \, Y} \sim \as^2 + \as^3.
\end{align}
It is natural to assume that NNLO BFKL evolution, along with NNLO
corrections to the impact factors and daughter dipole scattering would
lead to the $\gamma^* \gamma^*$ cross section of the form
\begin{align}\label{NNLO_param}
  \sigma^{\gamma^* \gamma^*}_{LO+NLO+NNLO} \sim (\as^2 + \as^3 +
  \as^4) \, e^{\text{const} \, (\as + \as^2 + \as^3) \, Y} \sim \as^2
  + \as^3 + \as^4.
\end{align}

\begin{figure}[h]
  \includegraphics[width=3.9in]{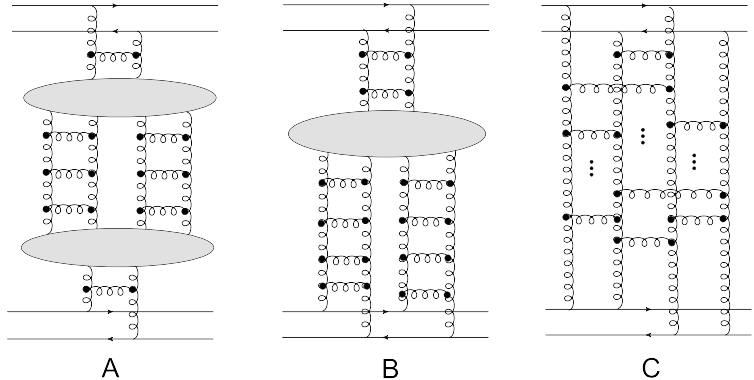}
  \caption{Examples of diagrams contributing to the dipole--dipole
    scattering cross section at the same order in $\as$ and $Y$ as the
    NNLO BFKL ladder exchange.}
\label{multi} 
\end{figure}

While this is the correct NNLO BFKL contribution to the $\gamma^*
\gamma^*$ cross section, there are in fact other corrections
contributing to the $\gamma^* \gamma^*$ cross section at the
order-$\as^4$ in our power counting. The corrections result from
multiple BFKL pomeron exchanges and from interactions mediated by the
Bartels-Kwiecinski-Praszalowicz (BKP) multi-reggeon states
\cite{Bartels:1980pe,Kwiecinski:1980wb}. These are schematically shown
in \fig{multi}, which depicts examples of a pomeron loop in panel A,
pomeron fan diagram in panel B, and a 4-reggeon BKP state in panel C.

Single-splitting fan diagram or a single-loop diagram in \fig{multi}
give the following parametric contribution to the onium--onium cross
section \cite{Mueller:1994jq,Mueller:1995gb}
\begin{align}\label{NNLO_param2}
  \sigma^{\gamma^* \gamma^*}_{fan \ or \ loop} \sim \left[ \as^2 \,
    e^{\text{const} \, \as \, Y} \right]^2 \sim \as^4,
\end{align}
which, in our counting, is of the same order as the NNLO BFKL
(single-ladder) contribution to the cross section. The diagrams in
\fig{multi} are, clearly, not included in the BFKL evolution
equation. We conclude that the NNLO BFKL description of the
dipole--dipole scattering receives order-$1$ corrections from the
diagrams in \fig{multi}, at least in terms of the $\as$ and $Y$ power
counting \eqref{counting}.

At the same time, one can define BFKL evolution with an arbitrary
order kernel in $\as$ by linearizing the arbitrary-order BK/JIMWLK
evolution equations for scattering of a dipole on a dense target. In
the large-$N_c$ limit, where BK equation for the dipole amplitude is
valid to arbitrary order, mathematically it can always be linearized
giving us the arbitrary-order BFKL equation. Beyond large-$N_c$ one
can use the (first) equation in the Balitsky-JIMWLK hierarchy, the
equation for the fundamental dipole: linearizing this equation one
would still recover the arbitrary-order BFKL equation, now without the
large-$N_c$ constraint.  We see that mathematically one can always
define BFKL evolution to an arbitrary order in $\as$ as a
linearization of the dipole evolution equation.

Note also that linearizing the higher-order evolution of the
correlators of more than two Wilson lines, such as the fundamental
quadrupole \cite{JalilianMarian:2004da,Dominguez:2011gc}, and
subtracting out the BFKL evolution pieces, allows one to also
construct the BKP evolution equation \cite{Altinoluk:2013rua} with the
kernel calculated to any higher order in the coupling. Hence, in the
$s$-channel language, the BFKL evolution is defined as the linearized
evolution of a dipole, while the BKP evolution results from the
linearized equation for correlators of higher number of Wilson lines
(with BFKL exchanges subtracted out).

Indeed a physical question arises concerning the existence of a region
where this linearization is justified, in the sense that the BFKL
evolution obtained by BK/JIMWLK dipole evolution linearization
dominates the scattering cross section. We can imagine two regimes in
which an arbitrary-order BFKL evolution dominates over non-linear
corrections.

Consider DIS on a dense target, in which saturation effects could be
important. According to the conventional wisdom, the nonlinearities
can be neglected for $Q \gg Q_s (Y)$, with $Q_s (Y)$ the saturation
scale describing the target for scattering with the total rapidity
interval $Y$. (Note that saturation effects may still come in through
the initial conditions to the linear BFKL evolution though
\cite{Gribov:1984tu,Mueller:2002zm,Kharzeev:2003wz}.) For BFKL to be
valid it is also important to make sure that the photon virtuality
$Q^2$ is not large enough for the DGLAP evolution effects to become
large. That is, we would want $\as \ln Q^2/Q_s^2 (Y) \ll 1$. In the
end we obtain the following range of $Q$ in which the arbitrary-order
BFKL evolution dominates the total DIS cross section
\begin{align}
Q_s^2 (Y) \ll Q^2 \ll Q_s^2 (Y) \, e^{\text{const} /\as}. 
\end{align}
For small enough $\as$ this region could be large. In the
dilute--dilute $\gamma^* \gamma^*$ scattering case it appears to be
less clear how to define a similar kinematic regime of BFKL dominance:
in this case the large-$N_c$ limit comes in handy.

The other regime in which the nonlinearities associated with multiple
pomeron exchanges and other multi-reggeon states could be neglected is
the very large-$N_c$ limit.  When $N_c$ is very large, all
interactions are suppressed, but the non-linear (or multi-reggeon)
interactions are more suppressed than the linear BFKL
interactions. Namely, a single-BFKL ladder exchange in the
dipole--dipole scattering is parametrically of the order
\begin{align}\label{LO_paramN}
  \sigma^{\gamma^* \gamma^*} \sim (\as^2 + \as^3 \, N_c + \ldots ) \,
  e^{\text{const} \, [\as \, N_c + (\as \, N_c)^2 + \ldots] \, Y} \sim
  \frac{1}{N_c^2},
\end{align}
where we now keep track of the powers of $\as$ and leading powers of
$Y$ and $1/N_c$. In the last step in \eq{LO_paramN} we have employed
the 't Hooft large-$N_c$ limit. One can show that the multiple
exchange diagrams, like those in \fig{multi}, are suppressed by at
least two powers of $N_c$ compared to the single BFKL ladder exchange
\eqref{LO_paramN}. This is clear for the pomeron loop and fan diagrams
(panels A and B) in \fig{multi}, since they are explicitly non-planar,
and, hence, $N_c$-suppressed. For the BKP evolution in panel C the
$N_c$ suppression becomes apparent after one subtracts out of it the
single- and double-pomeron exchange contributions (see
e.g. \cite{Altinoluk:2013rua}). In the end we see that
 \begin{align}\label{NNLO_param3}
\sigma^{\gamma^* \gamma^*}_{fan / loop / BKP} \sim \frac{1}{N_c^4},
\end{align}
and these corrections can be neglected when compared to the single
BFKL ladder exchange in the large (but finite) $N_c$ limit. This is
why in the large-$N_c$ limit BFKL equation is dominant up to any order
in the coupling $\as$. This conclusion is confirmed by the fact that
it makes sense to define BFKL equation even at strong coupling, as
shown in \cite{Brower:2006ea,Janik:1999zk} using the anti-de Sitter
space/conformal field theory (AdS/CFT) correspondence
\cite{Maldacena:1997re,Witten:1998qj}. (Note that BK evolution, while
being large-$N_c$ in the projectile wave function, does resum
subleading-$N_c$ corrections in the multiple interactions with the
target: while being $1/N_c^2$-suppressed, such corrections are
enhanced by powers of $A^{1/3}$ for a nuclear target with atomic
number $A$, or, equivalently, by powers of color charge density $\rho$
of partons in a proton or a nucleus.)

To summarize, we have argued that it is possible mathematically to
define the BFKL equation at any order of the perturbation theory as
the linearization of the dipole evolution equation. Moreover, we have
identified two physical situations in which the BFKL exchange
dominates over other interactions. This clarifies what we mean by
solving the BFKL equation at NNLO, as we will do in the remainder of
this Section.


\subsection{Eigenfunctions and Eigenvalues of the NNLO BFKL Equation}

In this Section we derive the eigenfunction of the NNLO BFKL equation
in the $n=0$ azimuthally-symmetric case. Generalization to $n\neq 0$
case can be easily accomplished along the lines of
Sec.~\ref{sec:azim}.

Similarly to the NLO eigenfunction case we obtained in our previous
paper \cite{Chirilli:2013kca}, we look for NNLO BFKL eigenfunctions by
adding perturbative corrections to the LO eigenfunction up to NNLO in
the following way:
\begin{align}
  H_{\frac{1}{2}+i \, \nu} (k) = k^{-1 + 2\, i \, \nu} \, \left[ 1 +
    \bar{\alpha}_\mu \,f^{(1)}_\nu (k) + \bar{\alpha}_\mu^2 \,
    f^{(2)}_\nu (k) \right].
      \label{Hnnlo}
\end{align}
The NLO correction to the BFKL eigenfunctions was found in
\cite{Chirilli:2013kca} yielding (cf. \eq{eigenf_n_nu})
\begin{align}
  \label{eq:f1}
  f^{(1)}_\nu(k) = \beta_2 \left[ i \, {\chi_0 (\nu) \over 2 \,
      \chi'_0 (\nu)} \, \ln^2{k^2\over \mu^2} + \frac{1}{2} \left(
      {\partial_\nu}\frac{\chi_0 (\nu)}{ \chi_0' (\nu)} \right)
    \ln{k^2\over \mu^2} \right].
\end{align}

Our goal is to construct the NNLO correction $f^{(2)}_\nu (k)$ by
making sure functions $H_{\frac{1}{2}+i \, \nu} (k)$ are
eigenfunctions of the LO+NLO+NNLO BFKL kernel (cf. \eq{eq:LONLO})
\begin{align}
  \label{eq:LONLONNLO}
  K^{{\rm LO}+ {\rm NLO} + {\rm NNLO}} (k,q) \equiv \bar{\alpha}_\mu
  \, K^{\rm LO} (k,q) + \bar{\alpha}_\mu^2 \, K^{\rm NLO} (k,q) +
  \bar{\alpha}_\mu^3 \, K^{\rm NNLO} (k,q)
\end{align}
to order-$\amu^3$. (The full BFKL kernel is $K(k,q) = K^{{\rm LO}+
  {\rm NLO} + {\rm NNLO}}(k,q) + {\cal O} (\amu^4)$.) Indeed the full
NNLO BFKL kernel is not known at the time of writing this
paper. However, we will not need to know the NNLO BFKL kernel in order
to find its eigenfunctions. In \cite{Chirilli:2013kca} we have
constructed the NLO BFKL kernel's eigenfunctions employing only the
projection of the NLO BFKL kernel onto the LO BFKL eigenfunctions
(powers of momentum). Similarly here we will only need a projection of
the NNLO BFKL kernel onto the LO BFKL eigenfunctions. While this
projection is also not known in general, its structure could be
inferred from that of the NLO kernel projection \eqref{conf-proj}: the
NNLO projection should consist of an order-$\amu^3$ conformal part
along with the $k$-dependent one- and two-loop running coupling
corrections to the LO+NLO projection \eqref{conf-proj}. We thus write
(cf. Eq.~(63) in \cite{Chirilli:2013kca})
\begin{align}
  \int d^2 q \, K^{{\rm LO}+ {\rm NLO}+ {\rm NNLO}}(k,q) \ q^{-1+2 i
    \nu} = \left\{ \bar{\alpha}_\mu \, \chi_0 (\nu) \left[ 1-
      \bar{\alpha}_\mu \, \beta_2 \, \ln{k^2\over \mu^2} +
      \bar{\alpha}_\mu^2 \beta_2^2 \, \ln^2 {k^2\over \mu^2} +
      \bar{\alpha}_\mu^2 \, \beta_3 \, \ln{k^2\over \mu^2} \right]
  \right. \notag \\ \left. + \bar{\alpha}_\mu^2 \, \left[ \frac{i}{2}
      \, \beta_2 \, \chi'_0 (\nu) + \chi_1 (\nu) \right] \left[ 1- 2\,
      \bar{\alpha}_\mu \, \beta_2 \, \ln{k^2\over \mu^2} \right] +
    \bar{\alpha}_\mu^3 \, \left[ \chi_2 (\nu) + i \, \delta_2 (\nu)
    \right] \right\} \, k^{-1 + 2 i \nu}.
      \label{prjectLOeigf}
\end{align}
Here the two-loop QCD beta-function is given by
\begin{align}
  \mu^2{d\bar{\alpha}_\mu\over d\mu^2} = -\beta_2 \,
  \bar{\alpha}_\mu^2 + \beta_3 \, \bar{\alpha}_\mu^3.
\end{align}

An important point in the determination of the NLO BFKL eigenfunctions
in \cite{Chirilli:2013kca} was to show that the corresponding
eigenvalues were real since the NLO kernel acts as a Hermitean
operator on its eigenfunctions. For this reason, although we do not
need to know exactly the expression of the NNLO BFKL kernel, we have
to split the conformal part of its projection on to the LO
eigenfunctions into the imaginary (odd under the $\nu\to -\nu$) and
real (even under $\nu\to -\nu$) contributions. Thus, we define the
real part of the conformal contribution as $\chi_2(\nu)$ and the
imaginary part one as $i \, \delta_2 (\nu)$ in \eq{prjectLOeigf}. Let
us stress again that at this point neither $\chi_2(\nu)$ nor $\delta_2
(\nu)$ are known: however, at the end of this calculation we will
obtain an explicit expression for $\delta_2 (\nu)$ (see \eq{delta}
below).

Let us consider the action of the NNLO BFKL kernel on the NNLO
eigenfunction (\ref{Hnnlo}). We require that the eigenfunction
condition is satisfied,
\begin{align}\label{eig_eq}
  \int d^2 q \, K^{{\rm LO}+ {\rm NLO}+ {\rm NNLO}}(k,q) \,
  H_{\frac{1}{2}+i \, \nu} (q) = \Delta (\nu) \, H_{\frac{1}{2}+i \,
    \nu} (k),
\end{align}
with $\Delta(\nu)$ denoting the BFKL eigenvalue. The LO and NLO
contributions to the eigenvalue are known
\cite{Balitsky:1978ic,Chirilli:2013kca} (see \eqref{eq:eig}). We
parametrize the NNLO contribution to the eigenvalue in terms of
another unknown function $c^{(2)}(\nu)$ such that
\begin{align}
  \label{eq:eig_NNLO}
  \Delta(\nu) = \bam \, \chi_0 (\nu) + \bam^2 \, \chi_1 (\nu) + \bam^3
  \, \left[ \chi_2 (\nu) + i \, \delta_2 (\nu) + c^{(2)}(\nu) \right]
  + {\cal O} (\bam^4).
\end{align}
Our task is to substitute Eqs.~\eqref{Hnnlo} and \eqref{eq:eig_NNLO}
into \eq{eig_eq} and find $f^{(2)}_\nu (k)$ and $c^{(2)}(\nu)$
satisfying the resulting equation at order-$\bam^3$.

We begin by evaluating the left-hand-side of \eq{eig_eq} at
order-$\bam^3$: using Eqs.~(\ref{Hnnlo}) and \eqref{eq:LONLONNLO} we
write
\begin{align}\label{lhs1}
  \int d^2 q \, & K^{{\rm LO}+ {\rm NLO}+ {\rm NNLO}}(k,q)
  H_{\frac{1}{2}+i \, \nu} (q) \, \bigg|_{{\cal O} (\bam^3)} \notag \\
  & = \bam^3 \, \int d^2 q \, \left[ K^{{\rm NNLO}} (k,q) \, q^{-1 + 2
      \, i \, \nu} + K^{{\rm NLO}} (k,q) \, q^{-1 + 2 \, i \, \nu} \,
    f^{(1)}_\nu (q) + K^{{\rm LO}} (k,q) \, q^{-1 + 2 \, i \, \nu} \,
    f^{(2)}_\nu (q) \right].
\end{align}
The first term on the right-hand-side of \eqref{lhs1} can be read off
from \eq{prjectLOeigf}:
\begin{align}
  \int d^2 q \, K^{{\rm NNLO}}(k,q) \ q^{-1+2 i \nu} = & \left\{
    \chi_0 (\nu) \left[ \beta_2^2 \, \ln^2 {k^2\over \mu^2} + \beta_3
      \, \ln{k^2\over \mu^2} \right] \right. \notag \\ & \left. - 2\,
    \beta_2 \, \ln{k^2\over \mu^2} \, \left[ \frac{i}{2} \, \beta_2 \,
      \chi'_0 (\nu) + \chi_1 (\nu) \right] + \chi_2 (\nu) + i \,
    \delta_2 (\nu) \right\} \, k^{-1 + 2 i \nu}.
      \label{NNLOonLO}
\end{align}
To find the second term on the right-hand-side of \eqref{lhs1} we
write, using \eq{eq:f1},
\begin{align}
  \label{lhs2}
  \int d^2 q \, & K^{{\rm NLO}} (k,q) \, q^{-1 + 2 \, i \, \nu} \,
  f^{(1)}_\nu (q) \notag \\ & = \beta_2 \left[ i \, {\chi_0 (\nu)
      \over 2 \, \chi'_0 (\nu)} \, ( - i \, \partial_\nu - \ln
    \mu^2)^2 + \frac{1}{2} \left( {\partial_\nu}\frac{\chi_0 (\nu)}{
        \chi_0' (\nu)} \right) ( - i \, \partial_\nu - \ln \mu^2)
  \right] \int d^2 q \, K^{{\rm NLO}} (k,q) \, q^{-1 + 2 \, i \, \nu}.
\end{align}
The projection of $K^{{\rm NLO}}$ onto $q^{-1 + 2 \, i \, \nu}$ can be
read off from the order-$\bam^2$ terms on the right side of
\eq{prjectLOeigf}. Using that in \eq{lhs2}, after some algebra one
arrives at
\begin{align}
  \label{lhs22}
  \int d^2 q \, K^{{\rm NLO}} (k,q) \, q^{-1 + 2 \, i \, \nu} \,
  f^{(1)}_\nu (q) & = \Bigg[ - i \, \frac{\beta_2^2}{2} \,
  \frac{\chi_0^2}{\chi'_0} \, \ln^3 \frac{k}{\mu} - i \,
  \frac{\beta_2}{2} \, \left( - \frac{7}{2} \, i \, \beta_2 \, \chi_0
    + i \, \beta_2 \, \frac{\chi_0^2 \, \chi''_0}{\chi_0^{\prime \,
        2}} - \frac{\chi_0 \, \chi_1}{\chi'_0} \right) \, \ln^2
  \frac{k}{\mu} \notag \\ & - i \, \frac{\beta_2}{2} \, \left( -
    \frac{\beta_2}{2} \, \frac{\chi_0 \, \chi''_0}{\chi'_0} -
    \frac{3}{2} \, \beta_2 \, \chi'_0 + 2 \, i \, \frac{\chi_0 \,
      \chi'_1}{\chi'_0} + i \, \chi_1 \, \left( 1 - \frac{\chi_0 \,
        \chi''_0}{\chi^{\prime \, 2}_0} \right) \right) \, \ln
  \frac{k}{\mu} \notag \\ & + \frac{\beta_2^2}{4} \, \left(
    \frac{\chi_0 \, \chi'''_0}{\chi'_0} + \chi''_0 - \frac{\chi_0 \,
      \chi^{\prime\prime \, 2}_0}{\chi^{\prime \, 2}_0} \right) - i \,
  \frac{\beta_2}{2} \, \left( \frac{\chi_0 \, \chi''_1}{\chi'_0} +
    \chi'_1 - \frac{\chi_0 \, \chi'_1 \,
      \chi^{\prime\prime}_0}{\chi^{\prime \, 2}_0} \right) \Bigg] \,
  k^{-1 + 2 \, i \, \nu}.
\end{align}
Henceforth, for brevity, we will use $\chi_0\equiv \chi_0(\nu)$,
$\chi_1\equiv \chi_1(\nu)$, $\chi'_0\equiv \partial_\nu \chi_0(\nu)$,
$\chi''_0\equiv \partial^2_\nu \chi_0(\nu)$, etc.

Substituting Eqs.~\eqref{NNLOonLO}, \eqref{lhs22}, \eqref{lhs1},
\eqref{eq:eig_NNLO} into \eq{eig_eq} we obtain the following equation
at the order $\bam^3$
\begin{align}
  & -i \, {\beta_2^2\over 2} \, {\chi_0^2\over \chi'_0} \,
  \ln^3{k^2\over \mu^2} + \left(-{3\over 4}\chi_0 +{\chi^2_0 \,
      \chi''_0\over 2 \chi'^2_0} \right) \, \beta_2^2 \,
  \ln{k^2\over\mu^2} + \left( - i \, \beta^2_2 \, {\chi'_0\over 4} + i
    \, \beta^2_2 \, {\chi_0 \, \chi''_0\over 4\chi'_0} -2 \, \beta_2
    \, \chi_1 + \beta_2 \, {\chi_0 \, \chi'_1\over\chi'_0 } +
    \beta_3\chi_0 \right) \, \ln{k^2\over\mu^2} \notag \\ & +
  \beta_2^2\Big({\chi_0 \chi'''_0\over 4\chi'_0} + {\chi_0''\over 4} -
  {\chi_0\chi''^2_0\over 4\chi'^2_0}\Big) -i \, {\beta_2\over 2} \,
  \Big({\chi_0\chi''_1\over \chi'_0}+\chi'_1 - {\chi_0\chi''_0
    \chi'_1\over \chi'^2_0}\Big) + k^{1 - 2 \, i \, \nu} \int d^2 q \,
  K^{\rm LO}(k,q) \, f^{(2)}_\nu(q) \, q^{-1 + 2 \, i \, \nu}
  \nonumber\\
  & = \chi_0 \, f^{(2)}_\nu(k) +c^{(2)}(\nu).
\label{cond2-order-2}
\end{align}
To proceed we now need an ansatz for $f^{(2)}_\nu(k)$: similarly to
the NLO case \cite{Chirilli:2013kca}, we expand it in the powers of
$\ln k^2/\mu^2$,
\begin{align}
  f^{(2)}_\nu(k) = \sum^\infty_{n=0} c_n^{(2)}(\nu) \ln^n{k^2\over
    \mu^2}.
\label{ansatz-order-2}
\end{align}
In the NLO case we truncated the series at $n=2$. What determines the
order at which we should truncate the series is the minimum number of
terms needed in order to make the terms proportional to powers of $\ln
k^2/\mu^2$ in equation (\ref{cond2-order-2}) disappear. 

A quick analysis of \eq{cond2-order-2} shows that in the NNLO case one
has to truncate the series at $n=4$. We thus write
\begin{align}
  f^{(2)}_\nu (k) = c_0^{(2)}(\nu) + c_1^{(2)}(\nu) \, \ln {k^2\over
    \mu^2} + c_2^{(2)}(\nu) \, \ln^2 {k^2\over \mu^2} + c_3^{(2)}(\nu)
  \, \ln^3 {k^2\over \mu^2} + c_4^{(2)}(\nu) \, \ln^4 {k^2\over
    \mu^2}.
\label{ansatz-order-22}
\end{align}
The action of the LO BFKL kernel on the NNLO part of the eigenfunction
can be found using the standard replacement of logarithms by
derivatives,
\begin{align}
  \label{eq:LOonNNLO}
  \int d^2 q \, K^{\rm LO}(k,q) \, f^{(2)}_\nu(q) \, q^{-1 + 2 \, i \,
    \nu} & = \left[ c_0^{(2)}(\nu) + c_1^{(2)}(\nu) \, (- i
    \, \partial_\nu - \ln \mu^2) + c_2^{(2)}(\nu) \, (- i
    \, \partial_\nu - \ln \mu^2)^2 \right. \notag \\ & \left. +
    c_3^{(2)}(\nu) \, (- i \, \partial_\nu - \ln \mu^2)^3 +
    c_4^{(2)}(\nu) \, \, (- i \, \partial_\nu - \ln \mu^2)^4 \right]
  \, \chi_0 (\nu) \, k^{-1 + 2 \, i \, \nu}.
\end{align}

Substituting Eqs.~\eqref{ansatz-order-22} and \eqref{eq:LOonNNLO} into
\eq{cond2-order-2} we obtain (with $c^{(2)}_n \equiv c^{(2)}_n(\nu)$)
\begin{align}
  & -i \, c^{(2)}_1\chi'_0 - c^{(2)}_2\chi''_0 +i \,
  c^{(2)}_3\chi'''_0+ c^{(2)}_4\chi''''_0 +\beta_2^2 \, \left({\chi_0
      \chi'''_0\over 4\chi'_0} + {\chi_0''\over 4} -
    {\chi_0\chi''^2_0\over 4\chi'^2_0}\right) -i \, {\beta_2\over
    2}\left({\chi_0\chi''_1\over \chi'_0}+\chi'_1 - {\chi_0\chi''_0
      \chi'_1\over \chi'^2_0}\right)
  \nonumber\\
  & +\left[-2\,i\,c^{(2)}_2\chi'_0 - 3 \, c^{(2)}_3\chi''_0 + 4\, i \,
    c^{(2)}_4\chi'''_0 + i \, \beta^2_2\Big(-{\chi'_0\over 4}+
    {\chi_0\, \chi''_0\over 4\chi'_0}\Big) + \beta_2 \, \Big(-2 \chi_1
    + {\chi_0 \, \chi'_1 \over \chi'_0} \Big)+ \beta_3 \,
    \chi_0\right] \ln \frac{k^2}{\mu^2}
  \nonumber\\
  & + \left(-3 \, i \, c^{(2)}_3 \, \chi'_0 -6 \, c^{(2)}_4 \chi''_0
    -\beta^2_2{3\over 4}\chi_0+ \beta^2_2 \, {\chi^2_0 \,
      \chi''_0\over 2\chi'^2_0} \right) \ln^2 \frac{k^2}{\mu^2}
  -i\Big(4\chi'_0 c^{(2)}_4 +{\beta^2_2\over
    2}\frac{\chi_0^2}{\chi'_0}\Big) \ln^3 \frac{k^2}{\mu^2} =
  c^{(2)}(\nu).
\label{cond3-order-2}
\end{align}
By equating the powers of $\ln ({k^2}/{\mu^2})$ in \eq{cond3-order-2}
we are able to determine $c^{(2)}_2$, $c^{(2)}_3$ and $c^{(2)}_4$,
\begin{align}
  & c^{(2)}_2 (\nu) = \beta_2^2\left({5\over 8}
    {\chi''^2_0(\nu)\chi_0^2(\nu)\over\chi'^4_0(\nu)}
    -{\chi_0(\nu)\chi''_0(\nu)\over 4\chi'^2_0(\nu)}
    -{\chi_0^2(\nu)\chi'''_0(\nu)\over 4\chi'^3_0(\nu)}-{1\over 8}
  \right) +i\beta_2\left({\chi_1(\nu)\over \chi'_0(\nu)}
    -{\chi_0(\nu)\chi'_1(\nu)\over 2\chi'^2_0(\nu)}\right)
  -i\beta_3{\chi_0(\nu)\over 2\chi'_0(\nu)}, \nonumber
  \\
  & c^{(2)}_3 (\nu) = i\beta_2^2\left(-{5\over 12}
    {\chi''_0(\nu)\chi_0^2(\nu)\over \chi'^3_0(\nu)} +
    {\chi_0(\nu)\over 4\chi'_0(\nu)}\right),  \nonumber
  \\
  & c_4^{(2)}(\nu) = -\beta_2^2{\chi_0^2(\nu)\over 8\chi'^2_0(\nu)},
\label{c2c3c4}
\end{align}
and also find an equation which relates the coefficient $c^{(2)}_1$ to
the function $c^{(2)}(\nu)$:
\begin{align}
  c^{(2)}(\nu) = & - i \, c^{(2)}_1\chi'_0 + \beta_2^2 \,
  \Big(-{5\over 8} {\chi''^3_0\chi^2_0\over \chi'^4_0}+{2\over 3}
  {\chi^2_0\chi''_0\chi'''_0\over \chi'^3_0} +{3\over 8}\chi''_0 -
  {\chi''''_0\chi^2_0\over 8\chi'^2_0}\Big)
  \nonumber \\
  & + i \beta_2\Big(-{\chi_1\chi''_0\over \chi'_0} +
  {\chi_0\chi''_0\chi'_1\over \chi'^2_0} -{\chi'_1\over 2}
  -{\chi_0\chi''_1\over 2 \chi'_0}\Big) +i\beta_3{\chi_0\chi''_0\over
    2\chi'_0}.
\label{cc1-cond}
\end{align}

To further fix $c^{(2)}_1$ and $c^{(2)}$ we need to make sure the
eigenfunctions $H_{\frac{1}{2}+i \, \nu} (k)$ form a complete set of
functions. As we have already shown in \cite{Chirilli:2013kca} the
functions $H_{\frac{1}{2}+i \, \nu} (k)$ have to satisfy the
completeness relation 
\begin{align}\label{compl3}
  \int\limits_{-\infty}^\infty {d\nu\over 2\pi} \, H_{\frac{1}{2}+i \,
    \nu} (k) \, \left[ H_{\frac{1}{2}+i \, \nu} (k)\right]^* =
  \delta(k^2-k'^2)
\end{align}
order-by-order in the perturbation theory. This means that since the
LO eigenfunctions already satisfy this condition, then all the
higher-order in $\bam$ corrections to the eigenfunctions have to
cancel on the left-hand--side of \eq{compl3}, possibly giving us an
extra constraint on $c^{(2)}_1$ and $c^{(2)}$.  At order-$\bam$, the
eigenfunctions $H_{\frac{1}{2}+i \, \nu} (k)$ from \eqref{Hnnlo}
already satisfy the completeness relation provided that
$f^{(1)}_\nu(k)$ is given by (\ref{eq:f1}). At order-$\bam^2$ the
completeness relation leads to the following constraint
\begin{align}
  \int\limits_{-\infty}^\infty {d\nu\over 2\pi} \left(
    \frac{k^2}{k'^2} \right)^{i\nu}
  \left[f^{(1)}_\nu(k)\big(f^{(1)}_\nu(k')\big)^* + f^{(2)}_\nu(k) +
    \big(f^{(2)}_\nu(k')\big)^*\right] = 0.
\label{f2cond}
\end{align}
Plugging in $f^{(1)}_\nu(k)$ from \eq{eq:f1} and $f^{(2)}_\nu(k)$ from
\eq{ansatz-order-22} with the coefficients from \eq{c2c3c4} we see
that the condition \eqref{f2cond} is satisfied if
\begin{align}
  \label{eq:c1}
  {\rm Re}[c_1^{(2)}(\nu)] = \partial_\nu \, {\rm Im}[c_2^{(2)}(\nu)].
\end{align}
To see this one has to again use the relation $\ln (k^2/k'^2) \,
(k^2/k'^2)^{i \, \nu} = -i \, \partial_\nu \, (k^2/k'^2)^{i \, \nu}$
together with integration by parts to eliminate terms containing
powers of $\ln (k^2/k'^2)$, such that only terms containing $\ln (k
k'/\mu^2)$ are left. Using this technique one then has to require that
the coefficient of each power $\ln^n\mu^2$ (with $n=1,~2,~3,~4$) is
zero independently; it turns out that all the terms proportional to
$\ln^4\mu^2$, $\ln^3\mu^2$, $\ln^2\mu^2$ are identically zero, while
the terms proportional to $\ln\mu^2$ give the condition \eqref{eq:c1}
for ${\rm Re}[c^{(2)}_1(\nu)]$, which, with the help of \eq{c2c3c4}
leads to
\begin{align}
  {\rm Re}[c_1^{(2)}(\nu)] = \beta_2\left({\chi'_1(\nu)\over
      2\chi'_0(\nu)}-{\chi_1(\nu)\chi''_0(\nu)\over \chi'^2_0(\nu)} -
    {\chi''_1(\nu)\chi_0(\nu)\over 2 \chi'^2_0(\nu)} +
    {\chi_0(\nu)\chi'_1(\nu)\chi''_0(\nu)\over
      \chi'^3_0(\nu)}\right)-\beta_3\left(\half -
    {\chi_0(\nu)\chi''_0(\nu)\over 2\chi'^2_0(\nu)}\right).
\label{rec1}
\end{align}
On the other hand the terms without $\ln \mu^2$ in (\ref{f2cond})
result in the following condition for ${\rm Re}[c^{(2)}_0(\nu)]$ and
${\rm Im}[c^{(2)}_1(\nu)]$,
\begin{align}
  &
  \partial_\nu{\rm Im}[c^{(2)}_1(\nu)] - 2{\rm Re}[c^{(2)}_0(\nu)]=
  \Bigg( -{5\over 2}{\chi_0^2\chi''^4_0\over \chi'^6_0} + {5\over
    4}{\chi_0 \chi''^3_0\over \chi'^4_0} + {7\over
    2}{\chi_0^2\chi'''_0\chi''^2_0\over \chi'^5_0}
  \nonumber\\
  & -{4\over 3}{\chi_0\chi'''_0\chi''_0\over \chi'^3_0} -{3\over
    4}{\chi^2_0\chi''''_0\chi''_0\over \chi'^4_0}
  -{\chi_0^2\chi'''^2_0\over 2\chi'^4_0} + {\chi_0 \chi''''_0\over
    4\chi^{\prime \, 2}_0} + {\chi^2_0 \, \chi^{(5)}_0\over 12\chi'^3_0}
  \Bigg) \, \beta_2^2, \label{c1c0}
\end{align}
where $\chi^{(5)}_0 = \partial_\nu^5 \chi_0 (\nu)$.  We can now use
the result (\ref{rec1}) for ${\rm Re}[c_1^{(2)}(\nu)]$ in
(\ref{cc1-cond}) to determine ${\rm Im}[c^{(2)}(\nu)]$
\begin{align} {\rm Im}[c^{(2)}(\nu)] = {\chi'_0\over 2}\beta_3 -
  \chi'_1 \beta_2
\label{imc}
\end{align}
and also obtain a relation between ${\rm Re}[c^{(2)}](\nu)$ and ${\rm
  Im}[c^{(2)}_1(\nu)]$
\begin{align}\label{rec-imc1}
  {\rm Re}[c^{(2)}(\nu)] = {\rm Im}[c^{(2)}_1(\nu)]\chi'_0 +
  \beta_2^2\Big(-{5\over 8} {\chi''^3_0\chi^2_0\over \chi'^4_0} +
  {2\over 3}{\chi^2_0\chi''_0\chi'''_0\over \chi'^3_0} +{3\over
    8}\chi''_0 - {\chi''''_0\chi^2_0\over 8\chi'^2_0}\Big).
\end{align}

Let us summarize what we have done so far. We started with the unknown
functions of $f^{(2)}_\nu(k)$ and $c^{(2)}(\nu)$, then introduced an
ansatz for $f^{(2)}_\nu(k)$ which effectively expressed all the
unknowns in terms of a total of six complex-valued functions:
$c^{(2)}_0(\nu)$, $c^{(2)}_1(\nu)$, $c^{(2)}_2(\nu)$,
$c^{(2)}_3(\nu)$, $c^{(2)}_4(\nu)$ and $c^{(2)}(\nu)$.  Then, by
requiring that the functions $H_{\frac{1}{2}+i \, \nu} (k)$ are
eigenfunctions of the NNLO BFKL kernel and that they satisfy the
completeness relation \eqref{compl3}, we were able to determine
$c^{(2)}_2(\nu)$, $c^{(2)}_3(\nu)$, $c^{(2)}_4(\nu)$ given by
\eq{c2c3c4}, along with ${\rm Re}[c^{(2)}_1(\nu)]$ given in \eq{rec1}
and ${\rm Im}[ c^{(2)}(\nu)]$ given in \eq{imc}. We have also obtained
equations (\ref{rec-imc1}) and \eqref{c1c0} relating ${\rm
  Re}[c^{(2)}(\nu)]$ to ${\rm Im}[ c^{(2)}_1(\nu)]$ and ${\rm
  Im}[c^{(2)}_1(\nu)]$ to ${\rm Re}[c^{(2)}_0(\nu)]$.  At this point
the only undetermined function is $c^{(2)}_0(\nu)$: note that it does
not enter in the completeness condition \eqref{compl3} and the
functions $H_{\frac{1}{2}+i \, \nu} (k)$ are the eigenfunctions of the
NNLO BFKL kernel for any $c^{(2)}_0(\nu)$. This freedom of choosing
any $c^{(2)}_0(\nu)$ is analogous to the same freedom observed in
constructing the NLO BFKL eigenfunctions in \cite{Chirilli:2013kca}:
as we argued in \cite{Chirilli:2013kca}, this is a consequence of the
freedom of choosing any overall phase for the eigenfunctions along
with the freedom to reparametrize the $\nu$ variable (by shifting it
by an ${\cal O} (\bam^2)$ correction at NNLO).

We can now write down the expression for the NNLO corrections
$f^{(2)}_\nu (k)$ to the BFKL kernel eigenfunction (we use, as usual,
the short hand notation where $\chi_0(\nu)\equiv \chi_0$,
$\partial_\nu \chi_0(\nu)\equiv \chi'_0$, etc.)
\begin{align}
  f^{(2)}_\nu (k) = c^{(2)}_0+ \left[\beta_2\left({\chi'_1\over
        2\chi'_0}-{\chi_1\chi''_0\over \chi'^2_0} -
      {\chi''_1\chi_0\over 2 \chi'^2_0} + {\chi_0\chi'_1\chi''_0\over
        \chi'^3_0}\right)-\beta_3\left(\half - {\chi_0\chi''_0\over
        2\chi'^2_0}\right) +i \, {\rm
      Im[c^{(2)}_1]}\right]\ln{k^2\over \mu^2}
  \nonumber\\
  +\left[\beta_2^2\left({5\over 8} {\chi''^2_0\chi_0^2\over\chi'^4_0}
      -{\chi_0\chi''_0\over 4\chi'^2_0} -{\chi_0^2\chi'''_0\over
        4\chi'^3_0}-{1\over 8} \right) +i\beta_2\left({\chi_1\over
        \chi'_0} -{\chi_0\chi'_1\over 2\chi'^2_0}\right)
    -i\beta_3{\chi_0\over 2\chi'_0}\right]\ln^2{{k^2\over \mu^2}}
  \nonumber\\
  +\left[i \, \beta_2^2\left(-{5\over 12}
      {\chi''_0(\nu)\chi_0^2(\nu)\over \chi'^3_0(\nu)} +
      {\chi_0(\nu)\over 4\chi'_0(\nu)}\right) \right]\ln^3{k^2\over
    \mu^2} + \left[-\beta_2^2{\chi_0^2(\nu)\over
      8\chi'^2_0(\nu)}\right]\ln^4{k^2\over \mu^2}.
\label{f2}
\end{align}
The NNLO eigenfunctions are then given by
\begin{align}
  H_{\frac{1}{2}+i \, \nu} (k) = k^{-1 + 2 \, i \, \nu} \left[1 + \bam
    \, f^{(1)}_\nu(k) + \bam^2 \, f^{(2)}_\nu(k) \right]
  \label{NNLOeigef}
\end{align}
where $f^{(1)}_\nu(k)$ is given in Eq.~(\ref{eq:f1}) and
$f^{(2)}_\nu(k)$ is given in Eq.~(\ref{f2}).  

Before we proceed, let us note that the eigenfunctions
$H_{\frac{1}{2}+i \, \nu} (k)$ satisfy the orthogonality expression
\begin{align}
  \label{eq:ortho2}
  \int d^2 k \, H_{\frac{1}{2}+i \, \nu} (k) \, \left[
    H_{\frac{1}{2}+i \, \nu'} (k) \right]^* = 2 \, \pi^2 \, \delta
  (\nu - \nu')
\end{align}
as well, as can be verified explicitly to order $\bam^2$.

The BFKL kernel eigenvalues up to NNLO corrections are
\begin{align}
  \Delta(\nu) = & \bam \, \chi_0(\nu) + \bam^2 \, \chi_1(\nu) + \bam^3
  \, [\chi_2(\nu) + i \, \delta_2 (\nu) + {\rm Re}[c^{(2)}(\nu)] + i
  \,
  {\rm Im} [c^{(2)}]] + {\cal O} (\bam^4) \nonumber \\
  = & \bam \, \chi_0(\nu) + \bam^2 \, \chi_1(\nu) + \bam^3 \,
  \Bigg[\chi_2(\nu) + i \, \delta_2 (\nu) + {\rm Im}[c^{(2)}_1(\nu)]
  \, \chi'_0
  \nonumber\\
  & + \beta_2^2 \, \Big(-{5\over 8} {\chi''^3_0\chi^2_0\over
    \chi'^4_0} + {2\over 3}{\chi^2_0\chi''_0\chi'''_0\over \chi'^3_0}
  +{3\over 8}\chi''_0 - {\chi''''_0\chi^2_0\over 8\chi'^2_0}\Big)
  +i\left({\chi'_0\over 2}\beta_3 - \chi'_1 \beta_2\right)\Bigg] +
  {\cal O} (\bam^4) , \label{eigNNLO1}
\end{align}
where we have used \eq{rec-imc1} for Re$[c^{(2)}(\nu)]$ and \eq{imc}
for Im$[c^{(2)}(\nu)]$. The BFKL kernel has to be hermitean at any
order in the coupling constant: therefore, the BFKL kernel eigenvalues
have to be real. Requiring the eigenvalues \eqref{eigNNLO1} to be real
we fix the imaginary part $i \, \delta_2 (\nu)$ of the conformal piece
of the projection of the NNLO BFKL kernel onto the LO
eigenfunctions. We thus predict that
\begin{align}\label{delta}
  \delta_2 (\nu) = - {\chi'_0 (\nu) \over 2} \, \beta_3 + \chi'_1
  (\nu) \, \beta_2.
\end{align}
Note that this result is in complete agreement with our NNLO ansatz in
\cite{Chirilli:2013kca} (see Eq. (B6) there).

Thus, the NNLO BFKL kernel eigenvalues reduce to
\begin{align}
  \Delta(\nu) = \bam \, \chi_0(\nu) + \bam^2 \, \chi_1(\nu) + \bam^3
  \, \left[ \chi_2(\nu) + {\rm Im}[c^{(2)}_1(\nu)] \, \chi'_0 +
    \beta_2^2 \, \left(-{5\over 8} {\chi''^3_0\chi^2_0\over \chi'^4_0}
      + {2\over 3}{\chi^2_0\chi''_0\chi'''_0\over \chi'^3_0} +{3\over
        8}\chi''_0 - {\chi''''_0\chi^2_0\over 8\chi'^2_0}\right)
  \right] + {\cal O} (\bam^4).
\label{NNLOeigev}
\end{align}

At this point we have NNLO BFKL eigenfunctions \eqref{f2} and the
eigenvalues \eqref{NNLOeigev} for any function $c^{(2)}_0(\nu)$. One
could use the phase-choice and $\nu$-reparametrization freedoms to
select a convenient value of $c^{(2)}_0(\nu)$. Alternatively one can
argue that, since the functions $H_{\frac{1}{2}+i \, \nu} (k)$ form a
complete and orthogonal set of eigenfunctions of NNLO BFKL kernel for
any $c^{(2)}_0(\nu)$, a particularly simple choice is to put
\begin{align}
  \label{eq:choice}
  {\rm Im}[c^{(2)}_0(\nu)] = 0, \ \ \  {\rm Im}[c^{(2)}_1(\nu)] = 0,
\end{align}
with ${\rm Re}[c^{(2)}_0(\nu)]$ easily found from \eq{c1c0}. We then
obtain our final expression for the NNLO BFKL eigenvalues
\begin{align}
  \Delta(\nu) = \bam \, \chi_0(\nu) + \bam^2 \, \chi_1(\nu) + \bam^3
  \, \left[ \chi_2(\nu) + \beta_2^2 \, \left(-{5\over 8}
      {\chi''^3_0\chi^2_0\over \chi'^4_0} + {2\over
        3}{\chi^2_0\chi''_0\chi'''_0\over \chi'^3_0} +{3\over
        8}\chi''_0 - {\chi''''_0\chi^2_0\over 8\chi'^2_0}\right)
  \right] + {\cal O} (\bam^4)
\label{eigNNLO}
\end{align}
and eigenfunctions
\begin{align}
  f^{(2)}_\nu (k) = & - \frac{\beta_2^2}{2} \, \Bigg( -{5\over
    2}{\chi_0^2\chi''^4_0\over \chi'^6_0} + {5\over 4}{\chi_0
    \chi''^3_0\over \chi'^4_0} + {7\over
    2}{\chi_0^2\chi'''_0\chi''^2_0\over \chi'^5_0} -{4\over
    3}{\chi_0\chi'''_0\chi''_0\over \chi'^3_0} -{3\over
    4}{\chi^2_0\chi''''_0\chi''_0\over \chi'^4_0}
  -{\chi_0^2\chi'''^2_0\over 2\chi'^4_0} + {\chi_0 \chi''''_0\over
    4\chi^{\prime \, 2}_0} + {\chi^2_0 \, \chi^{(5)}_0\over
    12\chi'^3_0} \Bigg) \nonumber \\ & +
  \left[\beta_2\left({\chi'_1\over 2\chi'_0}-{\chi_1\chi''_0\over
        \chi'^2_0} - {\chi''_1\chi_0\over 2 \chi'^2_0} +
      {\chi_0\chi'_1\chi''_0\over \chi'^3_0}\right)-\beta_3\left(\half
      - {\chi_0\chi''_0\over 2\chi'^2_0}\right) \right]\ln{k^2\over
    \mu^2}
  \nonumber\\
  & + \left[\beta_2^2\left({5\over 8}
      {\chi''^2_0\chi_0^2\over\chi'^4_0} -{\chi_0\chi''_0\over
        4\chi'^2_0} -{\chi_0^2\chi'''_0\over 4\chi'^3_0}-{1\over 8}
    \right) +i\beta_2\left({\chi_1\over \chi'_0} -{\chi_0\chi'_1\over
        2\chi'^2_0}\right) -i\beta_3{\chi_0\over
      2\chi'_0}\right]\ln^2{{k^2\over \mu^2}}
  \nonumber\\
  & +\left[i \, \beta_2^2\left(-{5\over 12}
      {\chi''_0(\nu)\chi_0^2(\nu)\over \chi'^3_0(\nu)} +
      {\chi_0(\nu)\over 4\chi'_0(\nu)}\right) \right]\ln^3{k^2\over
    \mu^2} + \left[-\beta_2^2{\chi_0^2(\nu)\over
      8\chi'^2_0(\nu)}\right]\ln^4{k^2\over \mu^2}.
\label{f2fin}
\end{align}
The NNLO BFKL eigenfunctions are given by \eq{NNLOeigef} with
$f^{(1)}_\nu(k)$ given by Eq.~(\ref{eq:f1}) and $f^{(2)}_\nu(k)$ given
by Eq.~(\ref{f2fin}). Together with \eq{eigNNLO}, these are the main
results of this Section.

Notice that the only unknown function in (\ref{eigNNLO}) is
$\chi_2(\nu)$ which can be found only after performing the calculation
of the NNLO BFKL kernel (with NNLO BFKL evolution defined in
Sec.~\ref{sec:bey}). Steps in this direction were done in
\cite{Marzani:2007gk,Ball:2005mj}.

Another important observation is that to determine the structure of
higher-order corrections to the BFKL eigenfunctions and eigenvalues,
one only needs to know the running coupling part of the projection of
the corresponding higher-order BFKL kernel onto the LO eigenfunctions
(see, e.g., \eq{prjectLOeigf}), which can always be obtained as
running-coupling corrections to the lower-order projections. Hence our
procedure can be used to construct NNNLO and higher-order BFKL
eigenfunctions, and is limited only by our knowledge of the QCD
beta-function.


\subsection{General Form of the NNLO BFKL Equation Solution}

In \cite{Chirilli:2013kca} we have obtained a solution of the NNLO
BFKL equation for the Green function \eqref{eq:BFKL} (provided that
$\chi_2(\nu)$ is explicitly calculated), in an indirect way: we
assumed that the NNLO solution could be obtained by a simple
generalization of the NLO expression for the BFKL Green function to
the NNLO order. It turned out that such an ansatz did not solve the
NNLO BFKL equation and had to be augmented to solve NNLO BFKL with a
particular choice of $\delta_2 (\nu)$, incidentally the same as in
\eq{delta}. Our goal now is to construct the NNLO expression for the
BFKL Green function and to verify the ansatz proposed in
\cite{Chirilli:2013kca} (see Eq.~(B7) there).

The BFKL Green function can be written in terms of eigenfunctions as
\begin{align}
  G(k,k',Y) = \int\limits_{-\infty}^\infty {d\nu\over 2\pi^2} \,
  e^{\Delta(\nu) \, Y} \, H_{\frac{1}{2}+i \, \nu} (k) \,
  H_{\frac{1}{2}-i \, \nu} (k')
\label{NNLOsolu1}
\end{align}
with $H_{\frac{1}{2}+i \, \nu} (k)$ given by Eqs.~(\ref{eq:f1}),
(\ref{f2fin}), (\ref{NNLOeigef}), and $\Delta(\nu)$ given in
Eq.~(\ref{eigNNLO}) at NNLO. Substituting those results in
\eq{NNLOsolu1}, employing the same trick of turning each $\ln
(k^2/k'^2)$ into $- i \, \partial_\nu$ acting on the powers of
transverse momentum and integrating by parts, after a lengthy
calculation one can rewrite Eq.~(\ref{NNLOsolu1}) as
\begin{align}
  & G(k,k',Y) = \int {d\nu\over 2\pi^2 kk'} \, \left({k^2\over
      k'^2}\right)^{i\nu} \, e^{\Delta(\nu) Y} \Bigg\{ -\bam^2\beta_2
  Y\chi_0\ln{kk'\over \mu^2} - \bam^3 \, Y \, \left[ 2\beta_2 \chi_1 -
    \beta_3 \chi_0 \right] \, \ln{kk'\over \mu^2}
  \nonumber\\
  & + \bam^2\beta_2^2\left(\bam \chi_0 Y + \half(\bam \chi_0 Y)^2
  \right)\ln^2{kk'\over \mu^2} - \bam^3 \beta_2^2 Y \Big(-{5\over 8}
  {\chi''^3_0\chi^2_0\over \chi'^4_0} + {2\over
    3}{\chi^2_0\chi''_0\chi'''_0\over \chi'^3_0} +{3\over 8}\chi''_0 -
  {\chi''''_0\chi^2_0\over 8\chi'^2_0}\Big)
  \nonumber\\
  & + (\bam \beta_2)^2\left[-{1\over 24}(\bam
    Y)^3\chi_0(\nu)^2\chi''_0(\nu) + {1\over 4}(\bam
    Y)^2\chi_0(\nu)\left({\chi'_0(\nu)^2\over 2\chi_0(\nu)} -
      \chi''_0(\nu)\right) +\bam Y {\chi_0''(\nu)\over 4}\right]
  \Bigg\}
\end{align}
with $\Delta(\nu)$ given in Eq.~(\ref{eigNNLO}).  Using the fact that,
at NNLO accuracy,
\begin{align}
  \label{eq:delta_approx}
  e^{\Delta(\nu) Y} = e^{[\bam \, \chi_0 + \bam^2 \, \chi_1 + \bam^3
    \, \chi_2]Y} \, \left[ 1 + \bam^3 \beta_2^2 Y \Big(-{5\over 8}
    {\chi''^3_0\chi^2_0\over \chi'^4_0} + {2\over
      3}{\chi^2_0\chi''_0\chi'''_0\over \chi'^3_0} +{3\over 8}\chi''_0
    - {\chi''''_0\chi^2_0\over 8\chi'^2_0}\Big)\right] + {\cal O}
  (\as^3)
\end{align}
along with
\begin{align}
  e^{[\bas(kk')\chi_0(\nu) +\bas^2(kk')\chi_1(\nu)
    +\bas^3(kk')\chi_2(\nu)]Y} = & \, e^{[\bam \, \chi_0 + \bam^2 \,
    \chi_1 + \bam^3 \, \chi_2]Y} \Big[ 1 -\bam^2\beta_2
  Y\chi_0\ln{kk'\over \mu^2} - \bam^3 \, Y \, \left[ 2\beta_2 \chi_1 -
    \beta_3 \chi_0 \right] \, \ln{kk'\over \mu^2}
  \nonumber\\
  & + \bam^2\beta_2^2\left(\bam \chi_0 Y + \half(\bam \chi_0 Y)^2
  \right)\ln^2{kk'\over \mu^2} \Big] + {\cal O} (\as^3)
\end{align} 
we can write the NNLO BFKL solution as
\begin{align}
  & G(k,k', Y) = \int {d\nu\over 2\pi^2 k k'} e^{[\bas(kk')\chi_0(\nu)
    +\bas^2(kk')\chi_1(\nu) +\bas^3(kk')\chi_2(\nu)]Y} \left(k^2\over
    k'^2\right)^{i\nu}
  \nonumber\\
  & \times\left\{ 1+(\bam \beta_2)^2\left[-{1\over 24}(\bam
      Y)^3\chi_0(\nu)^2\chi''_0(\nu) + {1\over 4}(\bam
      Y)^2\chi_0(\nu)\left({\chi'_0(\nu)^2\over 2\chi_0(\nu)} -
        \chi''_0(\nu)\right) +\bam Y {\chi_0''(\nu)\over 4}\right]
  \right\}. 
\label{NNLOsolu2}
\end{align}
This result exactly coincides with, and, therefore, proves the ansatz
suggested in Eq.~(B7) of our previous paper \cite{Chirilli:2013kca}.

Note that the first term in the square brackets of \eq{NNLOsolu2}
taken in the saddle-point approximation of LO BFKL equation (that is,
at $\nu =0$), gives exactly the $\sim \, \as^5 \, Y^3$ term in the
exponent found in \cite{Kovchegov:1998ae}. With the accuracy of our
NNLO approximation we can not distinguish the ${\cal O} (\as^2)$-terms
in the exponent from those in the prefactor: it is, therefore,
possible that the first term (along with the other terms) in the
square brackets of \eq{NNLOsolu2} would exponentiate when high-order
corrections are included. An exponential form of the first term in the
square brackets of \eq{NNLOsolu2} without the saddle point
approximation was obtained in \cite{Grabovsky:2013gta}.


\section{Conclusions}
\label{sec:concl}

There are two main results presented in this paper. The first one is
the analytic NLO $\gamma^*\gamma^*$ cross section given in
Eqs.~(\ref{ggNLOcrossec}) and (\ref{ggNLOcrossecLL}) and plotted in
Figs.~\ref{NLOsigma1LT} and \ref{NLOsigma5LT}; the second one is the
NNLO BFKL Green function given in Eq.~(\ref{NNLOsolu2}) along with the
NNLO BFKL eigenfunctions (\ref{NNLOeigef}), (\ref{eq:f1}),
(\ref{f2fin}) and eigenvalue \eqref{eigNNLO}.

The structure of the NLO $\gamma^* \gamma^*$ cross sections
(\ref{ggNLOcrossec}) and (\ref{ggNLOcrossecLL}) is the same as the one
at LO: the two impact factors, the dipole-dipole cross section and the
energy-dependent exponential factor each receive corrections at NLO;
in addition each coupling constant is replaced by a running
coupling. The factorization of the scattering amplitude is represented
in \fig{gg_graph} and is likely to persist at higher orders (in the
linear approximation). The corresponding expression
\eqref{gg-analytic} for the forward $\gamma^* \gamma^*$ scattering
amplitude, involving projections of the impact factors on the NLO
eigenfunctions, is also likely to be true to all orders.

The $\gamma^*\gamma^*$ cross sections (\ref{ggNLOcrossec}) and
(\ref{ggNLOcrossecLL}) plotted in Figs.~\ref{NLOsigma1LT} and
\ref{NLOsigma5LT} are negative at small values of rapidity $Y$. This
issue may be resolved by adding the energy-suppressed contributions to
the cross sections, which become important at low energy (rapidity
$Y$) making the net cross section positive for all $Y$.

An ansatz for the structure of the solution of the NNLO BFKL equation
was constructed \cite{Chirilli:2013kca}, but it was obtained in an
indirect way by attempting to guess the NNLO BFKL solution using an
analogy to the NLO solution. In addition the ansatz from
\cite{Chirilli:2013kca} relied on a particular expression for
$\delta_2 (\nu)$. In this work we have obtained the solution of the
NNLO BFKL equation by explicitly constructing the NNLO eigenfunctions
in (\ref{NNLOeigef}). This procedure yielded the NNLO BFKL Green
function which confirmed the ansatz from our previous paper
\cite{Chirilli:2013kca}.


\section*{Acknowledgments}

The authors are grateful to Ian Balitsky, Dmitry Ivanov, Alessandro
Papa and Douglas Ross for encouraging discussions and to Dmitry Ivanov
for directing our attention to a missing overall factor in the
$\gamma^*\gamma^*$ cross section in the earlier version of this
paper. This material is based upon work supported by the
U.S. Department of Energy, Office of
Science, Office of Nuclear Physics under Award Number DE-SC0004286. \\


\section*{Appendix A: Projection of the dipole operator onto the LO
  eigenfunctions}
\renewcommand{\theequation}{A\arabic{equation}}
  \setcounter{equation}{0}
\label{A}

In this section we obtain the projection on to the LO eigenfunction of
the dipole operator in momentum space. The results apply both to the
``standard'' and composite dipole operators.

The decomposition of the product of two $2$-dimensional delta
functions using the conformal eigenfunctions of the LO BFKL kernel (in
transverse coordinate space represented by a complex plane) is
\begin{align}\label{Ecomp}
  \delta^{(2)}(z_1-z_{1'}) \, \delta^{(2)}(z_2 -z_{2'}) =
  \sum\limits_{n=-\infty}^{\infty} \int d^2z_0 \,
  \int\limits_{-\infty}^\infty {d\nu\over \pi^4} \, {\nu^2+{n^2\over
      4}\over z^2_{12} \, z^2_{1'2'}} \, E^{n, \nu} (z_{10},z_{20}) \,
  E^{n, \nu *} (z_{1'0},z_{2'0})
\end{align}
where
\begin{align}
  E^{n, \nu}(z_{10},z_{20}) = \left[{z_{12}\over z_{10}
    \, z_{20}}\right]^{\half+i\nu+{n\over 2}}
  \left[{z^*_{12}\over z^*_{10} \,
    z^*_{20}}\right]^{\half+i\nu-{n\over 2}}.
\label{eigenfunctions}
\end{align}
(Here $z \equiv z_x+i \, z_y, z^* \equiv z_x-iz_y$, $z_{ij}\equiv z_i-z_j$,
etc., where ${\vec z}_\perp = (z_x, z_y)$.)

%
Using the completeness relation \eqref{Ecomp} the dipole operator
defined in Eq. (\ref{col-dipo}) can be decomposed in the following way
\begin{align}\label{UEnn}
  \calu^\eta (z_1,z_2) & = \int d^2z_{1'} \, d^2z_{2'} \, \calu^\eta (z_{1'},
  z_{2'}) \, \delta^{(2)}(z_1-z_{1'}) \, \delta^{(2)}(z_2 - z_{2'})
  \notag \\ & = \int d^2z_0 \sum\limits_{n=-\infty}^{\infty} \,
  \int\limits_{-\infty}^\infty {d\nu\over \pi^2} \, \left( \nu^2 +
    \frac{n^2}{4} \right) \, E^{n, \nu}(z_{10},z_{20}) \, \calu_n^\eta
  (\nu,z_0)
\end{align}
where we have defined
\begin{align}
  \calu_n^\eta (\nu,z_0) \equiv \int {d^2z_{1'} \, d^2z_{2'} \over
    \pi^2 \, |z_{1'2'}|^4} \, E^{n, \nu *} (z_{1'0},z_{2'0}) \,
  \calu^\eta (z_{1'},z_{2'}).
\label{caluz0nu}
\end{align}

In high-energy scattering the $n=0$ mode dominates in the BFKL
evolution since it corresponds to the largest intercept. Concentrating
on the $n=0$ contribution we integrate Eq.~(\ref{caluz0nu}) over $z_0$
to obtain
\begin{align}
  \calu^\eta (\nu) & \equiv \int d^2 z_0\, \calu^\eta_0 (\nu,z_0)
  \nonumber\\
  & = \frac{1}{\pi} \, \int d^2 z_1 \, d^2 z_2 \, |z_{12}|^{- 3 + 2 \,
    i \, \nu} \, \frac{\Gamma^2 \left( \frac{1}{2} + i \, \nu \right)
    \, \Gamma \left( - 2 \, i \, \nu\right)}{\Gamma^2 \left(
      \frac{1}{2} - i \, \nu \right) \, \Gamma \left( 1 + 2 \, i \,
      \nu\right)}\, \calu^\eta (z_1, z_2).
\label{Unu1}
\end{align}
Writing for the $z_{12}$-direction-independent integrand $d^2 z_1 \,
d^2 z_2 = \pi \, d x_\perp^2 \, d^2 z$ with ${\vec x}_\perp = {\vec
  z}_{1\perp} - {\vec z}_{2\perp}$, $x_\perp = |{\vec x}_\perp|$ and
${\vec z}_\perp = {\vec z}_{2\perp})$ we recast \eq{Unu1} as
(cf. \eq{Utr})
\begin{align}
  \calu^\eta (\nu) = \int d^2 z \, \calu^\eta (\nu, {\vec z}_\perp) =
  \int d x_\perp^2 \, d^2 z \ x_\perp^{- 3 + 2 \, i \, \nu} \,
  \frac{\Gamma^2 \left( \frac{1}{2} + i \, \nu \right) \, \Gamma
    \left( - 2 \, i \, \nu\right)}{\Gamma^2 \left( \frac{1}{2} - i \,
      \nu \right) \, \Gamma \left( 1 + 2 \, i \, \nu\right)}\,
  \calu^\eta ({\vec x}_\perp + {\vec z}_\perp, {\vec z}_\perp),
\label{Unu2}
\end{align}
where we have switched back to the transverse vector notation (from
the complex plane one). Note again that $\calu^\eta ({\vec x}_\perp +
{\vec z}_\perp, {\vec z}_\perp)$ on the right-hand-side of \eq{Unu2}
now represents the leading-energy component of the dipole amplitude
\eqref{col-dipo}, with $\int d^2 z \, \calu^\eta ({\vec x}_\perp +
{\vec z}_\perp, {\vec z}_\perp)$ independent of the angle of ${\vec
  x}_\perp$.

Inverting \eq{Unu2} we write
\begin{align}
  \int d^2 z \, \calu^\eta ({\vec x}_\perp + {\vec z}_\perp, {\vec z}_\perp) =
  \int\limits_{-\infty}^\infty {d\nu\over 2 \, \pi} \, x_\perp^{1 - 2
    \, i \, \nu} \, \frac{\Gamma^2 \left( \frac{1}{2} - i \, \nu
    \right) \, \Gamma \left( 1 + 2 \, i \, \nu\right)}{\Gamma^2 \left(
      \frac{1}{2} + i \, \nu \right) \, \Gamma \left( - 2 \, i \,
      \nu\right)} \, \calu^\eta (\nu).
\label{Uxz1}
\end{align}
Finally, Fourier-transforming \eq{Uxz1} into transverse momentum space with
the help of \eq{Utr} yields
\begin{align}
  \int d^2 z \, \calu^\eta ({k}_\perp, {\vec z}_\perp) =
  \int\limits_{-\infty}^\infty d\nu \, k_\perp^{-3 + 2 \, i \, \nu} \,
  2^{3 - 2 \, i \, \nu} \, i \, \nu \, \left( \frac{1}{2} - i \, \nu
  \right)^2 \, \frac{\Gamma^3 \left( \frac{1}{2} - i \, \nu \right) \,
    \Gamma \left( 1 + 2 \, i \, \nu\right)}{\Gamma^3 \left(
      \frac{1}{2} + i \, \nu \right) \, \Gamma \left( 1- 2 \, i \,
      \nu\right)} \ \calu^\eta (\nu).
\label{Ukz1}
\end{align}

For the use in the main text, let us define 
\begin{align}
  f(\nu) = 2^{3 - 2 \, i \, \nu} \, i \, \nu \, \left( \frac{1}{2} - i
    \, \nu \right)^2 \, \frac{\Gamma^3 \left( \frac{1}{2} - i \, \nu
    \right) \, \Gamma \left( 1 + 2 \, i \, \nu\right)}{\Gamma^3 \left(
      \frac{1}{2} + i \, \nu \right) \, \Gamma \left( 1- 2 \, i \,
      \nu\right)}
\label{fnu}
\end{align}
and observe that
\begin{align}
  f(\nu) \, f(-\nu) = 4 \, \nu^2 \, (4 \, \nu^2+1)^2.
\label{ffnu}
\end{align}
With the help of definition \eqref{fnu} we rewrite \eq{Ukz1} as
\begin{align}
  \int d^2 z \, \calu^\eta ({k}_\perp, {\vec z}_\perp) =
  \int\limits_{-\infty}^\infty d\nu \, k_\perp^{-3 + 2 \, i \, \nu} \,
  f(\nu) \ \calu^\eta (\nu).
\label{Ukz2}
\end{align}


\section*{Appendix B: NLO $\gamma^*$$\gamma^*$ cross section}
\renewcommand{\theequation}{B\arabic{equation}}
  \setcounter{equation}{0}
\label{B}

In this section we provide details of the calculation leading to
Eqs.~\eqref{L0}, \eqref{IprojH} and \eqref{gg-analytic}.
In the derivation we will make use of the solution of the NLO BFKL equation 
\begin{align}
  G(k,k',Y) = \int\limits_{-\infty}^{+\infty} {d\nu\over 2\pi^2 kk'}\,
  e^{\Delta(\nu) \, Y}\, H_\nu(k) H_{-\nu}(k')
\end{align}
which, replacing the logarithm $\ln k$ with $-i\partial_\nu$ acting on
$k^{i\nu}$ and integrating by parts, can be written as
\begin{align}
  G(k,k',Y) = \int\limits_{-\infty}^{+\infty} {d\nu\over 2\pi^2 kk'}\,
  e^{[\bam \chi_0(\nu) +\bam^2\chi_1(\nu)]Y}\, \left(k^2\over
    k'^2\right)^{i\nu} \Big(1-\bam^2\beta_2\chi_0(\nu) Y\ln {kk'\over
    \mu^2}\Big)
\label{NLOsol}
\end{align}
Eq. (\ref{NLOsol}) is a result we obtained in
Ref. \cite{Chirilli:2013kca}.

We proceed with \eq{L0}. Using Eqs.~\eqref{Leige} and \eqref{L-dipole}
we write
\begin{align}
  \label{eq:B1}
  \left\langle {\cal L}^{a_0} (\nu_1) \, {\cal L}^{a_0} (\nu_2)
  \right\rangle/S_\perp = \int\limits_0^\infty d k^2_1 H_{-\nu_1}
  (k_1) \, \int\limits_0^\infty d k^2_2 H_{-\nu_2} (k_2) \, {1\over
    S_\perp}\left\langle \frac{k^2_1 \, [{\cal U}^{a_0} ({\vec
        k}_{1\perp})]^{\rm comp}}{\as (k^2_1)} \, \frac{k^2_2 \,
      [{\cal U}^{a_0} ({\vec k}_{2\perp})]^{\rm comp}}{\as (k^2_2)}
  \right\rangle .
\end{align}
Employing Eqs.~\eqref{Ukz2} and \eqref{eq:Ukdef} we write
\begin{align}
  \label{eq:B2}
  \left\langle \call^{a_0}(k_1) \, \call^{a_0}(k_2)
  \right\rangle/S_\perp = &\int\limits_{-\infty}^\infty d\nu \, d \nu'
  \, f(\nu) \, f (\nu') \,
  k_1^{-1 + 2 \, i \, \nu} \, k_2^{-1 + 2 \, i \, \nu'}\notag\\
  ~~~&\times \frac{1}{\as (e^{i \partial_\nu}) \, \as
    (e^{i \partial_\nu'})} {1\over S_\perp}\left\langle [{\cal
      U}^{\eta_0} (\nu)]^{\rm comp} \, [{\cal U}^{\eta_0} (\nu')]^{\rm
      comp} \right\rangle \, .
\end{align}
Using the result of \cite{Babansky:2002my} (or the calculation of
\cite{Balitsky:2009yp} modified for QCD) we write (cf. \eq{scadiponu})
\begin{align}
  \label{eq:B3}
  {1\over S_\perp}\langle [\calu^{a_{1}}(\nu)]^{\rm comp} \,
  [\calu^{a_{2}}(\nu')]^{\rm comp}\rangle = -{4 \, \pi^2(N^2_c -
    1)\over N^2_c} \, \as (e^{i \partial_\nu}) \, \as
  (e^{i \partial_\nu'}) \, {1 \over \nu^2 \, (1+4 \, \nu^2)^2} \,
  \delta(\nu+\nu') \, \Big( 1 + \bam \, F(\nu) \Big)
\end{align}
with ${\rm Re}[F(\nu)]$ given by \eq{Fnu} and the partial derivatives
acting on everything to their right. Plugging Eqs.~\eqref{eq:B3} and
\eqref{eq:B2} into \eq{eq:B1} we obtain
\begin{align}
  \label{eq:B4}
  \left\langle {\cal L}^{a_0} (\nu_1) \, {\cal L}^{a_0} (\nu_2)
  \right\rangle /S_\perp = \int\limits_0^\infty d k^2_1 H_{-\nu_1}
  (k_1) \, \int\limits_0^\infty d k^2_2 H_{-\nu_2} (k_2) \,
  \int\limits_{-\infty}^\infty d\nu \, k_1^{-1 + 2 \, i \, \nu} \,
  k_2^{-1 - 2 \, i \, \nu} \left[-{16 \, \pi^2(N^2_c - 1)\over
      N^2_c}\right] \, \left[ 1 + \bam \, F(\nu) \right].
\end{align}
With the order-$\as$ precision of our NLO calculation, the $\bam \,
F(\nu)$-term only multiplies the leading terms in functions $H_{-\nu}
(k)$, which are simply powers of momentum (see \eq{eigenf_n_nu}) and
can be easily integrated over $k_1$ and $k_2$. We thus write
\begin{align}
  \label{eq:B5}
  \left\langle {\cal L}^{a_0} (\nu_1) \, {\cal L}^{a_0} (\nu_2)
  \right\rangle /S_\perp = -{16 \, \pi^2(N^2_c - 1)\over N^2_c} \,
  \left[ 1 + \bam \, F(\nu_1) \right] \int\limits_0^\infty d k^2_1
  H_{-\nu_1} (k_1) \, \int\limits_0^\infty d k^2_2 H_{-\nu_2} (k_2) \,
  \int\limits_{-\infty}^\infty d\nu \, k_1^{-1 + 2 \, i \, \nu} \,
  k_2^{-1 - 2 \, i \, \nu} .
\end{align}
Making use of the following relation
\begin{align}
  \int\limits_{-\infty}^\infty d\nu \int\limits_0^\infty \, dk^2_1 \,
  dk^2_2 \, (k^2_1)^{-\half+i\nu} \, (k^2_2)^{-\half-i\nu} \,
  H_{-\nu_1}(k_1) \, H_{-\nu_2}(k_2) = (2\pi)^2 \,
  \delta(\nu_1+\nu_2),
\end{align}
which follows from completeness of powers and orthogonality of
functions $H_{-\nu} (k)$, we arrive at
\begin{align}
  \label{eq:B6}
  \left\langle {\cal L}^{a_0} (\nu_1) \, {\cal L}^{a_0} (\nu_2)
  \right\rangle /S_\perp = -{64 \, \pi^4 \, (N^2_c - 1)\over N^2_c} \,
  \left[ 1 + \bam \, F(\nu_1) \right] \, \delta(\nu_1+\nu_2),
\end{align}
which is precisely \eq{L0}. 

Substituting \eq{eq:B6} into \eq{NLO_A2} and integrating over $\nu_2$
yields
\begin{align}
  {\cal A}_{\lambda_1, \lambda_2} (q_1, q_2) = i \, 8 \, s \, \pi^2 \,
  \varepsilon^{\lambda_1 \, *}_{\rho_1} (q_1) \,
  \varepsilon^{\lambda_1}_{\sigma_1} (q_1) \, \varepsilon^{\lambda_2
    \, *}_{\rho_2} (q_2) \, \varepsilon^{\lambda_2}_{\sigma_2} (q_2)
  \, {N^2_c - 1 \over N^2_c} \! \int\limits_{-\infty}^\infty d \nu \,
  I^{\rho_1\sigma_1} (q_1,\nu) \, I^{\rho_2\sigma_2} (q_2, - \nu) \,
  e^{\frac{\Delta(\nu)}{2} \, \ln (a_1 \, a_2)} \notag \\ \times \,
  \left[ 1 + \bam \, F(\nu) \right].
\label{eq:B8}
\end{align}

Employing Eqs.~\eqref{eq:Iqnu_def} and \eqref{IF_Mellin} we write
\begin{align}
  \label{eq:B9}
  I^{\rho\sigma} (q,\nu) = \int \frac{d^2 k}{(2\pi)^2\,k^2} \, \as (k^2) \,
  H_\nu (k) \, \int\limits_{-\infty}^\infty d \nu' \, \left(
    \frac{k^2}{Q^2} \right)^{\half - i \, \nu'} \, {\tilde
    I}^{\rho\sigma}_{\rm LO+NLO} (q, \nu').
\end{align}
Using Eq. (\ref{eq:B9}) in Eq. (\ref{eq:B8}), keeping only terms up to
the right order, we obtain
\begin{align} 
{\cal A}_{\lambda_1, \lambda_2} (q_1, q_2) = & i \, 8 \,
  s \, \pi^2 \, \varepsilon^{\lambda_1 \, *}_{\rho_1} (q_1) \,
  \varepsilon^{\lambda_1}_{\sigma_1} (q_1) \, \varepsilon^{\lambda_2\,
    *}_{\rho_2} (q_2) \, \varepsilon^{\lambda_2}_{\sigma_2} (q_2) \,
  {N^2_c - 1 \over N^2_c}
  \nonumber\\
  &\times \int {d^2 k_1\over
    (2\pi)^2\,k^2_1}\alpha_s(k_1)\int\limits_{-\infty}^{+\infty} d\nu_1
  \left({k^2_1\over Q^2_1}\right)^{\half - i\nu_1}
  \tilde{I}^{\rho_1\sigma_1}_{\rm LO+NLO}(q_1,\nu_1)
  \nonumber\\
  &\times \int {d^2 k_2\over
    (2\pi)^2\,k^2_2}\alpha_s(k_2)\int\limits_{-\infty}^{+\infty} d\nu_2
  \left({k^2_2\over Q^2_2}\right)^{\half - i\nu_2}
  \tilde{I}^{\rho_2\sigma_2}_{\rm LO+NLO}(q_2,\nu_2)
  \nonumber\\
  &\times \int\limits_{-\infty}^{+\infty}d\nu \,
  e^{\frac{\Delta(\nu)}{2} \, \ln (a_1 \, a_2)}
  \bigg[H_\nu(k_1)H_{-\nu}(k_2) + \bam \, F(\nu) \, k_1^{-1+2 i
    \nu}k_2^{-1-2 i \nu}\bigg].
\end{align} 
We now use the solution for the Green function of the NLO BFKL
equation given in Eq. (\ref{NLOsol}) along with the trick of writing
$\ln k$ as $-i \, \partial_\nu$ acting on $k^{-1+2 i \nu}$ and the
orthogonality of the power-like LO BFKL eigenfunctions to arrive at
\begin{align}
  &{\cal A}_{\lambda_1, \lambda_2} (q_1, q_2) = i \, 2 \, s \, \pi^2
  \, \varepsilon^{\lambda_1 \, *}_{\rho_1} (q_1) \,
  \varepsilon^{\lambda_1}_{\sigma_1} (q_1) \, \varepsilon^{\lambda_2\,
    *}_{\rho_2} (q_2) \, \varepsilon^{\lambda_2}_{\sigma_2} (q_2) \,
  {N^2_c - 1 \over N^2_c} \,\bam^2 \int\limits_{-\infty}^{+\infty}
  d\nu[1+\bam F(\nu)]e^{\frac{\Delta(\nu)}{2} \, \ln (a_1 \, a_2)}
  \nonumber\\
  & \times \int\limits_{-\infty}^{+\infty} {d\nu_1\over
    Q_1^{1-2i\nu_1}}\tilde{I}^{\rho_1\sigma_1}_{\rm LO+NLO} (q_1,
  \nu_1) \Bigg[\Bigg(1-\bam\beta_2(i\partial_{\nu_1}-\ln\mu^2)
  -\bam^2\beta_2{\chi_0(\nu_1)\over 4}\ln (a_1 \,
  a_2)(i\partial_{\nu_1}-\ln\mu^2)\Bigg)\delta(\nu-\nu_1)\Bigg]
  \nonumber\\
  &\times \int\limits_{-\infty}^{+\infty} {d\nu_2\over
    Q_2^{1-2i\nu_2}}\tilde{I}^{\rho_2\sigma_2}_{\rm LO+NLO} (q_2,
  \nu_2) \Bigg[\Bigg(1-\bam\beta_2(i\partial_{\nu_2}-\ln\mu^2)
  -\bam^2\beta_2{\chi_0(\nu_2)\over 4}\ln (a_1 \,
  a_2)(i\partial_{\nu_2}-\ln\mu^2)\Bigg)\delta(\nu +\nu_2)\Bigg].
\end{align}
Carrying out integrations over $\nu_1$ and $\nu_2$ by parts to
eliminate the derivatives of the delta functions we obtain
\begin{align}
  &{\cal A}_{\lambda_1, \lambda_2} (q_1, q_2) = i \, 2 \, s \, \pi^2
  \, \varepsilon^{\lambda_1 \, *}_{\rho_1} (q_1) \,
  \varepsilon^{\lambda_1}_{\sigma_1} (q_1) \, \varepsilon^{\lambda_2\,
    *}_{\rho_2} (q_2) \, \varepsilon^{\lambda_2}_{\sigma_2} (q_2) \,
  {N^2_c - 1 \over N^2_c} \, \bam^2 \int\limits_{-\infty}^{+\infty}
  {d\nu\over Q_1Q_2}\,\left({Q_1^2\over Q_2^2}\right)^{i\nu}
  e^{\frac{\Delta(\nu)}{2} \, \ln (a_1 \, a_2)}
  \nonumber\\
  &\times \Bigg\{ \tilde{I}^{\rho_1\sigma_1}_{\rm
    LO+NLO}(q_1,\nu)\tilde{I}^{\rho_2\sigma_2}_{\rm LO+NLO}(q_2,-\nu)
  \Big(1+\bam \, F(\nu) \Big) -2 \, \bam \, \beta_2
  \tilde{I}^{\rho_1\sigma_1}_{\rm LO}(q_1,\nu) \,
  \tilde{I}^{\rho_2\sigma_2}_{\rm LO}(q_2, - \nu)\ln{Q_1Q_2\over
    \mu^2}
  \nonumber\\
  &-\bam^2\beta_2 \frac{\chi_0(\nu)}{2} \ln (a_1 \, a_2)
  \tilde{I}^{\rho_1\sigma_1}_{\rm LO}(q_1,\nu) \,
  \tilde{I}^{\rho_2\sigma_2} (q_2, - \nu)\ln{Q_1Q_2\over \mu^2}
  +i\bam\beta_2 \bigg[\tilde{I}^{\rho_2\sigma_2}_{\rm LO}(q_2, -\nu)
  \partial_\nu\tilde{I}^{\rho_1\sigma_1}_{\rm LO}(q_1,\nu)
  \nonumber\\
  & - \tilde{I}^{\rho_1\sigma_1}_{\rm LO}(q_1,\nu)
        \partial_\nu\tilde{I}^{\rho_2\sigma_2}_{\rm LO}(q_2,
        -\nu)\bigg] \left( 1 + \bam{\chi_0(\nu)\over 4} \ln (a_1 \,
          a_2) \right) \Bigg\}.
\end{align}
Finally, we exponantiate the terms proportional to $\bam\ln (a_1 \,
a_2)$ containing $\ln \mu^2$ and absorb the terms with $\beta_2$ into
the running coupling obtaining (with $\tilde{I}^{\rho\sigma} (\nu)
\equiv \tilde{I}^{\rho\sigma}(q,\nu)$ for brevity)
\begin{align}
      \label{eq:B14}
      & {\cal A}_{\lambda_1, \lambda_2} (q_1, q_2) = i \, 2 \, s \,
      \pi^2 \, \varepsilon^{\lambda_1 \, *}_{\rho_1} (q_1) \,
      \varepsilon^{\lambda_1}_{\sigma_1} (q_1) \,
      \varepsilon^{\lambda_2 \, *}_{\rho_2} (q_2) \,
      \varepsilon^{\lambda_2}_{\sigma_2} (q_2) \, \frac{\as (Q_1^2) \,
        \as (Q_2^2)}{Q_1 \, Q_2} \, {N^2_c - 1 \over N^2_c} \!
      \int\limits_{-\infty}^\infty d \nu \, \left( \frac{Q_1^2}{Q_2^2}
      \right)^{i \, \nu} \\ & \times \, e^{\left[ \bas (Q_1 \, Q_2) \,
          \chi_0 (\nu) + \bas^2 (Q_1 \, Q_2) \, \chi_1 (\nu) \right]
        \, \half \, \ln (a_1 \, a_2)} \, \Bigg\{ {\tilde I}_{\rm
        LO+NLO}^{\rho_1\sigma_1} (\nu) \, {\tilde I}_{\rm
        LO+NLO}^{\rho_2\sigma_2} (- \nu) \, \Big( 1 + \bas (Q_1 \,
      Q_2) \, F(\nu) \Big) \notag \\ & + i \, \bam \, \beta_2 \,
      \left( {\tilde I}_{\rm LO}^{\rho_2\sigma_2} (\nu)
        \, \partial_{\nu} {\tilde I}_{\rm LO}^{\rho_1\sigma_1} (\nu) -
        {\tilde I}_{\rm LO}^{\rho_1\sigma_1} (\nu) \, \partial_{\nu}
        {\tilde I}_{\rm LO}^{\rho_2\sigma_2} (\nu)\right) \, \left( 1
        + \bas (Q_1 \, Q_2) \, \frac{\chi_0 (\nu)}{4} \, \ln (a_1 \,
        a_2) \right) \Bigg\}, \notag
\end{align}
which is exactly \eq{gg-analytic} with the NLO accuracy.



\begin{thebibliography}{10}

\bibitem{Balitsky:2001gj}
I.~Balitsky, {\it {High-energy QCD and Wilson lines}},
  \href{http://xxx.lanl.gov/abs/hep-ph/0101042}{{\tt hep-ph/0101042}}.

\bibitem{Jalilian-Marian:2005jf}
J.~Jalilian-Marian and Y.~V. Kovchegov, {\it Saturation physics and deuteron
  gold collisions at {RHIC}},  {\em Prog. Part. Nucl. Phys.} {\bf 56} (2006)
  104--231, [\href{http://xxx.lanl.gov/abs/hep-ph/0505052}{{\tt
  hep-ph/0505052}}].

\bibitem{Weigert:2005us}
H.~Weigert, {\it Evolution at small {$x_{\text{bj}}$: The Color Glass
  Condensate}},  {\em Prog. Part. Nucl. Phys.} {\bf 55} (2005) 461--565,
  [\href{http://xxx.lanl.gov/abs/hep-ph/0501087}{{\tt hep-ph/0501087}}].

\bibitem{Iancu:2003xm}
E.~Iancu and R.~Venugopalan, {\it The color glass condensate and high energy
  scattering in {QCD}},  \href{http://xxx.lanl.gov/abs/hep-ph/0303204}{{\tt
  hep-ph/0303204}}.

\bibitem{Gelis:2010nm}
F.~Gelis, E.~Iancu, J.~Jalilian-Marian, and R.~Venugopalan, {\it {The Color
  Glass Condensate}},  {\em Ann.Rev.Nucl.Part.Sci.} {\bf 60} (2010) 463--489,
  [\href{http://xxx.lanl.gov/abs/1002.0333}{{\tt arXiv:1002.0333}}].

\bibitem{KovchegovLevin}
Y.~V. Kovchegov and E.~Levin, {\em Quantum Chromodynamics at High Energy}.
\newblock Cambridge University Press, 2012.

\bibitem{Gribov:1984tu}
L.~V. Gribov, E.~M. Levin, and M.~G. Ryskin, {\it {Semihard Processes in QCD}},
   {\em Phys. Rept.} {\bf 100} (1983) 1--150.

\bibitem{Mueller:1989st}
A.~H. Mueller, {\it {Small x Behavior and Parton Saturation: A QCD Model}},
  {\em Nucl. Phys.} {\bf B335} (1990) 115.

\bibitem{McLerran:1993ni}
L.~D. McLerran and R.~Venugopalan, {\it Computing quark and gluon distribution
  functions for very large nuclei},  {\em Phys. Rev.} {\bf D49} (1994)
  2233--2241, [\href{http://xxx.lanl.gov/abs/hep-ph/9309289}{{\tt
  hep-ph/9309289}}].

\bibitem{McLerran:1993ka}
L.~D. McLerran and R.~Venugopalan, {\it Gluon distribution functions for very
  large nuclei at small transverse momentum},  {\em Phys. Rev.} {\bf D49}
  (1994) 3352--3355, [\href{http://xxx.lanl.gov/abs/hep-ph/9311205}{{\tt
  hep-ph/9311205}}].

\bibitem{McLerran:1994vd}
L.~D. McLerran and R.~Venugopalan, {\it Green's functions in the color field of
  a large nucleus},  {\em Phys. Rev.} {\bf D50} (1994) 2225--2233,
  [\href{http://xxx.lanl.gov/abs/hep-ph/9402335}{{\tt hep-ph/9402335}}].

\bibitem{Balitsky:1996ub}
I.~Balitsky, {\it Operator expansion for high-energy scattering},  {\em Nucl.
  Phys.} {\bf B463} (1996) 99--160,
  [\href{http://xxx.lanl.gov/abs/hep-ph/9509348}{{\tt hep-ph/9509348}}].

\bibitem{Balitsky:1998ya}
I.~Balitsky, {\it Factorization and high-energy effective action},  {\em Phys.
  Rev.} {\bf D60} (1999) 014020,
  [\href{http://xxx.lanl.gov/abs/hep-ph/9812311}{{\tt hep-ph/9812311}}].

\bibitem{Kovchegov:1999yj}
Y.~V. Kovchegov, {\it Small-x {$F_2$} structure function of a nucleus including
  multiple pomeron exchanges},  {\em Phys. Rev.} {\bf D60} (1999) 034008,
  [\href{http://xxx.lanl.gov/abs/hep-ph/9901281}{{\tt hep-ph/9901281}}].

\bibitem{Kovchegov:1999ua}
Y.~V. Kovchegov, {\it Unitarization of the {BFKL} pomeron on a nucleus},  {\em
  Phys. Rev.} {\bf D61} (2000) 074018,
  [\href{http://xxx.lanl.gov/abs/hep-ph/9905214}{{\tt hep-ph/9905214}}].

\bibitem{Jalilian-Marian:1997dw}
J.~Jalilian-Marian, A.~Kovner, and H.~Weigert, {\it The {Wilson}
  renormalization group for low x physics: Gluon evolution at finite parton
  density},  {\em Phys. Rev.} {\bf D59} (1998) 014015,
  [\href{http://xxx.lanl.gov/abs/hep-ph/9709432}{{\tt hep-ph/9709432}}].

\bibitem{Jalilian-Marian:1997gr}
J.~Jalilian-Marian, A.~Kovner, A.~Leonidov, and H.~Weigert, {\it The {Wilson}
  renormalization group for low x physics: Towards the high density regime},
  {\em Phys. Rev.} {\bf D59} (1998) 014014,
  [\href{http://xxx.lanl.gov/abs/hep-ph/9706377}{{\tt hep-ph/9706377}}].

\bibitem{Iancu:2001ad}
E.~Iancu, A.~Leonidov, and L.~D. McLerran, {\it {The renormalization group
  equation for the color glass condensate}},  {\em Phys. Lett.} {\bf B510}
  (2001) 133--144.

\bibitem{Iancu:2000hn}
E.~Iancu, A.~Leonidov, and L.~D. McLerran, {\it Nonlinear gluon evolution in
  the color glass condensate. {I}},  {\em Nucl. Phys.} {\bf A692} (2001)
  583--645, [\href{http://xxx.lanl.gov/abs/hep-ph/0011241}{{\tt
  hep-ph/0011241}}].

\bibitem{Albacete:2010sy}
J.~L. Albacete, N.~Armesto, J.~G. Milhano, P.~Quiroga-Arias, and C.~A. Salgado,
  {\it {AAMQS: A non-linear QCD analysis of new HERA data at small-x including
  heavy quarks}},  {\em Eur. Phys. J.} {\bf C71} (2011) 1705,
  [\href{http://xxx.lanl.gov/abs/1012.4408}{{\tt arXiv:1012.4408}}].

\bibitem{ALbacete:2010ad}
J.~L. Albacete and A.~Dumitru, {\it {A model for gluon production in heavy-ion
  collisions at the LHC with rcBK unintegrated gluon densities}},
  \href{http://xxx.lanl.gov/abs/1011.5161}{{\tt arXiv:1011.5161}}.

\bibitem{Lappi:2013zma}
T.~Lappi and {H. M\"{a}ntysaari}, {\it {Single inclusive particle production at
  high energy from HERA data to proton-nucleus collisions}},
  \href{http://xxx.lanl.gov/abs/1309.6963}{{\tt arXiv:1309.6963}}.

\bibitem{Balitsky:2006wa}
I.~I. Balitsky, {\it {Quark Contribution to the Small-$x$ Evolution of Color
  Dipole}},  {\em Phys. Rev. D} {\bf 75} (2007) 014001,
  [\href{http://xxx.lanl.gov/abs/hep-ph/0609105}{{\tt hep-ph/0609105}}].

\bibitem{Gardi:2006rp}
E.~Gardi, J.~Kuokkanen, K.~Rummukainen, and H.~Weigert, {\it Running coupling
  and power corrections in nonlinear evolution at the high-energy limit},  {\em
  Nucl. Phys.} {\bf A784} (2007) 282--340,
  [\href{http://xxx.lanl.gov/abs/hep-ph/0609087}{{\tt hep-ph/0609087}}].

\bibitem{Kovchegov:2006vj}
Y.~Kovchegov and H.~Weigert, {\it {Triumvirate of Running Couplings in
  Small-$x$ Evolution}},  {\em Nucl. Phys. {\bf A}} {\bf 784} (2007) 188--226,
  [\href{http://xxx.lanl.gov/abs/hep-ph/0609090}{{\tt hep-ph/0609090}}].

\bibitem{Kovchegov:2006wf}
Y.~V. Kovchegov and H.~Weigert, {\it {Quark loop contribution to BFKL
  evolution: Running coupling and leading-N(f) NLO intercept}},  {\em Nucl.
  Phys.} {\bf A789} (2007) 260--284,
  [\href{http://xxx.lanl.gov/abs/hep-ph/0612071}{{\tt hep-ph/0612071}}].

\bibitem{Albacete:2007yr}
J.~L. Albacete and Y.~V. Kovchegov, {\it Solving high energy evolution equation
  including running coupling corrections},  {\em Phys. Rev.} {\bf D75} (2007)
  125021, [\href{http://xxx.lanl.gov/abs/0704.0612}{{\tt 0704.0612}}].

\bibitem{Balitsky:2008zz}
I.~Balitsky and G.~A. Chirilli, {\it {Next-to-leading order evolution of color
  dipoles}},  {\em Phys. Rev.} {\bf D77} (2008) 014019,
  [\href{http://xxx.lanl.gov/abs/0710.4330}{{\tt arXiv:0710.4330}}].

\bibitem{Balitsky:2013fea}
I.~Balitsky and G.~A. Chirilli, {\it {Rapidity evolution of Wilson lines at the
  next-to-leading order}},  {\em Phys.Rev.} {\bf D88} (2013) 111501,
  [\href{http://xxx.lanl.gov/abs/1309.7644}{{\tt arXiv:1309.7644}}].

\bibitem{Kovner:2013ona}
A.~Kovner, M.~Lublinsky, and Y.~Mulian, {\it {Complete JIMWLK Evolution at
  NLO}},  \href{http://xxx.lanl.gov/abs/1310.0378}{{\tt arXiv:1310.0378}}.

\bibitem{Kuraev:1977fs}
E.~A. Kuraev, L.~N. Lipatov, and V.~S. Fadin, {\it {The Pomeranchuk
  singlularity in non-Abelian gauge theories}},  {\em Sov. Phys. JETP} {\bf 45}
  (1977) 199--204.

\bibitem{Kuraev:1976ge}
E.~A. Kuraev, L.~N. Lipatov, and V.~S. Fadin, {\it Multi - reggeon processes in
  the yang-mills theory},  {\em Sov. Phys. JETP} {\bf 44} (1976) 443--450.

\bibitem{Balitsky:1978ic}
I.~Balitsky and L.~Lipatov, {\it {The Pomeranchuk Singularity in Quantum
  Chromodynamics}},  {\em Sov.J.Nucl.Phys.} {\bf 28} (1978) 822--829.

\bibitem{Fadin:1998py}
V.~S. Fadin and L.~N. Lipatov, {\it {BFKL} pomeron in the next-to-leading
  approximation},  {\em Phys. Lett.} {\bf B429} (1998) 127--134,
  [\href{http://xxx.lanl.gov/abs/hep-ph/9802290}{{\tt hep-ph/9802290}}].

\bibitem{Ciafaloni:1998gs}
M.~Ciafaloni and G.~Camici, {\it {Energy scale(s) and next-to-leading BFKL
  equation}},  {\em Phys. Lett.} {\bf B430} (1998) 349--354,
  [\href{http://xxx.lanl.gov/abs/hep-ph/9803389}{{\tt hep-ph/9803389}}].

\bibitem{Kovchegov:1998ae}
Y.~V. Kovchegov and A.~H. Mueller, {\it {Running coupling effects in BFKL
  evolution}},  {\em Phys. Lett.} {\bf B439} (1998) 428--436,
  [\href{http://xxx.lanl.gov/abs/hep-ph/9805208}{{\tt hep-ph/9805208}}].

\bibitem{Levin:1998pka}
E.~Levin, {\it {The BFKL high-energy asymptotics in the next-to-leading
  approximation}},  \href{http://xxx.lanl.gov/abs/hep-ph/9806228}{{\tt
  hep-ph/9806228}}.

\bibitem{Armesto:1998gt}
N.~Armesto, J.~Bartels, and M.~Braun, {\it {On the second order corrections to
  the hard pomeron and the running coupling}},  {\em Phys.Lett.} {\bf B442}
  (1998) 459--469, [\href{http://xxx.lanl.gov/abs/hep-ph/9808340}{{\tt
  hep-ph/9808340}}].

\bibitem{Ciafaloni:2001db}
M.~Ciafaloni, M.~Taiuti, and A.~H. Mueller, {\it {Diffusion corrections to the
  hard pomeron}},  {\em Nucl.Phys.} {\bf B616} (2001) 349--366,
  [\href{http://xxx.lanl.gov/abs/hep-ph/0107009}{{\tt hep-ph/0107009}}].

\bibitem{Ross:1998xw}
D.~A. Ross, {\it {The effect of higher order corrections to the BFKL equation
  on the perturbative pomeron}},  {\em Phys. Lett.} {\bf B431} (1998) 161--165,
  [\href{http://xxx.lanl.gov/abs/hep-ph/9804332}{{\tt hep-ph/9804332}}].

\bibitem{Andersen:2003an}
J.~R. Andersen and A.~Sabio~Vera, {\it {Solving the BFKL equation in the
  next-to-leading approximation}},  {\em Phys.Lett.} {\bf B567} (2003)
  116--124, [\href{http://xxx.lanl.gov/abs/hep-ph/0305236}{{\tt
  hep-ph/0305236}}].

\bibitem{Andersen:2003wy}
J.~R. Andersen and A.~Sabio~Vera, {\it {The gluon Green's function in the BFKL
  approach at next-to-leading logarithmic accuracy}},  {\em Nucl.Phys.} {\bf
  B679} (2004) 345--362, [\href{http://xxx.lanl.gov/abs/hep-ph/0309331}{{\tt
  hep-ph/0309331}}].

\bibitem{Salam:1998tj}
G.~P. Salam, {\it A resummation of large sub-leading corrections at small x},
  {\em JHEP} {\bf 07} (1998) 019,
  [\href{http://xxx.lanl.gov/abs/hep-ph/9806482}{{\tt hep-ph/9806482}}].

\bibitem{Ciafaloni:1999yw}
M.~Ciafaloni, D.~Colferai, and G.~P. Salam, {\it Renormalization group improved
  small-x equation},  {\em Phys. Rev.} {\bf D60} (1999) 114036,
  [\href{http://xxx.lanl.gov/abs/hep-ph/9905566}{{\tt hep-ph/9905566}}].

\bibitem{Ciafaloni:2003rd}
M.~Ciafaloni, D.~Colferai, G.~Salam, and A.~Stasto, {\it {Renormalization group
  improved small x Green's function}},  {\em Phys.Rev.} {\bf D68} (2003)
  114003, [\href{http://xxx.lanl.gov/abs/hep-ph/0307188}{{\tt
  hep-ph/0307188}}].

\bibitem{Chirilli:2013kca}
G.~A. Chirilli and Y.~V. Kovchegov, {\it {Solution of the NLO BFKL Equation and
  a Strategy for Solving the All-Order BFKL Equation}},  {\em JHEP} {\bf 1306}
  (2013) 055, [\href{http://xxx.lanl.gov/abs/1305.1924}{{\tt
  arXiv:1305.1924}}].

\bibitem{Grabovsky:2013gta}
A.~Grabovsky, {\it {On the solution to the NLO forward BFKL equation}},  {\em
  JHEP} {\bf 1309} (2013) 098, [\href{http://xxx.lanl.gov/abs/1307.3152}{{\tt
  arXiv:1307.3152}}].

\bibitem{Lipatov:1985uk}
L.~Lipatov, {\it {The Bare Pomeron in Quantum Chromodynamics}},  {\em
  Sov.Phys.JETP} {\bf 63} (1986) 904--912.

\bibitem{Kotikov:2000pm}
A.~Kotikov and L.~Lipatov, {\it {NLO corrections to the BFKL equation in QCD
  and in supersymmetric gauge theories}},  {\em Nucl.Phys.} {\bf B582} (2000)
  19--43, [\href{http://xxx.lanl.gov/abs/hep-ph/0004008}{{\tt
  hep-ph/0004008}}].

\bibitem{Balitsky:2008rc}
I.~Balitsky and G.~A. Chirilli, {\it {Conformal kernel for NLO BFKL equation in
  N = 4 SYM}},  {\em Phys.Rev.} {\bf D79} (2009) 031502,
  [\href{http://xxx.lanl.gov/abs/0812.3416}{{\tt arXiv:0812.3416}}].

\bibitem{Balitsky:2009xg}
I.~Balitsky and G.~A. Chirilli, {\it {NLO evolution of color dipoles in N=4
  SYM}},  {\em Nucl.Phys.} {\bf B822} (2009) 45--87,
  [\href{http://xxx.lanl.gov/abs/0903.5326}{{\tt arXiv:0903.5326}}].

\bibitem{Balitsky:2009yp}
I.~Balitsky and G.~A. Chirilli, {\it {High-energy amplitudes in N=4 SYM in the
  next-to-leading order}},  {\em Int. J. Mod. Phys.} {\bf A25} (2010) 401--410,
  [\href{http://xxx.lanl.gov/abs/0911.5192}{{\tt arXiv:0911.5192}}].

\bibitem{Dokshitzer:1977sg}
Y.~L. Dokshitzer, {\it {Calculation of the Structure Functions for Deep
  Inelastic Scattering and $e^+ e^-$ Annihilation by Perturbation Theory in
  Quantum Chromodynamics}},  {\em Sov. Phys. JETP} {\bf 46} (1977) 641--653.

\bibitem{Gribov:1972ri}
V.~N. Gribov and L.~N. Lipatov, {\it {Deep inelastic e p scattering in
  perturbation theory}},  {\em Sov. J. Nucl. Phys.} {\bf 15} (1972) 438--450.

\bibitem{Altarelli:1977zs}
G.~Altarelli and G.~Parisi, {\it {Asymptotic Freedom in Parton Language}},
  {\em Nucl. Phys.} {\bf B126} (1977) 298.

\bibitem{Gross:1973id}
D.~J. Gross and F.~Wilczek, {\it {Ultraviolet Behavior of Non-Abelian Gauge
  Theories}},  {\em Phys. Rev. Lett.} {\bf 30} (1973) 1343--1346.

\bibitem{Georgi:1951sr}
H.~Georgi and H.~D. Politzer, {\it {Electroproduction scaling in an
  asymptotically free theory of strong interactions}},  {\em Phys.Rev.} {\bf
  D9} (1974) 416--420.

\bibitem{Christ:1972ms}
N.~H. Christ, B.~Hasslacher, and A.~H. Mueller, {\it {Light cone behavior of
  perturbation theory}},  {\em Phys.Rev.} {\bf D6} (1972) 3543.

\bibitem{Floratos:1977au}
E.~Floratos, D.~Ross, and C.~T. Sachrajda, {\it {Higher Order Effects in
  Asymptotically Free Gauge Theories: The Anomalous Dimensions of Wilson
  Operators}},  {\em Nucl.Phys.} {\bf B129} (1977) 66--88.

\bibitem{Vogt:2004mw}
A.~Vogt, S.~Moch, and J.~Vermaseren, {\it {The Three-loop splitting functions
  in QCD: The Singlet case}},  {\em Nucl.Phys.} {\bf B691} (2004) 129--181,
  [\href{http://xxx.lanl.gov/abs/hep-ph/0404111}{{\tt hep-ph/0404111}}].

\bibitem{Balitsky:2012bs}
I.~Balitsky and G.~A. Chirilli, {\it {Photon impact factor and
  $k_T$-factorization for DIS in the next-to-leading order}},  {\em Phys.Rev.}
  {\bf D87} (2013) 014013, [\href{http://xxx.lanl.gov/abs/1207.3844}{{\tt
  arXiv:1207.3844}}].

\bibitem{Babansky:2002my}
A.~Babansky and I.~Balitsky, {\it Scattering of color dipoles: From low to high
  energies},  {\em Phys. Rev.} {\bf D67} (2003) 054026,
  [\href{http://xxx.lanl.gov/abs/hep-ph/0212075}{{\tt hep-ph/0212075}}].

\bibitem{Balitsky:2010ze}
I.~Balitsky and G.~A. Chirilli, {\it {Photon impact factor in the
  next-to-leading order}},  {\em Phys.Rev.} {\bf D83} (2011) 031502,
  [\href{http://xxx.lanl.gov/abs/1009.4729}{{\tt arXiv:1009.4729}}].

\bibitem{Balitsky:2008zza}
I.~Balitsky and G.~A. Chirilli, {\it {Next-to-leading order evolution of color
  dipoles}},  {\em Phys.Rev.} {\bf D77} (2008) 014019,
  [\href{http://xxx.lanl.gov/abs/0710.4330}{{\tt arXiv:0710.4330}}].

\bibitem{Bartels:2004bi}
J.~Bartels and A.~Kyrieleis, {\it {NLO corrections to the $\gamma^*$ impact
  factor: First numerical results for the real corrections to $\gamma^*_L$}},
  {\em Phys.Rev.} {\bf D70} (2004) 114003,
  [\href{http://xxx.lanl.gov/abs/hep-ph/0407051}{{\tt hep-ph/0407051}}].

\bibitem{Bartels:2002uz}
J.~Bartels, D.~Colferai, S.~Gieseke, and A.~Kyrieleis, {\it {NLO corrections to
  the photon impact factor: Combining real and virtual corrections}},  {\em
  Phys.Rev.} {\bf D66} (2002) 094017,
  [\href{http://xxx.lanl.gov/abs/hep-ph/0208130}{{\tt hep-ph/0208130}}].

\bibitem{Bartels:2001mv}
J.~Bartels, S.~Gieseke, and A.~Kyrieleis, {\it {The Process $\gamma^*_L + q \to
  (q {\bar q} g) + q$: Real corrections to the virtual photon impact factor}},
  {\em Phys.Rev.} {\bf D65} (2002) 014006,
  [\href{http://xxx.lanl.gov/abs/hep-ph/0107152}{{\tt hep-ph/0107152}}].

\bibitem{Fadin:1998hr}
V.~S. Fadin, R.~Fiore, A.~Flachi, and M.~I. Kotsky, {\it Quark-antiquark
  contribution to the {BFKL} kernel},  {\em Phys. Lett.} {\bf B422} (1998)
  287--293, [\href{http://xxx.lanl.gov/abs/hep-ph/9711427}{{\tt
  hep-ph/9711427}}].

\bibitem{Costa:2013zra}
M.~S. Costa, J.~Drummond, V.~Goncalves, and J.~Penedones, {\it {The role of
  leading twist operators in the Regge and Lorentzian OPE limits}},
  \href{http://xxx.lanl.gov/abs/1311.4886}{{\tt arXiv:1311.4886}}.

\bibitem{Korchemsky:1987wg}
G.~P. Korchemsky and A.~V. Radyushkin, {\it Renormalization of the {Wilson}
  loops beyond the leading order},  {\em Nucl. Phys.} {\bf B283} (1987)
  342--364.

\bibitem{Korchemskaya:1994qp}
I.~Korchemskaya and G.~Korchemsky, {\it {High-energy scattering in QCD and
  cross singularities of Wilson loops}},  {\em Nucl.Phys.} {\bf B437} (1995)
  127--162, [\href{http://xxx.lanl.gov/abs/hep-ph/9409446}{{\tt
  hep-ph/9409446}}].

\bibitem{Kovchegov:2007vf}
Y.~V. Kovchegov and H.~Weigert, {\it {Collinear Singularities and Running
  Coupling Corrections to Gluon Production in CGC}},  {\em Nucl. Phys.} {\bf
  A807} (2008) 158--189, [\href{http://xxx.lanl.gov/abs/0712.3732}{{\tt
  arXiv:0712.3732}}].

\bibitem{Brodsky:1983gc}
S.~J. Brodsky, G.~P. Lepage, and P.~B. Mackenzie, {\it On the elimination of
  scale ambiguities in perturbative quantum chromodynamics},  {\em Phys. Rev.}
  {\bf D28} (1983) 228.

\bibitem{Caporale:2008is}
F.~Caporale, D.~Y. Ivanov, and A.~Papa, {\it {BFKL resummation effects in the
  $\gamma^* \gamma^*$ total hadronic cross section}},  {\em Eur.Phys.J.} {\bf
  C58} (2008) 1--7, [\href{http://xxx.lanl.gov/abs/0807.3231}{{\tt
  arXiv:0807.3231}}].

\bibitem{Ivanov:2006gt}
D.~Y. Ivanov and A.~Papa, {\it {Electroproduction of two light vector mesons in
  next-to-leading BFKL: Study of systematic effects}},  {\em Eur.Phys.J.} {\bf
  C49} (2007) 947--955, [\href{http://xxx.lanl.gov/abs/hep-ph/0610042}{{\tt
  hep-ph/0610042}}].

\bibitem{Ivanov:2005gn}
D.~Y. Ivanov and A.~Papa, {\it {Electroproduction of two light vector mesons in
  the next-to-leading approximation}},  {\em Nucl.Phys.} {\bf B732} (2006)
  183--199, [\href{http://xxx.lanl.gov/abs/hep-ph/0508162}{{\tt
  hep-ph/0508162}}].

\bibitem{Mueller:1994jq}
A.~H. Mueller and B.~Patel, {\it Single and double {BFKL} pomeron exchange and
  a dipole picture of high-energy hard processes},  {\em Nucl. Phys.} {\bf
  B425} (1994) 471--488, [\href{http://xxx.lanl.gov/abs/hep-ph/9403256}{{\tt
  hep-ph/9403256}}].

\bibitem{Mueller:1995gb}
A.~H. Mueller, {\it Unitarity and the {BFKL} pomeron},  {\em Nucl. Phys.} {\bf
  B437} (1995) 107--126, [\href{http://xxx.lanl.gov/abs/hep-ph/9408245}{{\tt
  hep-ph/9408245}}].

\bibitem{Braun:1995hh}
M.~Braun, {\it {The Bootstrap condition for many Reggeized gluons and the
  photon structure function at low $x$ and $N_c \to \infty$}},  {\em Z.Phys.}
  {\bf C71} (1996) 601--612,
  [\href{http://xxx.lanl.gov/abs/hep-ph/9502403}{{\tt hep-ph/9502403}}].

\bibitem{Salam:1995uy}
G.~Salam, {\it {Studies of unitarity at small $x$ using the dipole
  formulation}},  {\em Nucl.Phys.} {\bf B461} (1996) 512--538,
  [\href{http://xxx.lanl.gov/abs/hep-ph/9509353}{{\tt hep-ph/9509353}}].

\bibitem{Mueller:1996te}
A.~H. Mueller and G.~Salam, {\it {Large multiplicity fluctuations and
  saturation effects in onium collisions}},  {\em Nucl.Phys.} {\bf B475} (1996)
  293--320, [\href{http://xxx.lanl.gov/abs/hep-ph/9605302}{{\tt
  hep-ph/9605302}}].

\bibitem{Bartels:1980pe}
J.~Bartels, {\it {High-Energy Behavior in a Nonabelian Gauge Theory. 2. First
  Corrections to $T(n \to m)$ Beyond the Leading LNS Approximation}},  {\em
  Nucl.Phys.} {\bf B175} (1980) 365.

\bibitem{Kwiecinski:1980wb}
J.~Kwiecinski and M.~Praszalowicz, {\it {Three Gluon Integral Equation and Odd
  c Singlet Regge Singularities in QCD}},  {\em Phys.Lett.} {\bf B94} (1980)
  413.

\bibitem{JalilianMarian:2004da}
J.~Jalilian-Marian and Y.~V. Kovchegov, {\it {Inclusive two-gluon and valence
  quark-gluon production in DIS and p A}},  {\em Phys. Rev.} {\bf D70} (2004)
  114017, [\href{http://xxx.lanl.gov/abs/hep-ph/0405266}{{\tt
  hep-ph/0405266}}].

\bibitem{Dominguez:2011gc}
F.~Dominguez, A.~Mueller, S.~Munier, and B.-W. Xiao, {\it {On the small-x
  evolution of the color quadrupole and the Weizs\'acker-Williams gluon
  distribution}},  {\em Phys.Lett.} {\bf B705} (2011) 106--111,
  [\href{http://xxx.lanl.gov/abs/1108.1752}{{\tt arXiv:1108.1752}}].

\bibitem{Altinoluk:2013rua}
T.~Altinoluk, C.~Contreras, A.~Kovner, E.~Levin, M.~Lublinsky, {\em et.~al.},
  {\it {QCD Reggeon Calculus From KLWMIJ/JIMWLK Evolution: Vertices,
  Reggeization and All}},  {\em JHEP} {\bf 1309} (2013) 115,
  [\href{http://xxx.lanl.gov/abs/1306.2794}{{\tt arXiv:1306.2794}}].

\bibitem{Mueller:2002zm}
A.~H. Mueller and D.~N. Triantafyllopoulos, {\it The energy dependence of the
  saturation momentum},  {\em Nucl. Phys.} {\bf B640} (2002) 331--350,
  [\href{http://xxx.lanl.gov/abs/hep-ph/0205167}{{\tt hep-ph/0205167}}].

\bibitem{Kharzeev:2003wz}
D.~Kharzeev, Y.~V. Kovchegov, and K.~Tuchin, {\it Cronin effect and high-p(t)
  suppression in p a collisions},  {\em Phys. Rev.} {\bf D68} (2003) 094013,
  [\href{http://xxx.lanl.gov/abs/hep-ph/0307037}{{\tt hep-ph/0307037}}].

\bibitem{Brower:2006ea}
R.~C. Brower, J.~Polchinski, M.~J. Strassler, and C.-I. Tan, {\it {The Pomeron
  and Gauge/String Duality}},  {\em JHEP} {\bf 12} (2007) 005,
  [\href{http://xxx.lanl.gov/abs/hep-th/0603115}{{\tt hep-th/0603115}}].

\bibitem{Janik:1999zk}
R.~A. Janik and R.~B. Peschanski, {\it {High energy scattering and the AdS/CFT
  correspondence}},  {\em Nucl. Phys.} {\bf B565} (2000) 193--209,
  [\href{http://xxx.lanl.gov/abs/hep-th/9907177}{{\tt hep-th/9907177}}].

\bibitem{Maldacena:1997re}
J.~M. Maldacena, {\it The large {N} limit of superconformal field theories and
  supergravity},  {\em Adv. Theor. Math. Phys.} {\bf 2} (1998) 231--252,
  [\href{http://xxx.lanl.gov/abs/hep-th/9711200}{{\tt hep-th/9711200}}].

\bibitem{Witten:1998qj}
E.~Witten, {\it Anti-de sitter space and holography},  {\em Adv. Theor. Math.
  Phys.} {\bf 2} (1998) 253--291,
  [\href{http://xxx.lanl.gov/abs/hep-th/9802150}{{\tt hep-th/9802150}}].

\bibitem{Marzani:2007gk}
S.~Marzani, R.~D. Ball, P.~Falgari, and S.~Forte, {\it {BFKL at
  next-to-next-to-leading order}},  {\em Nucl.Phys.} {\bf B783} (2007)
  143--175, [\href{http://xxx.lanl.gov/abs/0704.2404}{{\tt arXiv:0704.2404}}].

\bibitem{Ball:2005mj}
R.~D. Ball and S.~Forte, {\it {All order running coupling BFKL evolution from
  GLAP (and vice-versa)}},  {\em Nucl.Phys.} {\bf B742} (2006) 158--175,
  [\href{http://xxx.lanl.gov/abs/hep-ph/0601049}{{\tt hep-ph/0601049}}].

\end{thebibliography}

\providecommand{\href}[2]{#2}\begingroup\raggedright\endgroup

\end{document}